\documentclass[12pt]{article}
\usepackage{putex}
\usepackage[pdftex]{graphicx}
\numberwithin{equation}{section}
\usepackage{cite}
\usepackage{hyperref}

\def\be{\begin{equation}}
\def\ee{\end{equation}}
\def\ba{\begin{eqnarray}}
\def\ea{\end{eqnarray}}

\begin{document}

\preprint{PUPT-2488\\UTTG-22-15\\TCC-010-15}
\title{Perturbations of vortex ring pairs}
\authors{Steven S. Gubser,$^\PU$ Bart Horn,$^{\UT,\CU}$ and Sarthak Parikh$^\PU$}
\institution{PU}{${}^1$Joseph Henry Laboratories, Princeton University, Princeton, NJ 08544, USA}
\institution{UT}{${}^2$Theory Group and Texas Cosmology Center, Department of Physics, \cr
University of Texas at Austin, Austin, TX 78712, USA} 
\institution{CU}{${}^3$Theory Group and Institute for Strings, Cosmology, and Astroparticle Physics, \cr  Department of Physics, Columbia University, New York, NY 10027, USA}

\abstract{We study pairs of co-axial vortex rings starting from the action for a classical bosonic string in a three-form background.  We complete earlier work on the phase diagram of classical orbits by explicitly considering the case where the circulations of the two vortex rings are equal and opposite.  We then go on to study perturbations, focusing on cases where the relevant four-dimensional transfer matrix splits into two-dimensional blocks.  When the circulations of the rings have the same sign, instabilities are mostly limited to wavelengths smaller than a dynamically generated length scale at which single-ring instabilities occur.  When the circulations have the opposite sign, larger wavelength instabilities can occur.}

\maketitle

\tableofcontents

\section{Introduction}
\label{INTRODUCTION}

Vortices are important excitations in fluid systems in two or more spatial dimensions. Diverse physical phenomena in nature admit a description in terms of quantized vortices in superfluids (see, for example~\cite{Huang:2015gza}).  In three spatial dimensions, which is our primary interest, vortices are extended in one spatial direction, i.e., they are string-like.  Recent work including \cite{Endlich:2013dma,Gubser:2014yma,Horn:2015zna}, following lines similar to \cite{Lund:1976ze}, has developed the view that methods of classical string theory can be usefully adapted to the study of vortex rings.  Salient features of this literature of particular relevance to the current work are:
 \begin{itemize}
  \item The background geometry is flat, but there is a constant Neveu-Schwarz (NS) three-form field strength $H_3 = dB_2$ with all its legs in spatial directions.  This field strength is dual to a density of the fluid or superfluid in which the vortices form.
  \item Vortices are treated in the limit where their motion is slow compared to the speed of sound.  Formally, this is achieved by taking a non-relativistic limit of classical string theory.
  \item Vortex motion is non-inertial.  The Lagrangian, in the limit of interest, is first-order in time derivatives, and the time derivative terms come from the $\int B_2$ coupling of the string to the background NS field strength.  Inertial terms in the Lagrangian are possible, but precisely because they are quadratic in time derivatives, their effects are suppressed at small velocities relative to the time derivative terms from $\int B_2$.
  \item A Nambu-Goto term is present, but its coefficient runs logarithmically due to divergences in the vortex self-interactions through exchange of excitations of $B_2$.  In the non-relativistic limit, these exchanges occur instantaneously and lead to a bilocal vortex-vortex interaction.  There is a dynamically generated length scale, call it $\ell_{n_\alpha}$, at which the coefficient of the Nambu-Goto term vanishes, and this length scale controls where much of the interesting dynamics happens.
  \item It is assumed throughout that the metric is non-dynamical, and that there is no dilaton, so that the main dynamics of interest comes from the strings and the NS two-form $B_2$.  Sound waves can be included---see in particular \cite{Endlich:2013dma,Horn:2015zna}---but we will not be concerned with their explicit effects in the current work.  It is explained for example in \cite{Gubser:2014yma} how the dynamics we study emerges from an approximate treatment of a Gross-Pitaevskii superfluid.
 \end{itemize}

The aim of the present work is to study motions of two vortex rings.  This subject has already been extensively studied in the literature, beginning with the work of Dyson \cite{Dyson1893} and Hicks \cite{Hicks1922}, and recently reviewed in \cite{vortexreview}.  See also \cite{Caplan2014} for recent investigations of the scattering and leapfrogging of vortex ring pairs, and \cite{Borisov2013, Shashikanth2003333} for analysis of the Hamiltonian dynamics. An exploration of leapfrogging vortices and their instabilities along lines similar to the current work can be found in \cite{NayarJP}.  Our methods provide an efficient route to a complete picture of the motion of unperturbed circular rings.  Also we give an analysis of linear perturbations around these motions which, though not complete, probably captures examples of most of the characteristic phenomena in this surprisingly complicated system.  We do not go beyond linear perturbation theory, but we do find assorted instabilities whose eventual fate would clearly be interesting to discover. 

We will start with perfectly circular rings whose centers move along the same axis: so-called co-axial vortex rings.  Co-axial vortex pairs are an integrable system, but they already exhibit a significant variety of phenomena: leapfrogging, pseudo-leapfrogging, chasing, nesting, attraction, repulsion and single passage.  Section~\ref{BASIC} is devoted to reviewing explicit examples of these phenomena.  The dynamics of co-axial vortex pairs has been studied almost exhaustively in \cite{Borisov2013}.  In section~\ref{BACKGROUND}, we will review the main results and fill in a small gap involving vortices with equal and opposite circulation.

In section~\ref{PERTURB}, we will study the stability of co-axial vortex rings.  Even a single circular vortex ring in isolation can be unstable.  We will refer to these single-ring instabilities as Widnall instabilities due to the works \cite{Widnall1973,Widnall1974,Widnall1975} of S.~Widnall in collaboration with J.~Sullivan, D.~Bliss, and C.-Y.~Tsai.  Pairs of vortex rings usually experience something similar to Widnall instabilities, and in addition they usually have further instabilities due to their influence on one another.  The analysis of these instabilities is somewhat complicated because the background motion is itself non-trivial, and a mix of analytic and numerical methods is necessary.  For example, when the background motion is periodic, as for leapfrogging vortices, then the natural framework to study the instabilities is Floquet theory, or some slight generalization of it to include forcing terms.  We will present the general framework for treating linearized perturbations of co-axial vortices, and then we will identify some situations in which the transfer matrix simplifies from its usual $4 \times 4$ form to a simpler block-diagonal form with $2 \times 2$ blocks.  

From our study of perturbations we find that while instabilities do usually arise, when both rings have circulations of the same sign, the instabilities tend always to occur at wavelengths comparable to or smaller than the ones involved in the Widnall instability---in other words, comparable to or smaller than the dynamically generated length scale $\ell_{n_\alpha}$ where the running tension of a given ring vanishes. We work in an approximation where vortex core-size is ignored, but in many physical systems the core-size is comparable to $\ell_{n_{\alpha}}$.  Thus our work in fact provides a check of {\it stability} for co-axial vortex pairs in a wide range of circumstances, provided finite core-size naturally cuts off instabilities at very small wavelengths, and does so in such a way as not to introduce new instabilities near the cutoff.  When the rings have circulations of opposite sign, however, the behavior of the perturbations can be quite different, and in this case there can be instabilities at large wavelengths. Indeed, the special case of colliding co-axial vortex rings with equal and opposite circulation was recently studied in this formalism~\cite{Gubser:2014yma}, where it was checked that instabilities similar to the ones studied by Lim and Nickels \cite{Lim1995} occur at wavelengths parametrically larger than the dynamical length scale.  We exhibit a large wavelength instability of a rather different sort which occurs on top of a periodic motion of a pair of vortex rings whose circulations have opposite signs.

\section{Basic setup and example behaviors}
\label{BASIC}

In this section we review the bi-local Lagrangian description for mutiple vortices moving slowly relative to one another and relative to the frame defined by the background value of the NS three-form.  We start with by considering a single string without the bi-local interactions that lead to the most interesting physics, and then we add in these interactions and explain how the Nambu-Goto action becomes a counterterm.  Finally, we show several examples of the motion of circular co-axial vortex rings.

\subsection{Free string}
\label{FREE}

Consider the standard Lagrangian
 \eqn{NGplusBtwo}{
  S = -\tau_1 \int_\Sigma d^2 \sigma \, \sqrt{-\det\left(g_{\mu\nu} 
    \partial_\alpha X^\mu \partial_\beta X^\nu \right)} + 
    \mu_1 \int_\Sigma {1 \over 2} B_{\mu\nu} \partial_\alpha X^\mu \partial_\beta X^\nu
     d\sigma^\alpha \wedge d\sigma^\beta
 }
for a single string with tension $\tau_1$ and charge $\mu_1$ moving in flat $3+1$ dimensions in the presence of a three-form field strength $H_3 = dB_2$ with
 \eqn{BtwoGauge}{
  B_2 = {\rho_0 \over 2} (X^1 dX^2 - X^2 dX^1) \wedge dX^3 \,,
 }
where we set $g^{\mu\nu} = \diag\{ -{1 \over c_s^2},1,1,1 \}$.  (We use $c_s$ instead of $c$ because physically, $c_s$ is the speed of sound in the fluid that supports the vortices.)  Taking the non-relativistic limit means sending $c_s \to \infty$ with $c_s \tau_1$ held fixed.  If we employ static gauge, $\sigma^0 = t$ and $\sigma^1 = \theta$, then \eno{NGplusBtwo} becomes
 \eqn{NGplusBtwoSimpler}{
  S = -c_s \tau_1 \int_\Sigma dt \, d\theta \, |\partial_\theta \vec{X}| + 
    \mu_1 \int_\Sigma B_2 \,,
 }
where $\vec{X}(t,\theta)$ specifies the location of the worldsheet in space at a given time $t$.  Let the string be circular with radius $r$, extended in the $X^1$-$X^2$ plane and centered on the origin.  Assume that the string moves in the $X^3$ direction.  In vector notation:
 \eqn{Xcircle}{
  \vec{X}(t,\theta) = \vec{Y}_{r,z}(\theta) \equiv
    \begin{pmatrix} r \cos\theta \\ r \sin\theta \\ z \end{pmatrix} \,,
 }
where $r$ and $z$ are allowed to depend on $t$.  Plugging \eno{Xcircle} into \eno{NGplusBtwoSimpler} leads immediately to 
 \eqn{Ssimplest}{
  S = 2\pi\rho_0 \mu_1 \int dt \, L
 }
where
 \eqn{Lsimplest}{
  L = -\eta r - {1 \over 2} r^2 \dot{z}
 }
and we have defined the ratio
 \eqn{etaDef}{
  \eta = {c_s \tau_1 \over \rho_0 \mu_1} \,.
 }
The equations of motion following from \eno{Lsimplest} are
 \eqn{eomsSimplest}{
  \dot{r} = 0 \qquad\qquad \dot{z} = -{\eta \over r} \,.
 }
So we see that circular vortex rings propagate at fixed size in a definite direction, related to their orientation, with a speed that increases as they become smaller.

\subsection{Including bi-local interactions}
\label{INTERACTING}

The modification of \eno{NGplusBtwoSimpler} which leads to most of the interesting dynamics is to consider how the string pulls on the NS two-form $B_2$.  We will state without proof, referring the interested reader to the derivation in \cite{Gubser:2014yma}, that one may use the following generalization of \eno{NGplusBtwoSimpler} to describe several vortices, labeled by an index $\alpha$, each with possibly a different charge $n_\alpha \mu_1$, moving as before much slower than the speed of sound:
 \eqn{Action}{
  S &= \sum_\alpha \left[ -c_s \tau_{n_\alpha,\rm bare} 
      \int_{\Sigma_\alpha} dt \, d\theta \, |\partial_\theta \vec{X}_\alpha| + 
     \mu_1 n_\alpha \int_{\Sigma_\alpha} B_2 \right]  \cr
    &\qquad{} 
     - {\lambda \over 2} \sum_{\alpha,\beta}
    n_\alpha n_\beta \int_{\rm reg} dt \, d\theta \, d\tilde \theta {\partial_\theta \vec{X}_\alpha \cdot 
       \partial_{\tilde \theta} \vec{X}_\beta \over |\vec{X}_\alpha(\theta) - 
         \vec{X}_\beta(\tilde \theta)|} \,.
 }
The last term in \eno{Action} is the bi-local interaction term.  It comes from integrating out the fluctuations of $B_2$ caused by the $\alpha$-th vortex and affecting the $\beta$-th vortex (or vice versa).  The coupling constant $\lambda$ is related to the normalization of the action for $H_3$, so it is for our purposes a free parameter.  The integral over $\theta$ and $\tilde\theta$ in the interaction term diverges when $\theta=\tilde\theta$, and we employ the standard regulator
 \eqn{StandardRegulator}{
  |\vec{X}_\alpha(\theta) - \vec{X}_\beta(\tilde\theta)| \to 
   \sqrt{a^2 + |\vec{X}_\alpha(\theta) - \vec{X}_\beta(\tilde\theta)|^2}
 }
where $a$ is a small length scale, essentially the core size of the vortices.  A logarithmic divergence as $a \to 0$ can be cured by adding in the Nambu term in \eno{Action} as a counterterm, with $\tau_{1,\rm bare} \sim \log a$.  To see this, first, note that 
 \eqn{dlda}{
  {d \over da} {1 \over \sqrt{a^2 + x^2}} = -{a \over (a^2+x^2)^{3/2}} \approx
    -{2 \over a} \delta(x) \,,
 }
where the approximate equality in \eno{dlda} is valid when $a \ll x$.  Differentiating \eno{Action} with respect to $a$ and using \eno{dlda} (under the assumption that the variation of all the $\vec{X}_\alpha$ is slow on the scale of $a$), we see that
 \eqn{dSda}{
  {dS \over da} &= -\sum_\alpha c_s {d\tau_{n_\alpha,\rm bare} \over da}
    \int_{\Sigma_\alpha} dt \, d\theta \, |\partial_\theta \vec{X}_\alpha|  \cr
   &\qquad{} +
    {\lambda \over a} \sum_{\alpha,\beta} n_\alpha n_\beta 
     \int dt \, d\theta \, d\tilde\theta \,
      \partial_\theta \vec{X}_\alpha \cdot \partial_{\tilde\theta} \vec{X}_\beta \,
      \delta(|\vec{X}_\alpha(\theta) - \vec{X}_\beta(\tilde\theta)|) \,.
 }
If we {\it assume} that $\vec{X}(\theta) - \vec{X}_\beta(\tilde\theta) = 0$ only when $\alpha=\beta$ and $\theta=\tilde\theta$, then the summation over $\beta$ and the integration over $\tilde\theta$ in \eno{dSda} can be done to obtain
 \eqn{dSdaSimpler}{
  {dS \over da} = \sum_\alpha \left( - c_s {d\tau_{n_\alpha,\rm bare} \over da} + 
    n_\alpha^2 {\lambda \over a} \right) \int_{\Sigma_\alpha} dt \, d\theta \, 
    |\partial_\theta \vec{X}_\alpha| \,.
 }
We want $dS/da = 0$ so that $S$ as a whole is invariant under the choice of cutoff $a$.  Referring to \eno{dSdaSimpler}, this implies
 \eqn{tauRuns}{
  a {d \over da} \left( c_s \tau_{n_\alpha,\rm bare} \right) =   n_\alpha^2 \lambda \,,
 }
so we conclude
 \eqn{tauForm}{
  c_s \tau_{n_\alpha,\rm bare} =  n_\alpha^2  \lambda \log {a \over {\rm a}_{n_\alpha}}
 }
for some constants ${\rm a}_{n_\alpha}$. Define the dynamical length scale $\ell_{n_\alpha}$, associated with a vortex with winding number $n_\alpha$ to be
 \eqn{ellnalphaDef}{
 \ell_{n_\alpha} \equiv {{\rm a}_{n_\alpha} e \over 8} = \frac{a}{8}e^{1-c_s \tau_{n_{\alpha},\rm bare}/n_{\alpha}^2 \lambda} \,.
 }
The bare tension term can be split into a finite physical tension term and a divergent term which includes the logarithmic divergence of the last term in \eno{Action},
 \eqn{tauSplit}{
 c_s \tau_{n_\alpha,\rm bare} = n_\alpha^2 \lambda \log {r_\alpha \over \ell_{n_\alpha}} + n_\alpha^2 \lambda \log { a e \over 8 r_\alpha}\,,
 }
where $r_\alpha$ is the radius of the $\alpha$-th vortex. Thus the dynamical length scale $\ell_{n_\alpha}$ corresponds to the length scale at which the first (physical) term in \eno{tauSplit} goes to zero.  There is some ambiguity to the splitting into physical and divergent contributions, which must be fixed by choosing a renormalization condition; more precisely, we will see that $\ell_{n_{\alpha}}$ is defined such that the velocity of a single vortex vanishes at this radius~\cite{Gubser:2014yma}.

The bare tension of a vortex ring with winding number\footnote{It is easy to show that a configuration with $|n_{\alpha}| > 1$ has higher energy than a configuration where the circulation is divided among vortices of unit winding, and in \cite{Pu:1999,Simula:2002,Mottonen:2003,Shin:2004} it was shown that a vortex with multiple units of winding may be unstable due to quantum mechanical effects at core sizes.  This will not affect our analysis at the level of the classical theory, however, and we will continue to take the philosophy that the winding number is an arbitrary parameter in our effective theory, and therefore we may choose it to have any integer value.} $n_2$ can be related to that of a vortex with winding number $n_1$, 
 \eqn{tauRelation}{
 c_s \tau_{n_2,\rm bare} = \left({n_2 \over n_1}\right)^2 c_s \tau_{n_1,\rm bare} + n_2^2 \lambda \log {\ell_{n_1} \over \ell_{n_2}}
 }
upon using the form of the tension in \eno{tauForm} and the definition of the dynamical length scale in \eno{ellnalphaDef}. The ratio of the dynamical length scales
 \eqn{chiDef}{
 \chi \equiv \ell_{n_2}/\ell_{n_1}
 }
must be determined by a more fundamental theory at short distances, and is to be understood as a free parameter in the effective field theory treatment presented here.

\subsection{Basic examples}
A pair of circular co-axial vortex rings exhibit a variety of motions, governed by the action in \eno{Action}. A brief description and a few select examples are presented below for each of the possible motions, and the explicit equations of motion will be presented in section \ref{PAIR}. Throughout the rest of this paper, the radii of vortices with winding numbers $n_1$ and $n_2$ will be $r_1$ and $r_2$, respectively with $\Delta r \equiv r_1 - r_2$. Correspondingly, the axial coordinates of the vortices will be $z_1$ and $z_2$, respectively with $\Delta z \equiv z_1 - z_2$.

 \begin{figure}
  \centerline{
 \includegraphics[width=0.33 \textwidth]{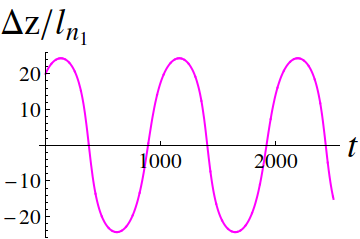}
 \includegraphics[width=0.33 \textwidth]{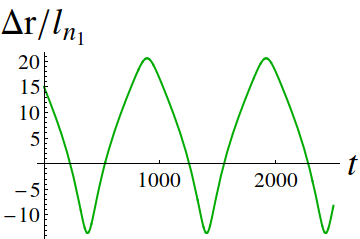}
 \includegraphics[width=0.33 \textwidth]{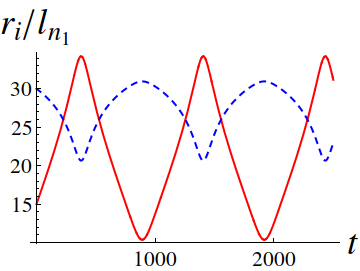}}
   \centerline{
 \includegraphics[width=0.33 \textwidth]{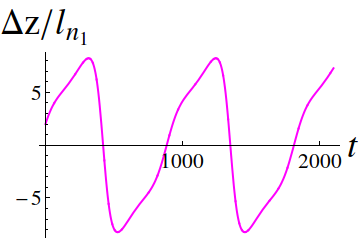}
 \includegraphics[width=0.33 \textwidth]{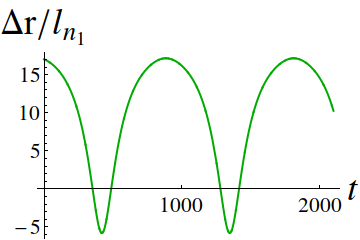}
 \includegraphics[width=0.33 \textwidth]{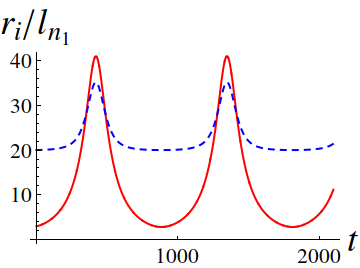}}
  \caption[Leapfrogging vortices]
    {\tabular[t]{@{}l@{}}Leapfrogging vortices. Initial conditions --  \\ Top: $r_1 = 30 \ell_{n_1}\,,\, r_2 = 15 \ell_{n_1}\,,\, \Delta z = 20 \ell_{n_1}\,,\, n_1=2\,,\, n_2=1\,,\, \chi=1/\log 2 \approx 1.44$.\\
Bottom: $r_1 = 20 \ell_{n_1}\,,\, r_2 = 3 \ell_{n_1}\,,\, \Delta z = 2 \ell_{n_1} \,,\, n_1=2\,,\, n_2=-1\,,\, \chi=1/\log 2$.\endtabular} 
\label{egLF}
 \end{figure} 

 \begin{figure}
  \centerline{
 \includegraphics[width=0.33 \textwidth]{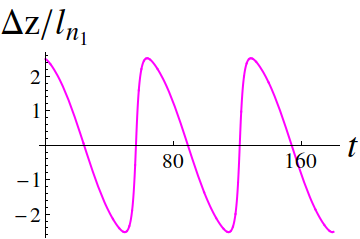}
 \includegraphics[width=0.33 \textwidth]{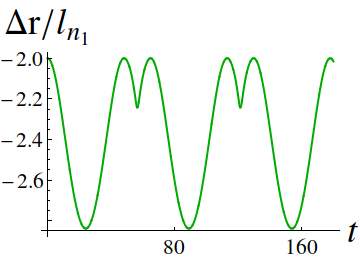}
 \includegraphics[width=0.33 \textwidth]{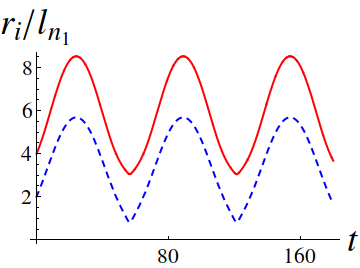}}
   \centerline{
 \includegraphics[width=0.33 \textwidth]{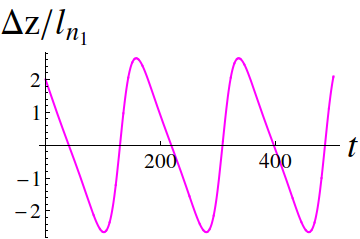}
 \includegraphics[width=0.33 \textwidth]{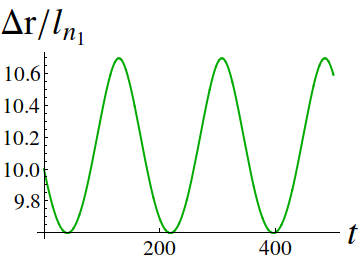}
 \includegraphics[width=0.33 \textwidth]{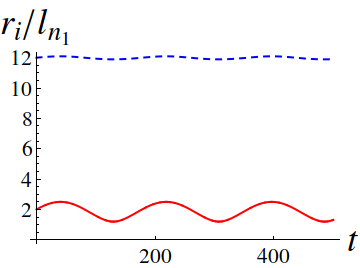}}
   \caption[Pseudo-leapfrogging vortices]
    {\tabular[t]{@{}l@{}}Pseudo-leapfrogging vortices. Initial conditions --  \\ Top: $r_1 = 2 \ell_{n_1}\,,\, r_2 = 4 \ell_{n_1}\,,\, \Delta z = 2.5 \ell_{n_1}\,,\, n_1 = 2\,,\, n_2 = -1\,,\, \chi= 1/\log 2 \approx 1.44$. \\
 Bottom: $r_1 = 12 \ell_{n_1}\,,\, r_2 = 2 \ell_{n_1}\,,\, \Delta z = 2 \ell_{n_1} \,,\, n_1= 1\,,\, n_2= -1\,,\, \chi= 1$.\endtabular} \label{egPLF}
 \end{figure} 

 \begin{figure}
  \centerline{
 \includegraphics[width=0.33 \textwidth]{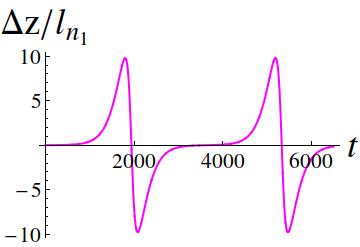}
 \includegraphics[width=0.33 \textwidth]{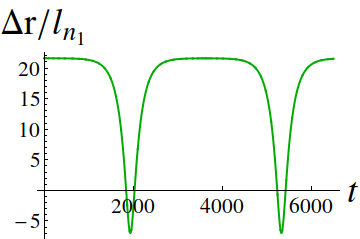}
 \includegraphics[width=0.33 \textwidth]{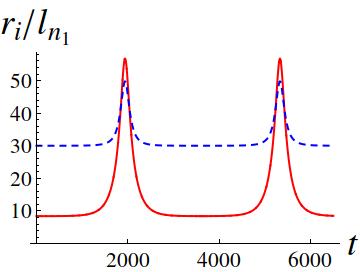}}
   \centerline{
 \includegraphics[width=0.33 \textwidth]{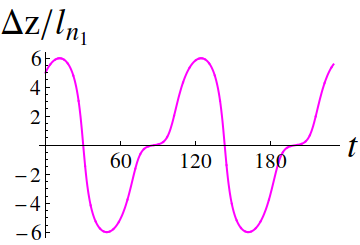}
 \includegraphics[width=0.33 \textwidth]{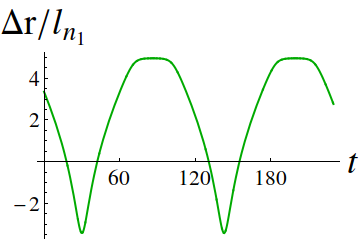}
 \includegraphics[width=0.33 \textwidth]{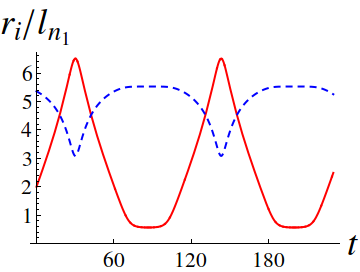}}
   \caption[Nesting vortices]
    {\tabular[t]{@{}l@{}}Nesting vortices. Initial conditions --  \\ Top: $r_1 = 30 \ell_{n_1}\,,\, r_2 = 8.4046 \ell_{n_1}\,,\, \Delta z = 0\,,\, n_1 = 2\,,\, n_2 = -1\,,\, \chi= 1/\log 2 \approx 1.44$. \\
 Bottom: $r_1 = 5.3555 \ell_{n_1}\,,\, r_2 = 2 \ell_{n_1}\,,\, \Delta z = 5 \ell_{n_1}\,,\, n_1 = 2\,,\, n_2 = 1\,,\, \chi= 1/\log 2$. \endtabular}  \label{egN}
 \end{figure} 
  
 \begin{figure}
  \centerline{
 \includegraphics[width=0.33 \textwidth]{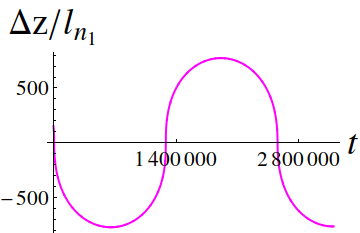}
 \includegraphics[width=0.33 \textwidth]{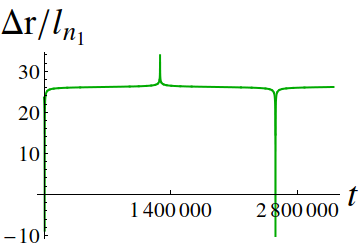}
 \includegraphics[width=0.33 \textwidth]{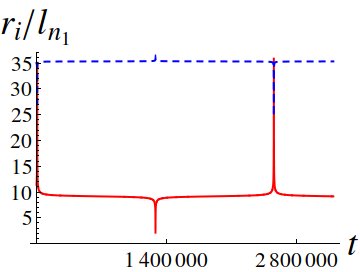}}
   \centerline{
 \includegraphics[width=0.33 \textwidth]{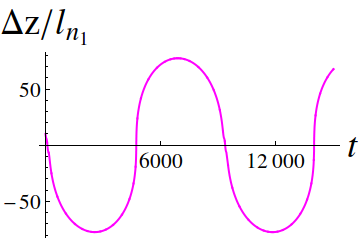}
 \includegraphics[width=0.33 \textwidth]{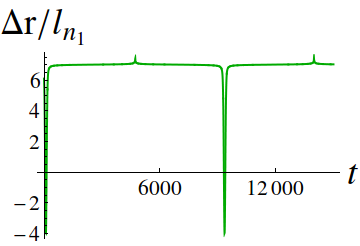}
 \includegraphics[width=0.33 \textwidth]{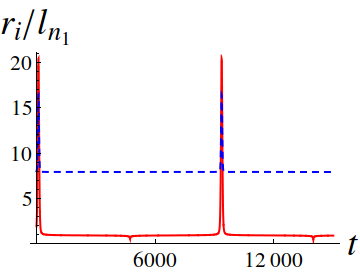}}
  \caption[Chasing vortices]
    {\tabular[t]{@{}l@{}}Chasing vortices. Initial conditions --  \\ Top: $r_1 = 35 \ell_{n_1}\,,\, r_2 = 11.41 \ell_{n_1}\,,\, \Delta z = 150 \ell_{n_1} \,,\, n_1 = 2\,,\, n_2 = 1\,,\, \chi= 1/\log 2$. \\
Bottom: $r_1 = 8 \ell_{n_1}\,,\, r_2 = 1.902 \ell_{n_1}\,,\, \Delta z = 10 \ell_{n_1} \,,\, n_1= 2\,,\, n_2= -1\,,\, \chi= 1/\log 2$. \endtabular}  \label{egCH}  
 \end{figure}    
   
 \begin{figure}
  \centerline{
 \includegraphics[width=0.33 \textwidth]{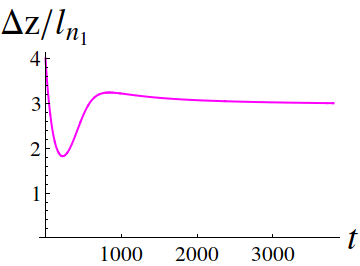}
 \includegraphics[width=0.33 \textwidth]{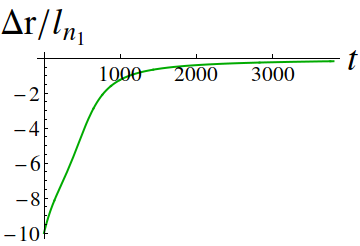}
 \includegraphics[width=0.33 \textwidth]{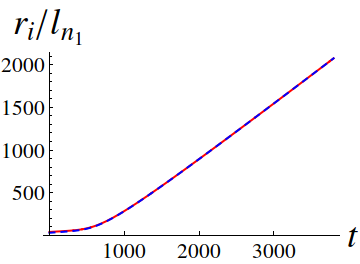}}
   \centerline{
 \includegraphics[width=0.33 \textwidth]{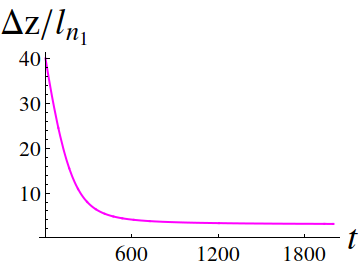}
 \includegraphics[width=0.33 \textwidth]{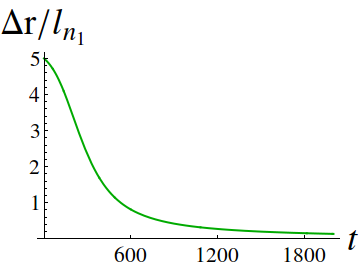}
 \includegraphics[width=0.33 \textwidth]{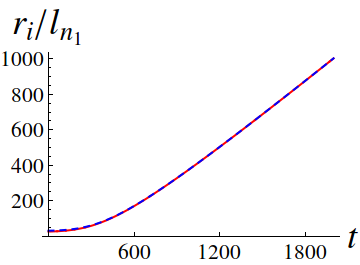}}
    \caption[Attracting vortices]
    {\tabular[t]{@{}l@{}}Attracting vortices. Initial conditions --  \\ Top: $r_1 = 30 \ell_{n_1}\,,\, r_2 = 40 \ell_{n_1}\,,\, \Delta z = 4 \ell_{n_1} \,,\, n_1 = 1\,,\, n_2 = -1\,,\, \chi= 1$. \\
Bottom: $r_1 = 30 \ell_{n_1}\,,\, r_2 = 25 \ell_{n_1}\,,\, \Delta z = 40 \ell_{n_1} \,,\, n_1= 1\,,\, n_2= -1\,,\, \chi= 1$. \endtabular}  \label{egATT}  
 \end{figure} 
 
 \begin{figure}
  \centerline{
 \includegraphics[width=0.33 \textwidth]{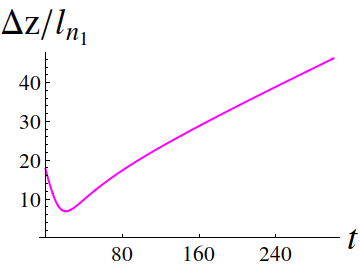}
 \includegraphics[width=0.33 \textwidth]{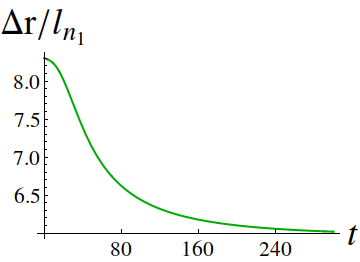}
 \includegraphics[width=0.33 \textwidth]{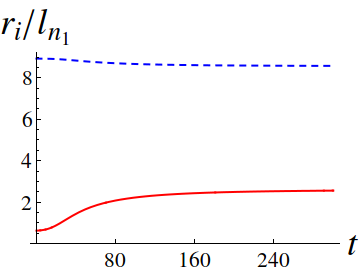}}
   \centerline{
 \includegraphics[width=0.33 \textwidth]{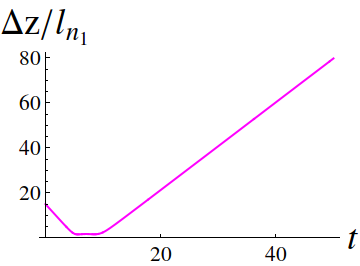}
 \includegraphics[width=0.33 \textwidth]{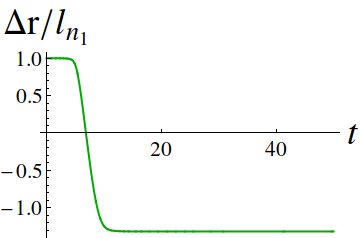}
 \includegraphics[width=0.33 \textwidth]{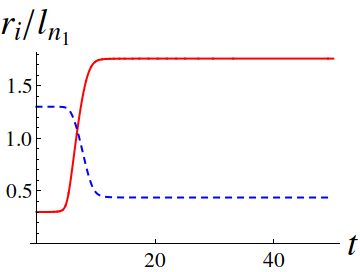}}
 \caption[Repelling vortices]
    {\tabular[t]{@{}l@{}}Repelling vortices. Initial conditions --  \\ Top: $r_1 = 9 \ell_{n_1}\,,\, r_2 = 0.8 \ell_{n_1}\,,\, \Delta z = 18 \ell_{n_1} \,,\, n_1 = 1\,,\, n_2 = 1\,,\, \chi= 1$. \\
Bottom: $r_1 = 1.3 \ell_{n_1}\,,\, r_2 = 0.3 \ell_{n_1}\,,\, \Delta z = 15 \ell_{n_1} \,,\, n_1= 2\,,\, n_2= 1\,,\, \chi= 1/\log 2$. \endtabular}  \label{egREP}    
 \end{figure} 
   
 \begin{figure}
  \centerline{
 \includegraphics[width=0.33 \textwidth]{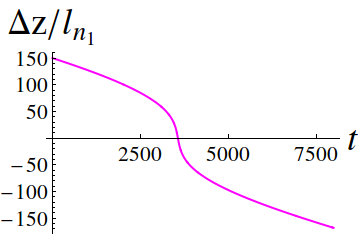}
 \includegraphics[width=0.33 \textwidth]{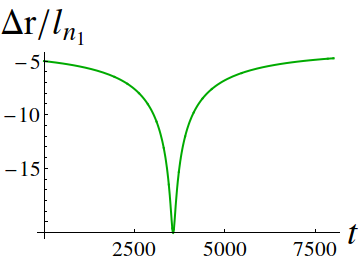}
 \includegraphics[width=0.33 \textwidth]{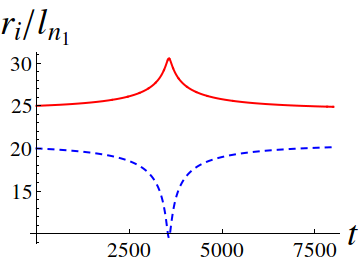}}
   \centerline{
 \includegraphics[width=0.33 \textwidth]{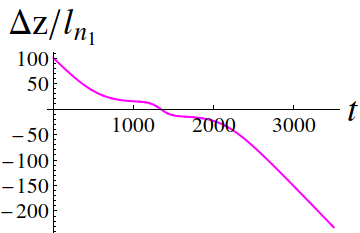}
 \includegraphics[width=0.33 \textwidth]{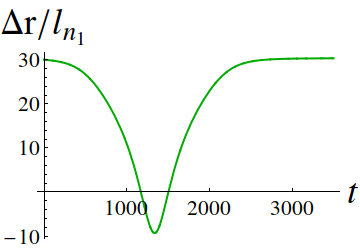}
 \includegraphics[width=0.33 \textwidth]{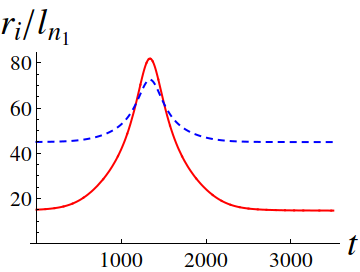}}
  \caption[Single passage of vortices]
    {\tabular[t]{@{}l@{}}Single passage of vortices. Initial conditions --  \\ Top: $r_1 = 20 \ell_{n_1}\,,\, r_2 = 25 \ell_{n_1}\,,\, \Delta z = 150 \ell_{n_1}\,,\, n_1 = 1\,,\, n_2 = 1\,,\, \chi= 1$.  \\
Bottom: $r_1 = 45 \ell_{n_1}\,,\, r_2 = 15 \ell_{n_1}\,,\, \Delta z = 100 \ell_{n_1} \,,\, n_1= 2\,,\, n_2= -1\,,\, \chi= 1/\log 2$. \endtabular}  \label{egSP}    
 \end{figure}
 
\begin{enumerate}
\item {\bf Leapfrogging.} Periodic motion of the vortices, where $\Delta z$ and $\Delta r$ oscillate periodically, assuming both positive and negative values in such a way that the vortices go around one another.  See figure \ref{egLF} for some examples.\footnote{Since the ratio of the dynamical length scales for vortices with unequal winding numbers is a free parameter in the effective field theory, we have chosen an arbitrary value for $\chi$ in examples where $n_2/n_1 \neq \pm 1$.}

\item {\bf Pseudo-leapfrogging.} Periodic motion of the vortices, where $\Delta z$ and $\Delta r$ oscillate periodically, but only $\Delta z$ runs over both positive and negative values. $\Delta r$ is either strictly positive or strictly negative. This motion can be thought of as a periodic motion where the vortices do not wind around one another, but instead one vortex is ``captured'' inside the other. See figure \ref{egPLF} for some examples.

\item {\bf Nesting.} A special case of periodic motion, where $\Delta z \approx 0$ and simultaneously $\Delta r \approx \rm constant$ over a finite period of time every cycle. See figure \ref{egN} for some examples.

\item {\bf Chasing.} A special case of periodic motion, where $\Delta z$ takes up arbitrarily large values and $\Delta r \approx \rm constant$ for most of the time during a cycle. See figure \ref{egCH} for some examples.

\item {\bf Attracting.} This aperiodic motion is only possible when $n_2/n_1 = -1$. It is characterized by $\Delta z \rightarrow 8 \ell_{n_1} /e$, and $\Delta r \rightarrow 0$ as $r_1, r_2 \rightarrow \infty$ at late times. See figure \ref{egATT} for some examples.

\item {\bf Repelling.} This aperiodic motion is characterized by a $\Delta z$ which doesn't change signs, and at late times $\left|\Delta z\right| \rightarrow \infty$ and $r_1, r_2, \Delta r \rightarrow \rm constants$, corresponding to two isolated vortices. See figure \ref{egREP} for some examples.

\item {\bf Single-passage.}  This aperiodic motion is characterized by $\Delta z$ changing signs exactly once, and at late times $\left|\Delta z\right| \rightarrow \infty$ and $r_1, r_2, \Delta r \rightarrow \rm constants$, corresponding to two isolated vortices. This corresponds to a pair of vortices with circulations of insufficient strength, which is required to form a bound state of leapfrogging vortices. See figure \ref{egSP} for some examples.
\end{enumerate}

In the next section we describe these motions in terms of trajectories in the phase space.

\section{Background Lagrangian and equations of motion}
\label{BACKGROUND}

\subsection{Single vortex with winding \texorpdfstring{$n_1$}{n1}}
\label{SINGLE}

Introduce a scaled Lagrangian
\eqn{LScaled}{
  S = 2\pi \rho_0 \mu_1 \int dt\, L_{\rm one\ vortex}\,,
  }  
and a rescaled tension and interaction strength
\eqn{ParamScaled}{
  \eta_{n_1, \rm bare} =  {c_s \tau_{n_1, \rm bare} \over \rho_0 \mu_1} \qquad \tilde\lambda = {\lambda \over \rho_0 \mu_1}\,.
  }
Then the Lagrangian for a single vortex, on substituting \eno{Xcircle} in \eno{Action} is found to be
\eqn{LDef0}{
  L_{\rm one \ vortex} = L_0 + n_1^2 \tilde\lambda  r_1 Q_0(q_1)
    }
where
\eqn{L0Def}{
  L_0 &= - \eta_{n_1, \rm bare} r_1 - {n_1\over 2} r_1^2 \dot{z_1}\,,
  }
and the second term comes from the self-interaction of the vortex, and was computed in Ref.~\cite{Gubser:2014yma} to be
\eqn{Q0Def}
{
Q_0(q_1) = \displaystyle{ q_1 E\left(-\frac{4}{q_1^2}\right)-\left(q_1+{2 \over q_1}\right) K\left(-\frac{4}{q_1^2}\right)}
} 
where $q_1 \equiv a/r_1$,  and $K(y)$ and $E(y)$ are the complete elliptic integrals of the first and second kind, respectively.\footnote{The complete elliptic integrals of the first and second kind, $K(y)$ and $E(y)$ respectively, are defined in terms of the parameter $y$ as 
\[ K(y) \equiv \int_0^{\pi/2} { d \alpha \over \sqrt{1-y \sin^2 \alpha }}\,,\qquad  E(y) \equiv \int_0^{\pi/2} { d \alpha \sqrt{1-y \sin^2 \alpha }}\,. \]} The subscripts on $r_1, z_1$ and $n_1$ label the vortex they describe. 
The analysis for a single circular vortex proceeds the same way as presented in Ref.~\cite{Gubser:2014yma}, except in terms of a rescaled time coordinate $n_1 t$. It follows from \eno{ellnalphaDef} that the dynamical length scale in this case is given by
 \eqn{ellZeroDef}{
  \ell_{n_1} = {a \over 8} e^{1-\eta_{n_1,\rm bare}/(n_1^2\tilde\lambda)}\,,
 }
and the vortex core size is understood to be of the order of the UV cutoff $a$. 

\subsection{Vortex pair with windings \texorpdfstring{$n_1$ and $n_2$}{n1 and n2}}
\label{PAIR}
The two-vortex action is described by
\eqn{LScaledTwoV}{
  S = 2\pi \rho_0 \mu_1 \int dt\, L_{\rm two\ vortex}\,.
  }  
The vortices are parametrized by \eno{Xcircle} with the subscript on $r_i, z_i$ and $n_i$ labelling the vortex they describe. Substituting the parametrization in \eno{Action} leads to
\eqn{LDefTwoV0}{
  L_{\rm two \ vortex} &= \left( L_{\rm one \ vortex} +    n_1 n_2{\tilde\lambda \over 2} \sqrt{r_1 r_2}  S_0  \right) +  \left( 1 \leftrightarrow 2 \right)\,,
    }
where $(1\leftrightarrow 2)$ stands for $(r_1 \leftrightarrow r_2,\, z_1 \leftrightarrow z_2,\, n_1 \leftrightarrow n_2)$, and
\eqn{S0Def}{
	S_0 = 2 Q_0 \left(\sqrt{q_p^2 + s_p^2} \right)\,,
			}
where 
\eqn{SpQpDef}{
	q_p = { a \over \sqrt{r_1  r_2}} \qquad  
	s_p =   \sqrt{\frac{(\Delta r)^2+(\Delta z)^2}{r_1 r_2}}\,,
	}
and we have used $\Delta r = r_1 - r_2$, and $\Delta z = z_1 - z_2$. Note that the vortex with winding number $n_2$ has a natural length scale $\ell_{n_2}$ determined by \eno{ellZeroDef} with $(n_1 \leftrightarrow n_2)$. 

Restricting to the small core limit $(q_1,\, q_2,\, q_p\rightarrow 0)$,\footnote{Here $q_1 = a/r_1$, $q_2 = a/r_2$.} the equations of motion for a pair of co-axial circular vortices in terms of a rescaled time coordinate $t \rightarrow \tilde\lambda n_1 t$ are
\eqn{UnpertEOMs1}{
   r_1\, \dot{r}_1
   &= -\gamma {2\Delta z \over \sqrt{(\Delta r)^2+(\Delta z)^2}}\, Q_0^\prime\left(
   s_p\right) \cr
    r_2\, \dot{r}_2
    &= {2\Delta z \over \sqrt{(\Delta r)^2+(\Delta z)^2}}\, Q_0^\prime\left(
    s_p\right) \,,
    }
and
\eqn{UnpertEOMs2}{
    r_1^2\,  \dot{z}_1
    &=  r_1 \log {\ell_{n_1} \over r_1} +{  \gamma \sqrt{r_1 r_2}   }\, Q_0\left(
    s_p \right) 
    - \gamma { \left(r_2^2 - r_1^2 +(\Delta z)^2\right)  \over  \sqrt{(\Delta r)^2+(\Delta z)^2}}\, Q_0'\left(
    s_p\right) \cr
    r_2^2 \, \dot{z}_2
       &=  \gamma  r_2 \log {\ell_{n_2} \over r_2} +{ \sqrt{r_1 r_2}  }\, Q_0\left(
        s_p \right) 
     + { \left(r_2^2 - r_1^2 - (\Delta z)^2\right)  \over  \sqrt{(\Delta r)^2+(\Delta z)^2}} \, Q_0'\left(
     s_p\right)\,,
  } 
where $Q_0'(s_p) \equiv d Q_0(s_p) / d s_p$, $s_p$ is defined in \eno{SpQpDef} and
\eqn{GammaDef}
{\gamma \equiv n_2/n_1\,.
}
The conserved energy of the vortex system is 
\eqn{EnergyTwoV}{
   \tilde\epsilon = -\tilde\lambda n_1^2\left(  r_1 \log {e \ell_{n_1} \over r_1} +  \gamma^2 r_2 \log {e \ell_{n_2} \over r_2} + 2 \gamma  \sqrt{r_1 r_2} Q_0(s_p) \right)\,,
  }  
and the conserved momentum along the $z$ direction is
\eqn{MomentumTwoV}{
   \tilde p_z = -{n_1\over 2} \left( r_1^2 + \gamma  r_2^2\right)\,.
  }  
This system is integrable because the four dimensional phase space, spanned by $\left(r_1,r_2,z_1,\Delta z\right)$ or equivalently by $\left(z_1\,,\Delta z\right)$ and their conjugate momenta,\footnote{These are related to the radii $r_1$ and $r_2$.} is constrained by two independent invariants, $\left(\tilde \epsilon\,, \tilde p_z\right)$. 

\subsubsection{Test limit: \texorpdfstring{$\left|\gamma\right| \ll 1$}{n2/n1 << 1}}

Consider the case where the two strings have very different winding numbers, with $\left|\gamma\right| \ll 1$.  In this case the first ring will move according to the solution for a single isolated ring.  Neglecting subleading terms in $\gamma$, the first ring obeys, in terms of a rescaled time coordinate $\tilde\lambda n_1 t$
\eqn{BgRingEom}
{
\dot{r}_1 = 0\, , \qquad \, \dot{z}_1 = {1 \over r_1}\log {\ell_{n_1} \over r_1}\,. 
}
For the second ring (referred to as the ``test'' ring from now on), the background equations of motion (in terms of the rescaled time) have to be solved numerically, although now there are only two remaining equations and two variables to be solved for:
\eqn{TestRingEom}
{
 r_2\, \dot{r}_2 &= {2\Delta z \over \sqrt{(\Delta r)^2+(\Delta z)^2}}\, Q_0^\prime\left( s_p\right) \cr
 r_2^2 \, \dot{z}_2    &=  
          { \sqrt{r_1 r_2}  }\, Q_0\left(s_p \right) 
     + { \left(r_2^2 - r_1^2 - (\Delta z)^2\right)  \over  \sqrt{(\Delta r)^2+(\Delta z)^2}} \, Q_0'\left(s_p\right)\,.
}
The reduction in the dimension of the system still leads to background dynamics just as rich as for the full four dimensional system (see the next subsection), but at the same time it makes studying the perturbations about the background more tractable, as we will see in section \ref{PERTURB}.

\subsection{Classification of the phase space}
\label{CLASSIFY}
   
\begin{figure}
  \centerline{\includegraphics[width=0.5\textwidth]{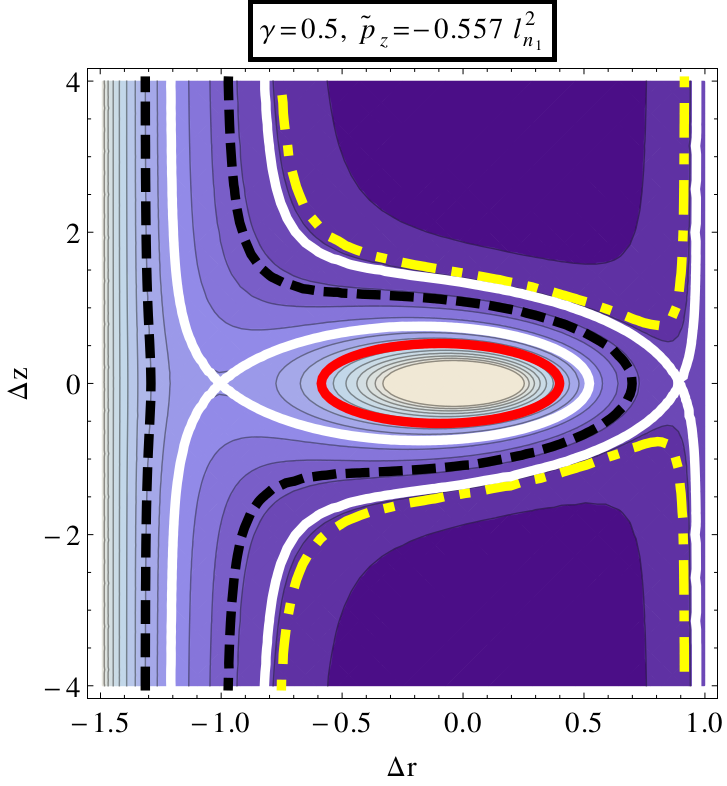}
  \includegraphics[width=0.5\textwidth]{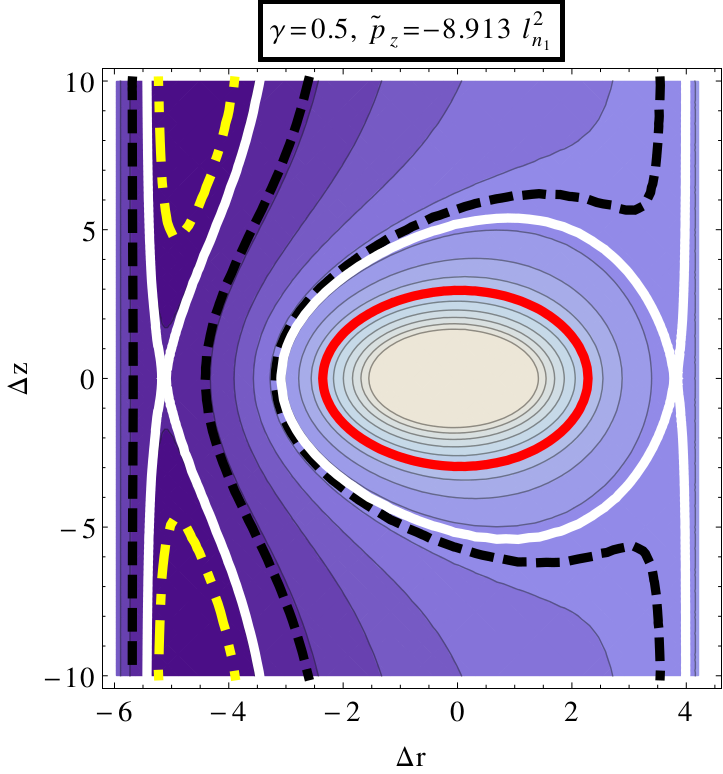}}
  \centerline{
 \includegraphics[width=0.5\textwidth]{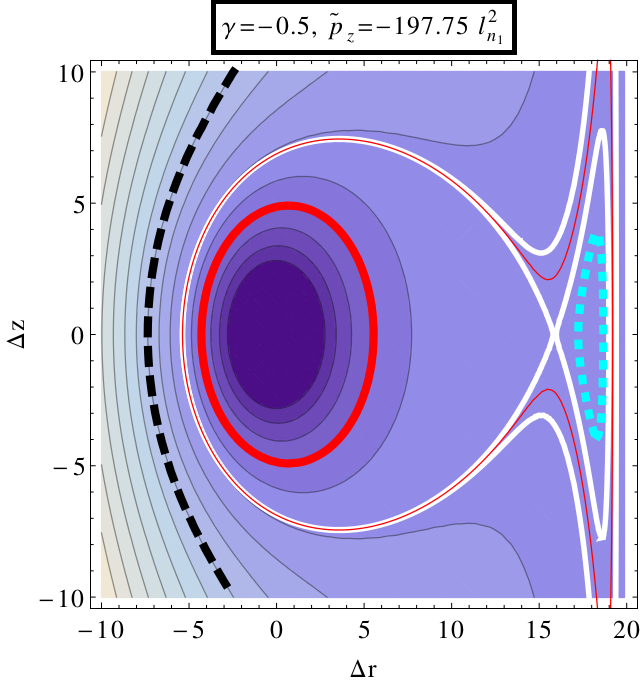}
  \includegraphics[width=0.5\textwidth]{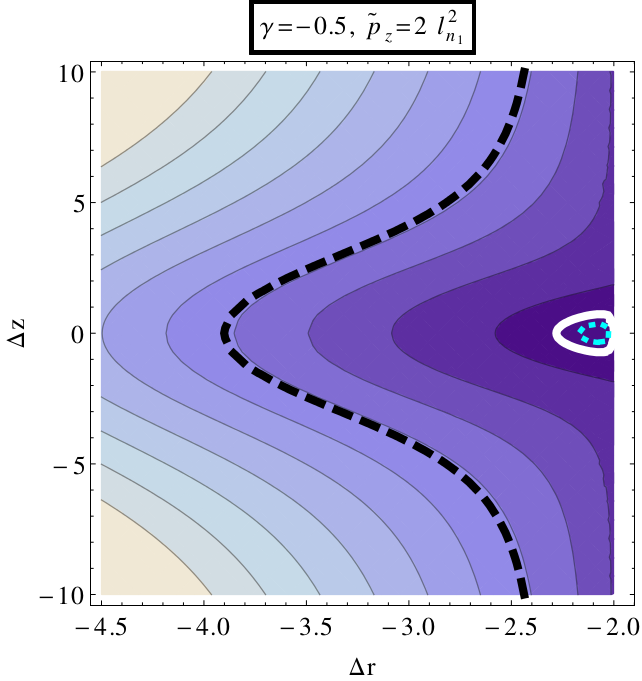}}
  \caption{(Color online.) Examples of phase diagrams. Each plot is evaluated at a fixed value of $\tilde{p}_z$. Top row: $\gamma > 0$. Bottom left: $\gamma <0\,,\, \tilde{p}_z < 0$. Bottom right: $\gamma <0\,,\, \tilde{p}_z >0$. We have chosen $\chi=1$ and the axes are in units of the dynamical length scale $\ell_{n_1}$. {\bf Color key} for typical examples: {\bf Solid red}: leapfrogging vortices. {\bf Dotted cyan}: pseudo-leapfrogging vortices. {\bf Dashed black}: single passage of vortices. {\bf Dot-dashed yellow}: repelling vortices. The {\bf solid white} curves represent the separatrix curves. The thin black curves mark out some possible phase space trajectories, corresponding to different values for the energy, denoted here as a contour gradient. }\label{GammaGrtLessZero}
 \end{figure}

\begin{figure}[t]
  \centerline{\includegraphics[width=0.5\textwidth]{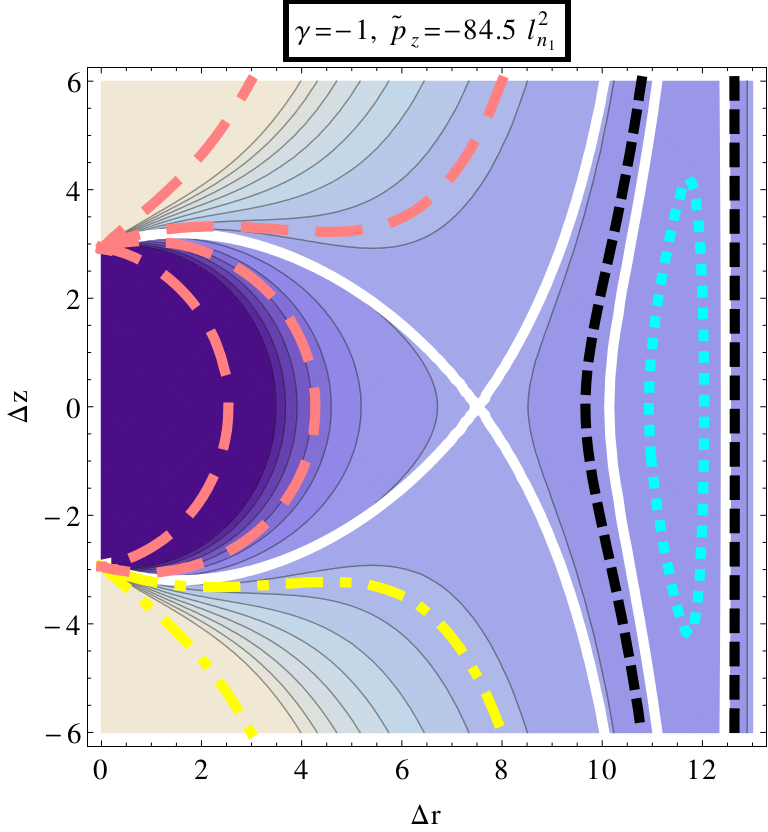}}
  \caption{(Color online.) Example of a phase diagram when $\gamma = -1$. 
Refer to figure \ref{GammaGrtLessZero} for the color key.
   In particular, note: {\bf Dot-dashed yellow} curves show examples of repulsion between vortices, each one originating from the repeller point at $\Delta r=0,\, \Delta z = -8\ell_{n_1}/e$. The {\bf Long-dashed pink} curves show examples of trajectories asymptotically approaching the attractor point at $\Delta r=0,\, \Delta z = 8\ell_{n_1}/e$.}\label{GammaMinusOne}
 \end{figure}

The classical phase space dynamics of a pair of vortex rings has been previously exhaustively classified~\cite{Borisov2013} for all values of $\gamma$ except for $\gamma=-1$, and in the restricted case when $\chi = \ell_{n_2}/ \ell_{n_1} = 1$, i.e.\ in the restricted case when the core sizes of vortices with different winding numbers are assumed to be equal. Several distinct regimes of behavior were found:

\begin{enumerate}
\item $\gamma > 0$.  In this regime there can be repulsion, single passage of one ring through another, or leapfrogging solutions. Refer to figure~\ref{GammaGrtLessZero} for two examples of typical phase diagrams in this case.

\item $\gamma \in (-1,0)$, $r_1^2 + \gamma r^2_2 > 0$.  In this regime there can be single passage, leapfrogging, or pseudo-leapfrogging, where the motion is periodic but the rings do not wind around one another. Refer to figure~\ref{GammaGrtLessZero} for an example of a typical phase diagram in this case.

\item $\gamma \in (-1,0)$, $r_1^2 + \gamma r^2_2 \leq 0$.  Single passage and pseudo-leapfrogging are the only possibilities. Refer to figure~\ref{GammaGrtLessZero} for an example of a typical phase diagram in this case.

\item $\gamma = -1$.  This regime was not investigated in Ref.~\cite{Borisov2013}. We present a phase space analysis of this case in Appendix \ref{FIXEDPOINTS}. In addition to repulsion and pseudo-leapfrogging, there are attractor solutions where  $\Delta r \rightarrow 0$ as $r_1, r_2 \to \infty$, and $\Delta z \to 8\ell_{n_1}/e$. Refer to figure~\ref{GammaMinusOne} for an example of a typical phase diagram in this case.
\end{enumerate}

 
\begin{figure}
  \centerline{\includegraphics[width=0.5\textwidth]{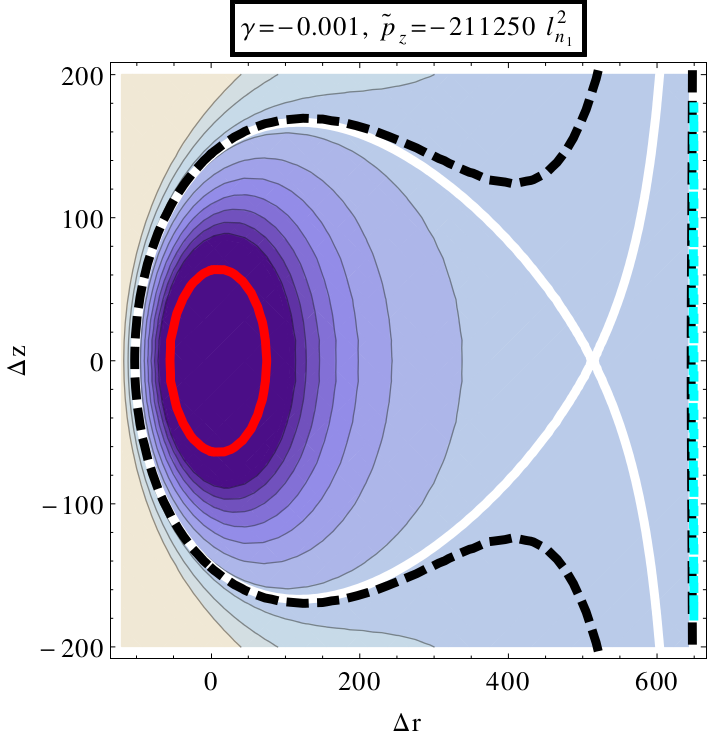}
  \includegraphics[width=0.5\textwidth]{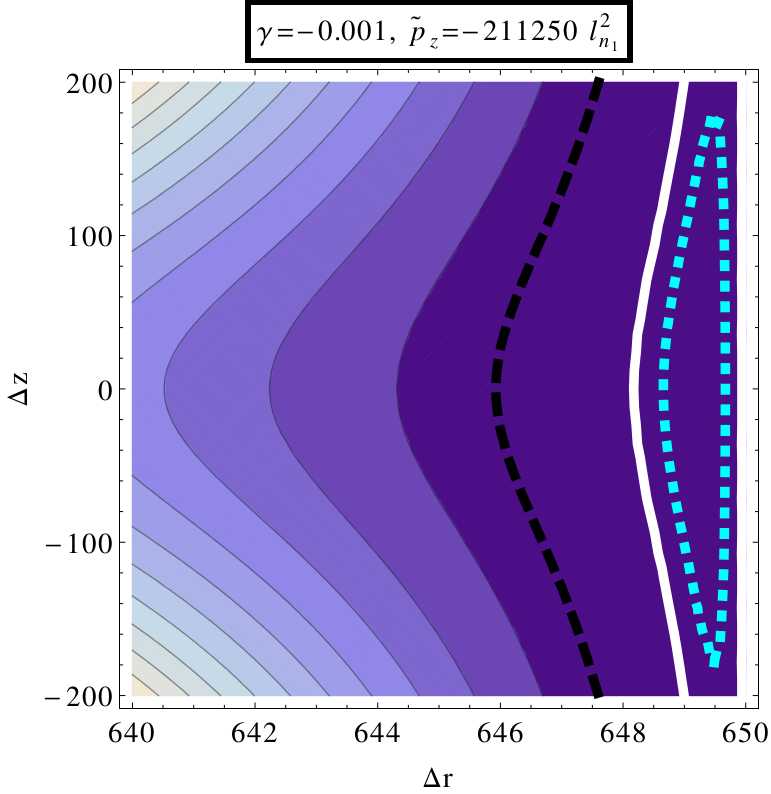}}
    \centerline{\includegraphics[width=0.5\textwidth]{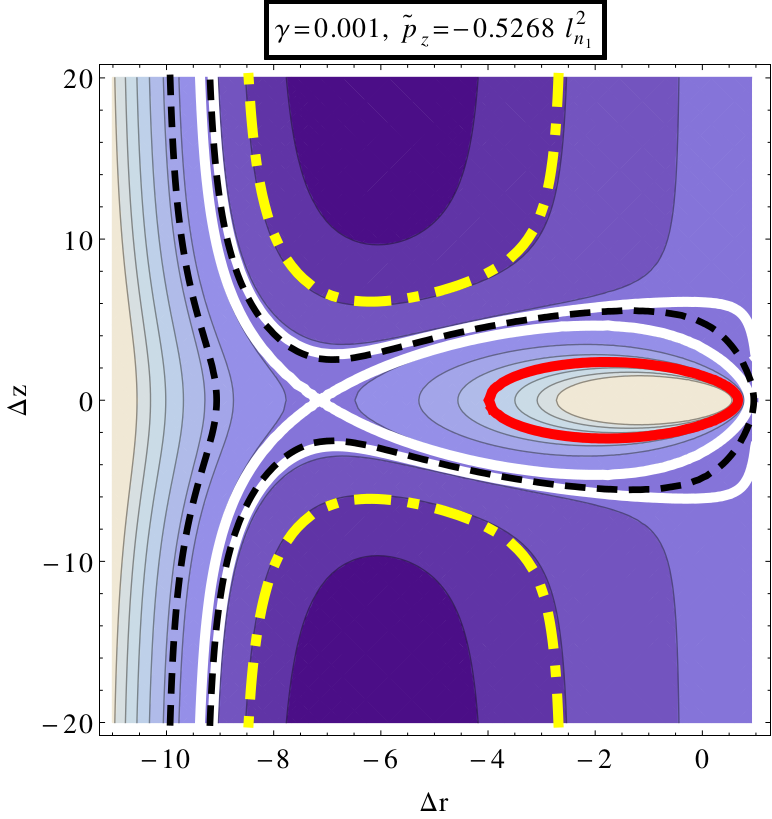}}
  \caption{(Color online.) Examples of phase diagrams when $\gamma \approx 0$. Top left: $\gamma < 0$. Top right: Zoom-in of figure on left. Bottom: $\gamma > 0$.
Refer to figure \ref{GammaGrtLessZero} for the color key.
  }\label{GammaZero}
 \end{figure}

The test ring undergoes an equally diverse array of motions as the pair of vortices for any value of $\gamma$, as demonstrated in figure \ref{GammaZero}.  
Generalising to a non-unit $\chi$ (defined in \eno{chiDef}) for $\gamma \neq \pm 1$, the qualitative classification presented above remains unaffected. However since the energy of the vortex system given in \eno{EnergyTwoV} depends on $\chi$, the quantitative details of the structure of the phase space such as the location of the fixed points and the locus of the separatrix curves, or equivalently the bifurcation curves of Ref.~\cite{Borisov2013}, would change with $\chi$.

The case of $\gamma = -1$, where the rings have equal and opposite winding, is studied in more detail in Appendix \ref{FIXEDPOINTS}, since this is the only case not analyzed in \cite{Borisov2013}.  In particular we give analytic expressions for the location of the saddle point where the attracting and repelling trajectories meet, and for the elliptic fixed point surrounded by pseudo-leapfrogging trajectories.

\begin{figure}
  \centerline{\includegraphics[width=0.5\textwidth]{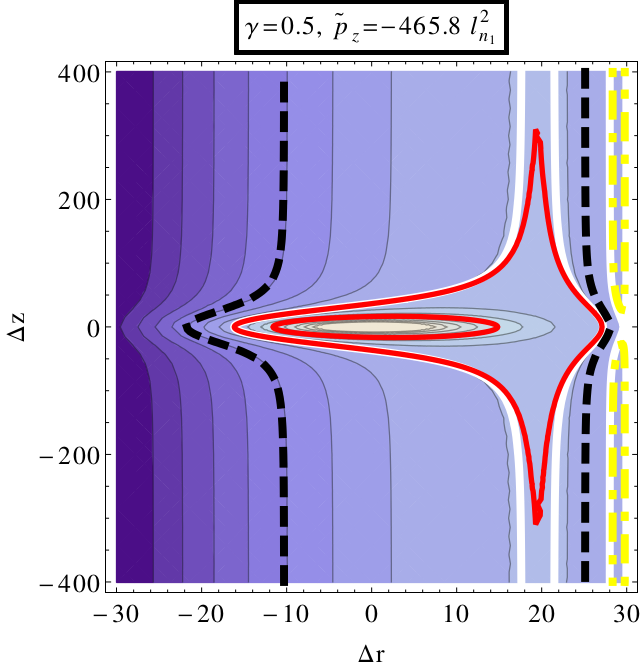}
  \includegraphics[width=0.5\textwidth]{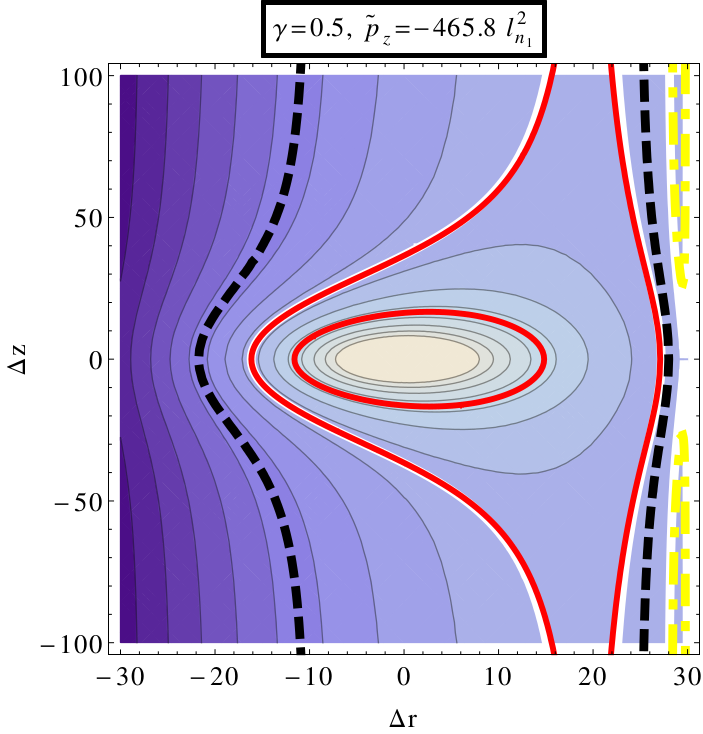}}
  \caption{(Color online.) Another example of a phase diagram when $\gamma > 0$, which exhibits chasing trajectories. Right: Zoom-in of figure on left. 
Refer to figure \ref{GammaGrtLessZero} for the color key.
  }\label{GammaGrtZero2}
 \end{figure}

Investigating the boundary between periodic and aperiodic solutions for general $\gamma$, we find limiting behavior that is one of three types:
\begin{itemize}
\item Chasing limit: in this regime the separation between the rings $\Delta z$ and the duration of a period can become arbitrarily large. An example of chasing vortices for $\gamma > 0$ is shown in figure \ref{GammaGrtZero2}, where chasing occurs as a limiting case of leapfrogging vortices. Chasing is also possible for $\gamma < 0$. In the bottom left panel of figure \ref{GammaGrtLessZero}, a (partial) trajectory corresponding to chasing vortices is shown in thin red.

\item Nesting limit: in this regime $\Delta z = \Delta \dot{z} = 0$.  The second derivative vanishes as well by the equations of motion, but there are not enough free parameters to make the third derivative vanish as well, so the duration of the period remains finite. Refer to the top left  and right panels of figure \ref{GammaGrtLessZero}: nesting happens along the trajectory which as a limiting case of leapfrogging vortices, approaches the saddle point on the left and the right, respectively.

\item Crushed limit: one of the rings shrinks to zero size at the moment where it passes through the other.  When $\tilde{p}_z < 0$, $r_2$ can shrink to zero if $\gamma < 0$, and either $r_1$ or $r_2$ can shrink to zero when $\gamma > 0$.  When $\tilde{p}_z > 0$, $r_1$ can vanish when $\gamma < 0$.  The crushed limit can be found e.g.\ at the right hand boundary of the plots in figures \ref{GammaGrtLessZero}-\ref{GammaZero}. 
\end{itemize}
These limits can be found by solving the equations of motion, which simplify for certain limiting values of $\gamma$.  In Appendix \ref{PhaseBoundaries} we use these to find the set of initial conditions in the parameters $r_1, r_2, \Delta z$ which correspond to periodic behavior for the special cases $\gamma = 1, 0, -1$.  These variables are perhaps more intuitive than the Hamiltonian formalism of \cite{Borisov2013} -- unfortunately, this formalism does not make the full phase diagram any simpler for general $\gamma$, and furthermore, without using the parameters $\tilde{\epsilon}, \tilde{p}_z$ it is difficult to specify a given trajectory uniquely.

\section{Perturbations}
\label{PERTURB}

We will now study the perturbations of the two-ring system and the stability of periodic background solutions, following the analysis of a single ring in \cite{Gubser:2014yma}.  When $\gamma > 0$, we will find that the system is usually stable when all relevant length scales (vortex radii, separation and wavelength of the perturbations) are much larger than the dynamical length scales $\ell_{n_1}, \ell_{n_2}$.  But when $\gamma < 0$, as we show in section~\ref{NEGATIVEGAMMA}, there can be large wavelength instabilities on top of a periodic background motion.  If it is assumed that the effective field theory is only good for wavelengths much larger than $\ell_{n_1},\ell_{n_2}$, then this is the end of the story. However, interesting structure in the perturbation equations arises at length scales somewhat smaller than $\ell_{n_1}$ and $\ell_{n_2}$.  In order to explore it, we assume that effective field theory is in fact good at wavelengths much greater than a physical core size $a \ll \ell_{n_1}, \ell_{n_2}$.  We will find that the time dependence of the background contributes several novel effects to the story, such as ``smearing'' the Widnall instability of a single ring over a larger range of possible modes, and a new class of instability modes due to parametric resonance with the background solution.  

We parametrize small perturbations around the circular vortex rings as follows,
 \eqn{VortexFourier}{
 \vec{X}_\alpha(t,\theta) =   \begin{pmatrix} r_\alpha(t) \cos\theta \\
    r_\alpha(t)  \sin\theta \\
    z_\alpha \end{pmatrix}
     + \epsilon \sum_{m=2}^{\infty} \vec{X}_{m\:\alpha}(t,\theta)
 }
where
 \eqn{VortexModesGen}{
     \vec{X}_{m\: \alpha}(t,\theta) = 
   \begin{pmatrix} (r_{m \alpha}(t) \cos m\theta +  s_{m \alpha}(t) \sin m\theta) \cos\theta \\
    (r_{m \alpha}(t) \cos m\theta +  s_{m \alpha}(t) \sin m\theta) \sin\theta \\
    z_{m \alpha}(t) \cos m\theta +  y_{m \alpha}(t) \sin m\theta \end{pmatrix} \,,
 }
and the index $\alpha$ labels the vortex. 
At the linearized level, the modes $m$ are independent of one another, and perturbations proportional to $\cos m \theta$ and $\sin m \theta$ decouple as well. Thus we are justified in considering each mode $m$ separately, and employing the following parametrization to describe each vortex,
\eqn{Xcircle2}{
  \vec{X}_\alpha(t,\theta) &= 
   \begin{pmatrix} \left(r_\alpha(t) + \epsilon r_{m \alpha}(t) \cos m\theta\right)\cos \theta  \\
    \left(r_\alpha(t) + \epsilon r_{m \alpha}(t) \cos m\theta\right)\sin \theta \\
    z_\alpha + \epsilon z_{m \alpha} \cos m\theta \end{pmatrix} \,.
 } 

Expanded to quadratic order in the perturbations, the scaled Lagrangian for a pair of vortices is then given by
\eqn{LDefTwoV}{
  L_{\rm two \ vortex} = \Big[ L_{\rm two \ vortex} \Big]_{O(\epsilon^0)} + \epsilon^2 \Big[ L_{\rm two \ vortex} \Big]_{O(\epsilon^2)}\,,
    }
where $\Big[ L_{\rm two \ vortex} \Big]_{O(\epsilon^0)}$ is given by \eno{LDefTwoV0} and 
 \eqn{LDefTwoV2}{
\hskip-0.21in\Big[ L_{\rm two \ vortex} \Big]_{O(\epsilon^2)} \!\! =&  \left(\!  \Big[ L_{\rm one\ vortex} \Big]_{O(\epsilon^2)} \!\! + n_1 n_2 {\tilde\lambda \over 2
  }  ( S_{rr} r_{m1}^2 + 2 S_{rz}  r_{m1} z_{m1} + S_{zz} z_{m1}^2 )
    +  ( 1 \leftrightarrow 2 ) \!\right) \cr
    &+ n_1 n_2   \tilde\lambda \Bigg( S_{rr2} r_{m1} r_{m2}  + S_{rz2} r_{m1} z_{m2} +  S_{zr2}   z_{m1} r_{m2} +  S_{zz2} z_{m1} z_{m2}  \Bigg)\, ,
}
where
\eqn{LOV2Def}{
\Big[ L_{\rm one\ vortex} \Big]_{O(\epsilon^2)} =& -\frac{1}{4} n_1 \dot{z}_1 r_{m1}^2 -\frac{1}{2} n_1 r_1 r_{m1}\dot{z}_{m1} \cr 
 + & \frac{\lambda}{2r_1} n_1^2 \left(\left(\frac{1}{8}R_{rr} - \frac{m^2}{2}\log\frac{r_1}{\ell_{n_1}}\right)r^2_{m1}+\left(\frac{1}{8}R_{zz} - \frac{m^2}{2}\log\frac{r_1}{\ell_{n_1}}\right)z^2_{m1}\right).
}
The $m$-dependent constants $R_{rr}$, $R_{zz}$ were given in Ref.~\cite{Gubser:2014yma} as
\begin{equation}
\begin{split}
R_{rr} &= (4m^2 - 1)\left(\frac{\Gamma'\left[m+ \frac{1}{2}\right]}{\Gamma[m + \frac{1}{2}]} - \frac{\Gamma'[1]}{\Gamma[1]} + 2 \log 2\right) - 2(m^2 + 2)\,, \\
R_{zz} &= (4m^2 - 3)\left(\frac{\Gamma'\left[m+ \frac{1}{2}\right]}{\Gamma[m + \frac{1}{2}]} - \frac{\Gamma'[1]}{\Gamma[1]} + 2 \log 2\right) - 2 m^2\,,
\end{split}
\end{equation}
where $\Gamma[m]$ is the Gamma function.
The functions $S_{rr},\, S_{rz}$, and $S_{zz}$ are independent of $m$, and in the small core limit ($a \rightarrow 0$) they are given by
\eqn{SijDef}{
	S_{rr} &= -{1 \over L_7} \left[ {\Delta z^2 \left(\Delta z^2+r_1^2+r_2^2\right) \left(\Delta z^2+(r_1+r_2)^2\right)  }\; K\left(-\frac{4}{s_p^2}\right) \right. \cr
	& \left. - {\left(\Delta z^6+2 \Delta z^4 r_1^2-2 \left(r_2^3-r_1^2 r_2\right)^2+\Delta z^2 \left(r_1^4+10 r_1^2 r_2^2-3 r_2^4\right)\right) } \; E\left(-\frac{4}{s_p^2}\right) \right] \cr
	S_{rz} &= {r_1\Delta z   \over L_7} \left[ { \left(\Delta z^2+r_1^2-r_2^2\right) \left(\Delta z^2+(r_1+r_2)^2\right)}\; K\left(-\frac{4}{s_p^2}\right)\right. \cr 
	&  -\left. {\left(\Delta z^4+r_1^4+2 r_1^2 \left(\Delta z^2+3 r_2^2\right)-7 r_2^4-6 \Delta z^2 r_2^2\right)}\;  E\left(-\frac{4}{s_p^2}\right)\right]\cr
	S_{zz} &= -{1 \over L_7} \left[  {\left(\Delta z^2+(r_1+r_2)^2\right) \left(\Delta z^2 \left(r_1^2+r_2^2\right)+\left(r_1^2-r_2^2\right)^2\right)  }\; K\left(-\frac{4}{s_p^2}\right) \right.  \cr 
	& \left. - {\left(\Delta z^4 \left(r_1^2+r_2^2\right)+\left(r_1^2-r_2^2\right)^2 \left(r_1^2+r_2^2\right)+2 \Delta z^2 \left(r_1^4-6 r_1^2 r_2^2+r_2^4\right)\right) } \;E\left(-\frac{4}{s_p^2}\right) \right]\,,
	}
where $L_7 \equiv \left(\Delta z^2+\Delta r^2\right)^{3/2} \left(\Delta z^2+(r_1+r_2)^2\right)^2$, and $s_p$ is defined in \eno{SpQpDef}.	 The corresponding functions needed in equation \eno{LDefTwoV2}, $S_{r2r2}\,,S_{r2z2}$ and $S_{z2z2}$ can be obtained from \eno{SijDef} simply by switching the indices $(1 \leftrightarrow 2)$.  The remaining functions, $S_{rr2},\, S_{rz2}$ and $S_{zz2}$, are $m$ dependent.  For any specific value of $m$, they can be evaluated by performing the following integrals:
\eqn{Sij2Def}{
	S_{rr2} &= {1 \over (r_1 r_2)^{3/2} } \int_{-1}^1 {du \over v} \left[ {n_{5/2} \over \left(q_p^2+s_p^2+4u^2\right)^{5/2}} 
	+ {n_{3/2} \over \left(q_p^2+s_p^2+4u^2\right)^{3/2}}
	+ {n_{1/2} \over \sqrt{q_p^2+s_p^2+4u^2}} \right]\cr
	S_{zr2} &= -{r_1 \Delta z \over (r_1 r_2)^{3/2}} \int_{-1}^1 {du \over v}\left[- \frac{ 3 \left( \left(u^2-v^2\right)+r_2/r_1\right)\left(u^2-v^2\right) T_{2 m}(v)}{ \left(q_p^2+s_p^2+4u^2\right)^{5/2}} \right. \cr 
	& \qquad \qquad \qquad \qquad + \left. \frac{ \left(u^2-v^2\right) T_{2 m}(v)+ 2 m  u^2 v\, U_{2 m-1}(v)}{\left(q_p^2+s_p^2+4u^2\right)^{3/2} } \right]\cr
		S_{rz2} & = S_{zr2}\left(r_1 \leftrightarrow r_2,\, z_1 \leftrightarrow z_2 \right) \cr
	S_{zz2} &= {1 \over (r_1 r_2)^{3/2}} \int_{-1}^1 {du \over v} \left[ -\frac{3 \Delta z^2 (u^2-v^2) T_{2 m}(v)}{\left(q_p^2+s_p^2+4u^2\right)^{5/2}} + \frac{r_1 r_2 \left(u^2-v^2\right) T_{2 m}(v)}{\left(q_p^2+s_p^2+4u^2\right)^{3/2} } \right. \cr
	 & \qquad \qquad \qquad \qquad \left. -\frac{m^2 r_1 r_2 T_{2 m}(v)}{\sqrt{q_p^2+s_p^2+4u^2} }\right]\,,
	}
where $q_p$ and $s_p$ are defined
in \eno{SpQpDef}, $T_n(v)$ and $U_n(v)$ are Chebyshev polynomials of the first and second kind, respectively, 
\eqn{uvDef}
{
v = \sqrt{1-u^2}\,,
}
and
\eqn{nDef}
{
 n_{5/2} &= 3 (u^2-v^2) (r_1 ( u^2-v^2 )+r_2 )  (r_1+r_2 ( u^2-v^2) )\, T_{2 m}(v) \cr
 n_{3/2} &= (v^2-u^2) (r_1^2+3 r_1 r_2  (u^2-v^2) + r_2^2 )\,  T_{2 m}(v) - 2 m u^2 v  (4 r_1 r_2 u^2 + \Delta r^2 )\, U_{2 m-1}(v)  \cr
n_{1/2} &=   (1+m^2 )  (u^2-v^2 ) r_1 r_2\, T_{2 m}(v)+ 4 m u^2 v \,r_1 r_2\, U_{2 m-1}(v)\,.
}
In the small core limit $q_p \rightarrow 0$, the integrals $S_{ij2}$ in \eno{Sij2Def} can be written as
\eqn{Sij2Form}{
S_{ij2} (r_1, r_2, \Delta z) = S_{ij2}^E (m;\: r_1, r_2, \Delta z) E\left(-{4\over s_p^2}\right) + S_{ij2}^K (m;\: r_1, r_2, \Delta z) K \left(-{4\over s_p^2}\right)\,,
}
where $S_{ij2}^E (m;\: r_1, r_2, \Delta z)$ and $S_{ij2}^K (m;\: r_1, r_2, \Delta z)$ are rational functions of $r_1,\, r_2$ and $\Delta z$ with integer coefficients which depend on $m$. Although we do not have general closed-form expressions for them for all $m$,\footnote{In appendix \ref{SIJ2FNS} we present explicit expressions for these functions in a discrete sum representation for all $m$.} upon evaluating the integrals at individual values of $m$, we find that with all other variables held fixed, the $S_{ij2}$ functions fall off exponentially fast at sufficiently large $m$.\footnote{Fixing the other variables to generic values, the $S_{ij2}$ functions may start exhibiting an exponential fall off starting with $m$ as small as $m=10$.}

The special case of a head-on collision between mirror vortices, i.e.~vortices of equal radii $r_1 = r_2$ with $\gamma = -1$ and ``mirrored'' perturbation amplitudes $r_{m1}=r_{m2}$ and $z_{m1}=-z_{m2}$, was studied in Ref.~\cite{Gubser:2014yma}. Using trigonometric identities, it is straightforward to check that in this special case the integrals in \eno{Sij2Def} and the expressions in \eno{SijDef} combine non-trivially to reproduce the integrals in equation (36) of Ref.~\cite{Gubser:2014yma}, as expected.\footnote{up to a factor of $2 r$, due to the different normalization of the $S_{ij}$ functions in the Lagrangian \eno{LDefTwoV2}.}

The equations of motion have the general form
\begin{equation}\label{ExplicitEOM}
\begin{split}
 \dot{z}_{m1} &= - \frac{\dot{z}_1}{r_1} r_{m1} + \frac{2}{r_1^2} \left(\frac{1}{8}R_{rr} - \frac{m^2}{2}\log \frac{r_1}{\ell_{n_1}}\right)r_{m1} \\ & \, + \frac{2 \gamma}{r_1} \big[S_{rr}r_{m1} + S_{rz}z_{m1} + S_{rr2}r_{m2} + S_{rz2}z_{m2}\big]\, ,\\
\dot{r}_{m1} &= -\frac{\dot{r}_1}{r_1}r_{m1} -\frac{2}{r_1^2} \left(\frac{1}{8}R_{zz} - \frac{m^2}{2}\log \frac{r_1}{\ell_{n_1}}\right)z_{m1}\\ & \,- \frac{2\gamma}{r_1} \big[S_{zz}z_{m1} + S_{rz} r_{m1} + S_{zz2}z_{m2} + S_{zr2}r_{m2}\big]\,,\\
\dot{z}_{m2} &= - \frac{\dot{z}_2}{r_2}r_{m2} + \frac{2\gamma}{r_2^2} \left(\frac{1}{8}R_{rr} - \frac{m^2}{2}\log \frac{r_2}{\ell_{n_2}}\right)r_{m2} \\ & \, + \frac{2}{r_2} \big[S_{r2r2}r_{m2} + S_{r2z2}z_{m2} + S_{rr2}r_{m1} + S_{zr2}z_{m1}\big]\, ,\\
\dot{r}_{m2} &= -\frac{\dot{r}_2}{r_2}r_{m2} -\frac{2\gamma}{r_2^2} \left(\frac{1}{8}R_{zz} - \frac{m^2}{2}\log \frac{r_2}{\ell_{n_2}}\right)z_{m2}\\ & \, - \frac{2}{r_2}\big[S_{z2z2}z_{m2} + S_{r2z2} r_{m2} + S_{zz2}z_{m1} + S_{rz2}r_{m1}\big]\,,
\end{split}
\end{equation}
where $\gamma = n_2/n_1$, we have absorbed a factor of $\tilde{\lambda}n_1$ into the definition of time, and $\dot{z}_i\,,\, \dot{r}_i$ are given by \eno{UnpertEOMs1}-\eno{UnpertEOMs2}.  This is a system of four coupled linear differential equations in four unknowns, with time-dependent coefficients.  Schematically, it has the form
\be
\label{EvolutionEqn}
\frac{d}{dt}\left(
\begin{array}{c}
z_{m1}\\
r_{m1}\\
z_{m2}\\
r_{m2}
\end{array}
\right) = M(t)\left(
\begin{array}{c}
z_{m1}\\
r_{m1}\\
z_{m2}\\
r_{m2}
\end{array}
\right)\,,
\ee
and in general we must solve for the evolution numerically.  We are often interested in studying the stability of periodic background solutions, which can be done using techniques of Floquet theory.  The quantity of interest is the transfer matrix $T$ and its eigenvalues, where $T$ is defined by
\be
\left(
\begin{array}{c}
z_{m1}(t_p)\\
r_{m1}(t_p)\\
z_{m2}(t_p)\\
r_{m2}(t_p)
\end{array}
\right) = T \left(
\begin{array}{c}
z_{m1}(0)\\
r_{m1}(0)\\
z_{m2}(0)\\
r_{m2}(0)
\end{array}
\right)\,,
\ee
where $t_{p}$ refers to the duration of a period of the background solution.  The stability of the perturbations will be determined by the absolute values of the eigenvalues of the transfer matrix -- if any are greater than one, the system will be unstable outside a measure zero subset of initial conditions, while if they are not, the system remains stable.  

  \begin{figure}
  \centerline{
  \includegraphics[width=0.4 \textwidth]{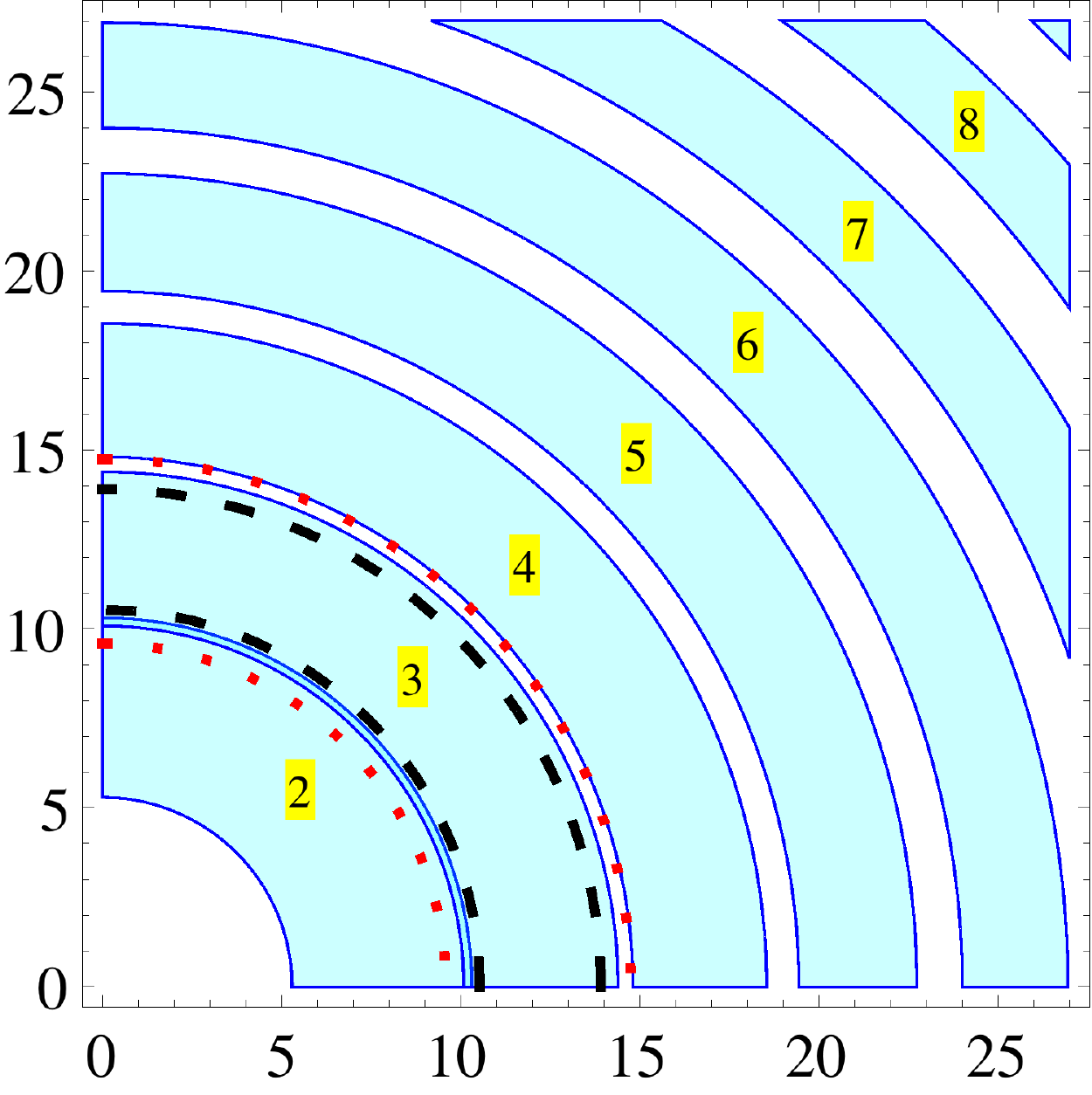}
  \includegraphics[width=0.6 \textwidth]{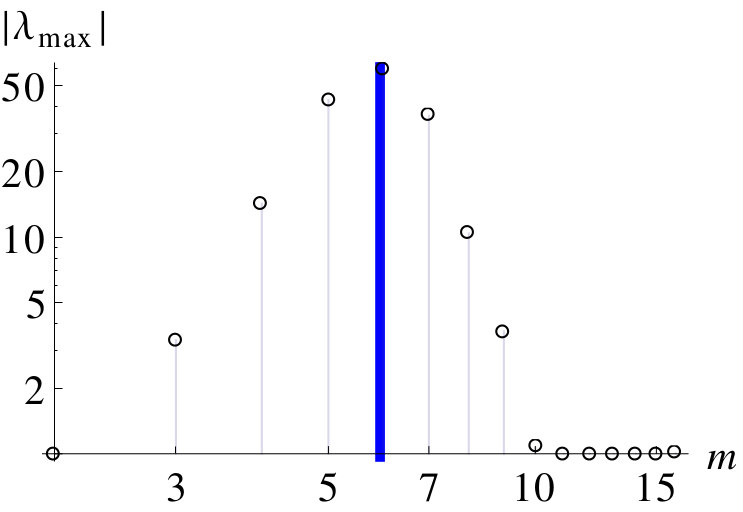}}
    \centerline{
 \includegraphics[width=0.4 \textwidth]{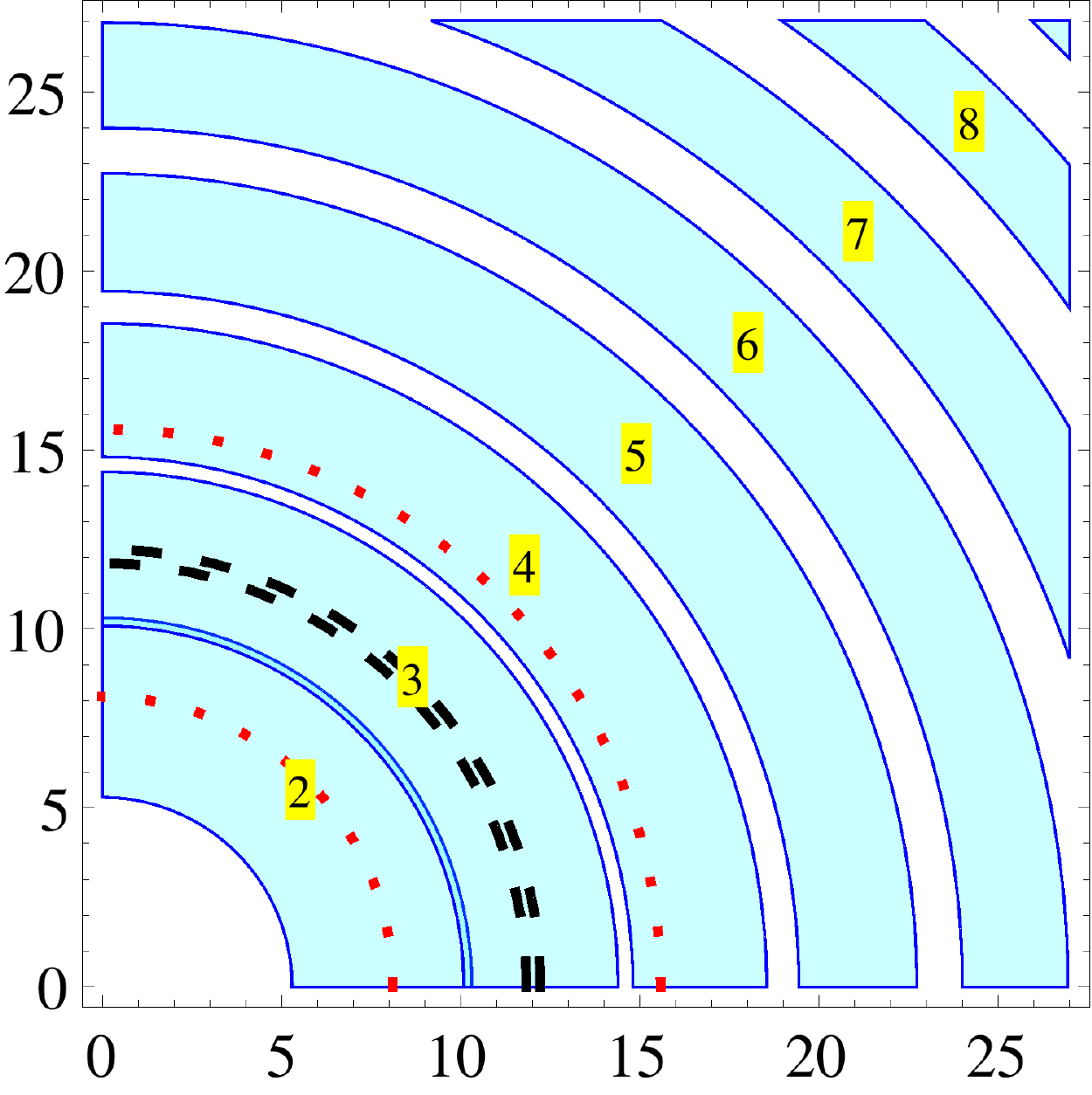}
  \includegraphics[width=0.6 \textwidth]{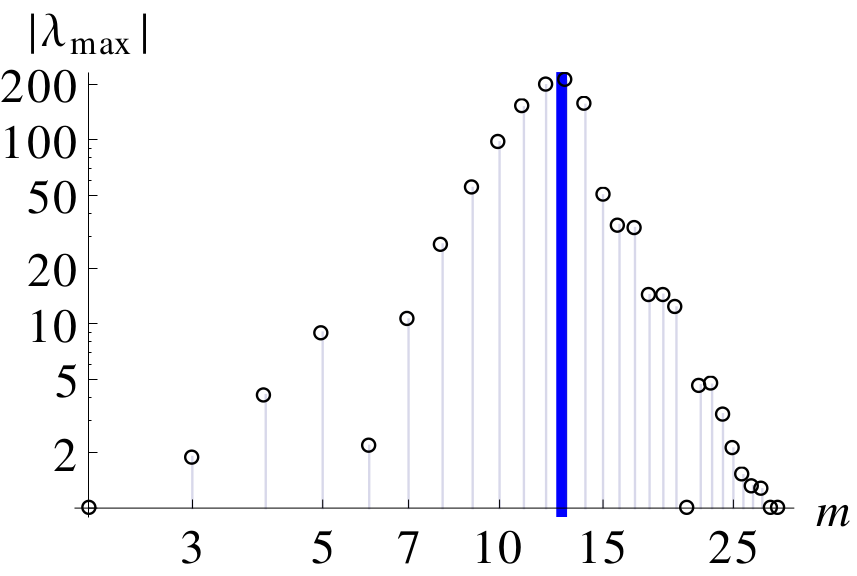}}
  \centerline{
 \includegraphics[width=0.4 \textwidth]{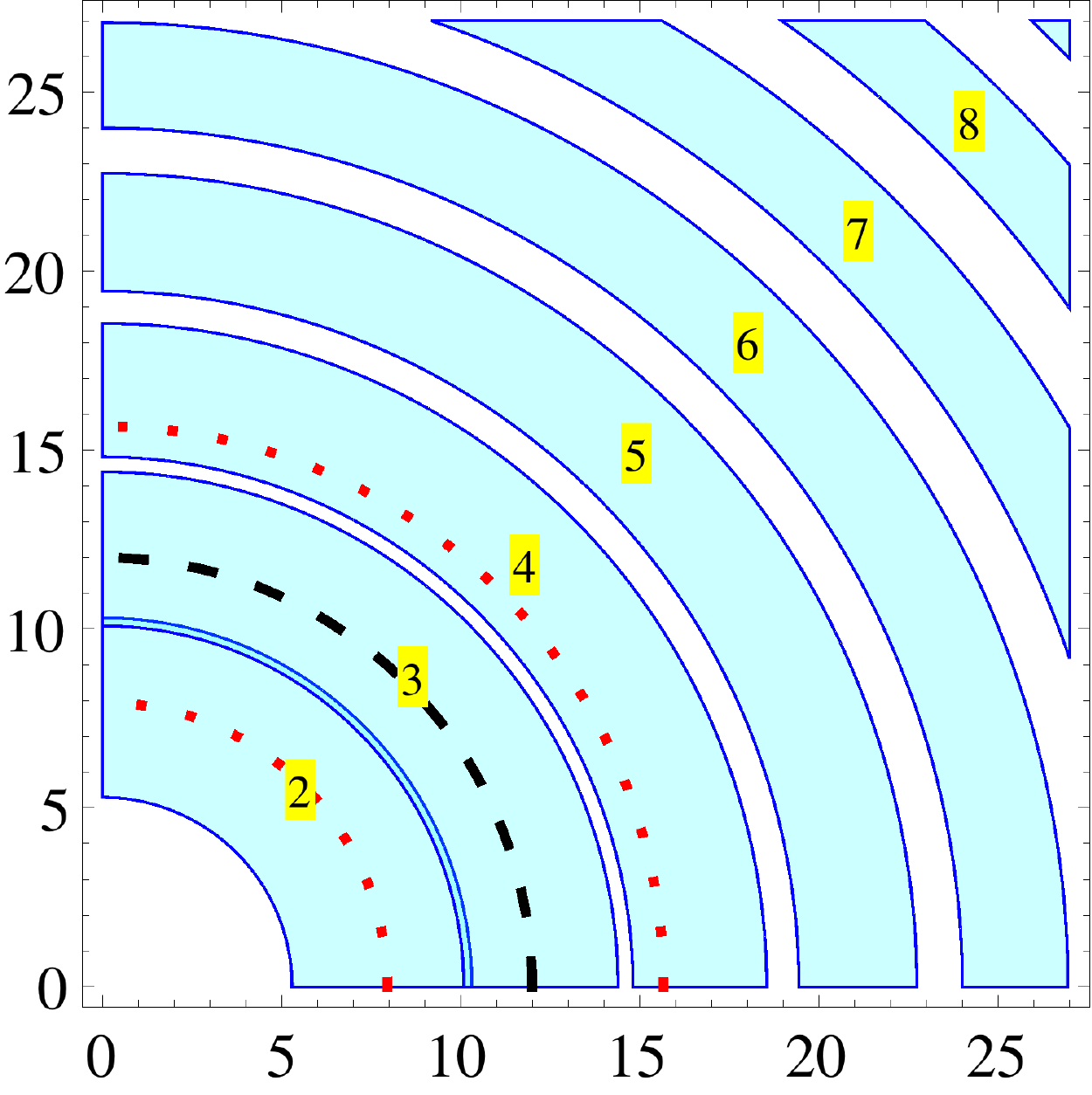}
  \includegraphics[width=0.6 \textwidth]{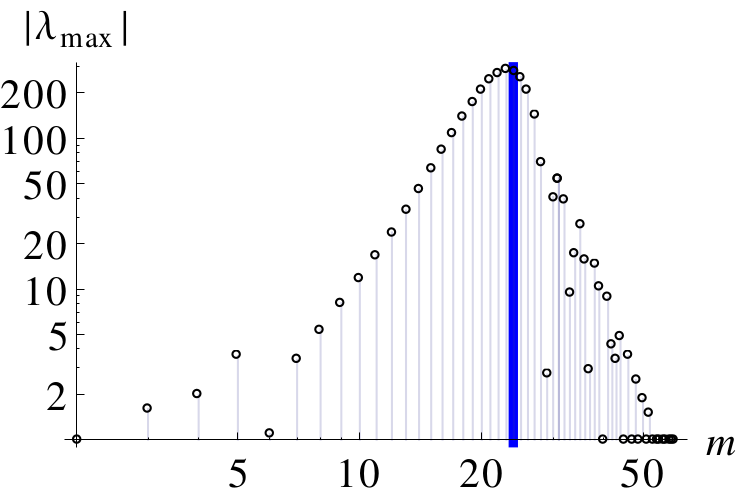}}
  \caption{(Color online.)  {\bf Left}: the radii corresponding to the Widnall instabilities are in cyan (with the unstable mode $m$ in yellow boxes), the radius of ring 1 varies within the dashed black circles, and the radius of ring 2 varies within the dotted red circles.  {\bf Right}: the maximum eigenvalue of the transfer matrix.  Initial conditions are $r_1 = 12 \ell_{n_1}\,,\, r_2 = 13 \ell_{n_1}\,,\, \Delta z = 4 \ell_{n_1}\,,\, \chi =1$. {\bf Top}: $\gamma = 0.5$. {\bf Middle}: $\gamma=0.05$. {\bf Bottom}: $\gamma = 0.01$. The blue vertical line marks the estimate for the most unstable mode $m_{\rm Biggest}$ as obtained from \eno{mBiggestEst}.}\label{pertNumerics}
 \end{figure}
  
A sample of numerical results are plotted in figure \ref{pertNumerics}, for a particular choice of initial conditions for $r_1,\, r_2$ and $\Delta z$, and different values of the parameters: the mode $m$ and the ratio of winding numbers $\gamma = n_2/n_1$.  A few general features to note are the following:
\begin{enumerate}

\item For $\gamma > 0$, the perturbations of the system tend to be stable at low $m$, when the wavelengths of the perturbations are larger than the dynamically generated length scales $\ell_{n_1},\, \ell_{n_2}$, which would correspond to the Widnall instability of a single ring \cite{Gubser:2014yma}.  For $\gamma \ll 1$, however, the length scales $\ell_{n_1},\, \ell_{n_2}$ may be quite different, parametrized by the ratio $\chi = \ell_{n_2} / \ell_{n_1}$ (which is a free parameter in the effective field theory description presented here). 

\item In the limit of large $m$ the system tends to be stable.  We will explain below why this is so, except for the possibility of narrow parametric resonances.

\item Notice that unlike the case of a single ring, a given set of initial conditions may allow more than one value of $m$ to be unstable -- we say that the Widnall instability windows of the single ring have been `smeared out'.

\item In fact, for $0 \leq \gamma \ll 1$, there is a wide range of values of $m$ such that the second ring becomes unstable, which we observe to begin at or around the value of $m$ corresponding to the Widnall instability, although stability is restored for very large values of $m$.  The range of the instability window and the most unstable mode $m_{\rm Biggest}$ grow as $\gamma$ becomes small.  We will find a quasi-analytic estimate for the mode corresponding to the fastest growing instability, as a function of $\gamma$ and $\chi$.

\item The behavior for $\gamma < 0$ and $|\gamma| \ll 1$ is very different, with the instability bulge replaced by a few disconnected bands, and with instability on large scales.  An example is illustrated in figure \ref{negativeGamma}, and we will say more about this case in section \ref{NEGATIVEGAMMA}.

\end{enumerate}

We can understand these behaviors in more detail by studying certain limits where the 4 by 4 matrix $M(t)$ reduces into 2 by 2 blocks.  In particular, four instances where this occurs are in the limits of large $m$, in the limit of small $\gamma$, at an elliptic fixed point for $\gamma < 0$, and in the limit where the separation between the rings is much smaller than the radius. We will discuss each of these limits in some detail.

To facilitate the discussion of these limits, we can decompose $M(t)$ into 2 by 2 blocks as
\be\label{MBlocks}
M(t) = \begin{pmatrix} I & II \\ III & IV \end{pmatrix}\,.
\ee
The blocks are given by
\eqn{MBlocksExplicit}{
\hskip-0.12in I &\!=\! \frac{2\gamma}{r_1}\! \begin{pmatrix} S_{rz} & S_{rr} \\ -S_{zz} & -S_{rz} \end{pmatrix} 
\!+\! \begin{pmatrix} 0 & \frac{2}{r_1^2} \! \left(\frac{1}{8}R_{rr} - \frac{m^2}{2}\log\frac{r_1}{\ell_{n_1}}\right) -\frac{\dot{z_1}}{r_1} \\
 \frac{-2}{r_1^2} \! \left(\frac{1}{8}R_{zz} - \frac{m^2}{2}\log\frac{r_1}{\ell_{n_1}}\right) & -\frac{\dot{r_1}}{r_1} \end{pmatrix}, \cr
\hskip-0.12in II &\!=\! \frac{2\gamma}{r_1} \! \begin{pmatrix} S_{rz2} & S_{rr2} \\ -S_{zz2} & -S_{zr2} \end{pmatrix}, \qquad \qquad
III \!=\! \frac{2}{r_2} \! \begin{pmatrix} S_{zr2} & S_{rr2} \\ -S_{zz2} & -S_{rz2} \end{pmatrix},\cr
\hskip-0.12in IV &\!=\! \frac{2}{r_2} \! \begin{pmatrix} S_{r2z2} & S_{r2r2} \\ -S_{z2z2} & -S_{r2z2} \end{pmatrix} \!\!+\!\! 
\begin{pmatrix} 0 & \frac{2\gamma}{r_2^2} \! \left(\frac{1}{8}R_{rr} - \frac{m^2}{2}\log\frac{r_2}{\ell_{n_2}}\right) -\frac{\dot{z_2}}{r_2} \\ \frac{-2\gamma}{r_2^2} \! \left(\frac{1}{8}R_{zz} - \frac{m^2}{2}\log\frac{r_2}{\ell_{n_2}}\right) & -\frac{\dot{r_2}}{r_2} \end{pmatrix}\!,
}
where $\dot{z}_i\,,\, \dot{r}_i$ are given by \eno{UnpertEOMs1}-\eno{UnpertEOMs2}. 
Using standard results of Floquet theory, the determinant of the transfer matrix can be calculated analytically:
\be
\det T = \exp\left(\int_{0}^{t_p} \tr M(t)\,dt\right) = \left(\frac{r_{1}(0)r_{2}(0)}{r_{1}(t_p)r_{2}(t_p)}\right) = 1\,.
\ee
and therefore the transfer matrix is always unitary.  

\subsection{Large \texorpdfstring{$m$}{m} limit}
\label{LARGEM} 

In the large $m$ limit, numerical results such as in figure \ref{pertNumerics} indicate that the system is stable.  This can be understood by first evaluating the integrals for each individual $m$ in \eno{Sij2Def} and noticing that the $m$-dependent terms in blocks $II$, $III$ fall off exponentially at large $m$. So the transfer matrix decomposes into 2 by 2 blocks.  For $|\gamma| m^2 \log m \gg 1$, the dominant terms are:
\begin{equation}
\begin{split}
\label{largeM}
I &=  \frac{m^2}{r_1^2}\left(\log \frac{4 m \ell_{n_1}}{r_1} + \gamma_{E} - \frac{1}{2}\right) \begin{pmatrix} 0 & 1 \\ -1 & 0 \end{pmatrix} \, \\
IV &=  \frac{\gamma m^2}{r_2^2}\left(\log \frac{4 m \ell_{n_2}}{r_2} + \gamma_{E} - \frac{1}{2}\right) \begin{pmatrix} 0 & 1 \\ -1 & 0 \end{pmatrix}\,.
\end{split}
\end{equation}
(Here $\gamma$ is a parameter, while $\gamma_{E} = 0.577\ldots$ is the Euler-Mascheroni constant.) The two rings decouple and the terms agree with the analysis of \cite{Gubser:2014yma}, albeit with different (non-constant) evolution of the background functions $r_1$, $r_2$.  

If the radius of a given ring is such that the Widnall instability occurs at a large value of $m$, there may now be multiple neighboring values of $m$ that are unstable as well, since the radius of each loop may scan over multiple instability bands during its evolution.  The smearing of the Widnall window is also observed in situations where the instability should correspond to a smaller value of $m$; in this case though, while the origin of the effect may be similar, the analytics are not as easy to understand.  

At any rate, at modes sufficiently larger than the ones corresponding to the Widnall instability, \eno{largeM} guarantees that the rings will be stable.  Exceptions may occur when one of the subleading time-dependent terms is in parametric resonance with the free oscillation.  Although these resonances will be very rare in parameter space, it is worth emphasizing that this is a possibility, and that this represents a novel class of instability sourced by the time-dependence.  We will return to this point in section \ref{LINES}.

\subsection{\texorpdfstring{$m \gg 1$ but $\gamma m^2 \log m \sim 1$}{Large but not too large m}}
\label{MSQRGAMMA1}

In the regime where $\gamma \ll 1$, it is possible to have $m$ large but still $\gamma m^2 \log m \sim 1$.  In figure \ref{pertNumerics} we see that when $\gamma > 0$ this corresponds to the formation of an instability bulge extending over many values of $m$.  The case $\gamma < 0$ is qualitatively different and will be discussed separately.  As we have just argued above, this bulge must end when $\gamma m^2 \log m \gg 1$.  For the values of $m$ in this `not-so-large' limit $I$, $II$ and $III$ are as described in section \ref{LARGEM}, but 
\be
\label{MatIV}
\hskip-0.12in IV \!=\! \frac{2}{r_2} \! \begin{pmatrix} S_{r2z2} & S_{r2r2} \\ -S_{z2z2} & -S_{r2z2} \end{pmatrix} \!\!+\!\! 
\begin{pmatrix} 0 & \frac{2\gamma}{r_2^2} \! \left(\frac{1}{8}R_{rr} - \frac{m^2}{2}\log\frac{r_2}{\ell_{n_2}}\right) -\frac{\dot{z_2}}{r_2} \\ \frac{-2\gamma}{r_2^2} \! \left(\frac{1}{8}R_{zz} - \frac{m^2}{2}\log\frac{r_2}{\ell_{n_2}}\right) & -\frac{\dot{r_2}}{r_2} \end{pmatrix}\!.
\ee
Ring 1 evolves independently with its perturbations governed by $I$, and is stable away from its Widnall bands of instability, while ring 2 will be stable or unstable depending on the 2 by 2 transfer matrix $T_2$ constructed from $IV$.  Just as for the full transfer matrix, it is easy to show that the product of the eigenvalues of $T_2$ is $1$.

To estimate how unstable a mode in the bulge can be, it is useful to consider the eigenvalues of the matrix $M(t)$ at a given instant in time.  It is consistent to take $r_1, r_2, \Delta r, \Delta z$ to be of the same parametric size $r$, or to take $r_1 \sim r_2 \sim r,$ $\Delta r, \Delta z \sim \Delta \ll r$.\footnote{The first case corresponds to an ordinary periodic solution, and the second corresponds to the limit of small separation between the rings.  The chasing limit exhibits seemingly different parametrics with $\Delta z \gg r_1 \sim r_2 \sim r$, but the interaction between the rings is negligible as long as the rings are far apart, so we do not need to worry about this case here.}  Two of the eigenvalues of $M(t)$ come from diagonalizing $I$ and correspond to the perturbations of ring 1; these are constant in time and are complex conjugates with absolute value $(m^2 \log m)/r^2$.  The eigenvalues $\lambda_3, \lambda_4$ determining the fate of the second ring are fixed by the trace and determinant of block $IV$.  Parametrically, these are of size:
\be
\lambda_3 + \lambda_4 \sim \max\Big\{\frac{1}{r^2}, \frac{1}{\Delta^2} \Big\}\, , \qquad \lambda_3 \lambda_4 \sim \max\Big\{\frac{1}{r^4}, \frac{1}{\Delta^4}\Big\} 
\ee
The absolute magnitude of the eigenvalues of $T_2$ must therefore obey
\be
|\lambda_{\rm max}| \leq \exp \left(\int_0^{t_p}|\lambda_{3,4}(t)|dt \right)
\ee
and since parametrically $t_p \sim \min\{\Delta^2, r^2\}$,  in the limit of small $\gamma$, we find $|\lambda_{\rm max}| \lesssim O(1)$ even when the instability bulge is at its largest.  More precisely, $|\lambda_{\rm max}|$ is bounded by a quantity which is parametrically independent of the parameters $\gamma$, $m$, though the bound may still be numerically large due to the exponential.  Indeed, numerical investigations indicate that values of $|\lambda_{\rm max}| \sim 200$ can be expected at the maximum of the bulge.

Note that the argument leading to the bound on $|\lambda_{\rm max}|$ does not explicitly depend on the sign of $\gamma$; since it is an inequality, however, it does not guarantee that the instability must occur.  As we will see in section \ref{NEGATIVEGAMMA} the instability bulge appears for small but positive $\gamma$ but is absent for small and negative $\gamma$.

\subsection{Small \texorpdfstring{$\gamma$}{n2/n1} and general \texorpdfstring{$m$}{m}}
 
In the limit where $\gamma \ll 1$ (but where $m$ is not necessarily large), the matrix $M(t)$ takes on the general form
\eqn{MBlocksExplicit3}{
\hskip-0.12in I & =  \begin{pmatrix} 0 & \frac{2}{r_1^2} \! \left(\frac{1}{8}R_{rr} - \frac{m^2}{2}\log\frac{r_1}{\ell_{n_1}}\right) -\frac{\dot{z_1}}{r_1} \\
 \frac{-2}{r_1^2} \! \left(\frac{1}{8}R_{zz} - \frac{m^2}{2}\log\frac{r_1}{\ell_{n_1}}\right) & -\frac{\dot{r_1}}{r_1} \end{pmatrix}, \cr
\hskip-0.12in II & = 0\,, \qquad \qquad
III =  \frac{2}{r_2} \! \begin{pmatrix} S_{zr2} & S_{rr2} \\ -S_{zz2} & -S_{rz2} \end{pmatrix},\cr
\hskip-0.12in IV &\!=\! \frac{2}{r_2} \! \begin{pmatrix} S_{r2z2} & S_{r2r2} \\ -S_{z2z2} & -S_{r2z2} \end{pmatrix} \!\!+\!\! 
\begin{pmatrix} 0 & \frac{2\gamma}{r_2^2} \! \left(\frac{1}{8}R_{rr} - \frac{m^2}{2}\log\frac{r_2}{\ell_{n_2}}\right) -\frac{\dot{z_2}}{r_2} \\ \frac{-2\gamma}{r_2^2} \! \left(\frac{1}{8}R_{zz} - \frac{m^2}{2}\log\frac{r_2}{\ell_{n_2}}\right) & -\frac{\dot{r_2}}{r_2} \end{pmatrix}\!.
}
In this case, just as in section \ref{MSQRGAMMA1}, ring 1 evolves independently\footnote{This follows from the fact that $II$ vanishes.} and $r_1$ is constant.\footnote{This follows from taking the $\gamma \ll 1$ limit in \eno{UnpertEOMs1}-\eno{UnpertEOMs2}.}  The perturbations obey
\begin{equation}
\begin{split}
\dot{r}_{m1} + \frac{z_{m1}}{4r_1^2} \left( R_{zz} - 4m^2 \log\frac{r_1}{\ell_{n_1}} \right) &= 0\\
\dot{z}_{m1} - \frac{r_{m1}}{4r_1^2}\left( R_{rr} - 4(m^2-1)\log\frac{r_1}{\ell_{n_1}} \right) &=0\,,
\end{split}
\end{equation}
and so they undergo simple harmonic motion with frequency
\be
\omega_m = \frac{1}{4r_1^2}\sqrt{  \left( R_{zz} - 4m^2 \log\frac{r_1}{\ell_{n_1}} \right) \left( R_{rr} - 4(m^2-1)\log\frac{r_1}{\ell_{n_1}} \right) }\;.
\ee
Remember that we have absorbed $\tilde{\lambda}n_1$ into the definition of time in this section.  For the second ring, the background equations of motion have to be solved numerically, but now there are only two remaining equations and two variables to be solved for.  The perturbation equations have the form
\begin{equation}\label{testLimit}
\begin{split}
\dot{z}_{m2} &= \alpha(t)r_{m2} + \beta(t)z_{m2} + \Gamma(t)\\
\dot{r}_{m2} &= a(t)z_{m2} + b(t) r_{m2} + c(t)
\end{split}
\end{equation}
where $\Gamma(t)$ has been capitalized to avoid confusion with the parameter $\gamma$, and the explicit expressions for the functions are given by
\eqn{}{
\hskip-0.11in \alpha &= \frac{2 S_{r2r2}}{r_2} + \frac{2\gamma}{r_2^2}\left(\frac{1}{8}R_{rr}-\frac{m^2}{2}\log \frac{r_2}{\ell_2}\right)- \frac{\dot{z}_2}{r_2}\, , \, \beta = \frac{2 S_{r2z2}}{r_2}\, , \, \Gamma  = {S_{rr2}r_{m1}+S_{zr2}z_{m1} \over r_2/2}\,,\cr
\hskip-0.11in a &= -\frac{2 S_{z2z2}}{r_2} - \frac{2\gamma}{r_2^2}\!\left(\frac{1}{8}R_{zz}-\frac{m^2}{2}\log \frac{r_2}{\ell_2}\right)\!, b = -\frac{2 S_{r2z2}}{r_2} - \frac{\dot{r}_2}{r_2}, \, c = -{S_{zz2}z_{m1}+S_{rz2}r_{m1} \over r_2/2}.
}
These equations \eqref{testLimit} can be written in terms of a pair of second-order ordinary uncoupled differential equations:
\begin{equation}
\begin{split}
\ddot{z}_{m2} + \left(-\frac{\dot{\alpha}}{\alpha} - \beta - b\right)\dot{z}_{m2} + \left(\frac{\dot{\alpha}\beta}{\alpha} - \dot{\beta} + b\beta - a \alpha \right)z_{m2} &= c \alpha + \alpha \left(\frac{d}{d t} - b\right)\left(\frac{\Gamma}{\alpha}\right)\,,\\
\ddot{r}_{m2} + \left(-\frac{\dot{a}}{a} - b - \beta\right)\dot{r}_{m2} + \left(\frac{\dot{a}b}{a} - \dot{b} + b\beta - a \alpha \right)r_{m2} &= \Gamma a + a \left(\frac{d}{d t} - \beta\right)\left(\frac{c}{a}\right)\,.
\end{split}
\end{equation}
These can also be expressed in terms of a sourced Hill equation.  For $z_{m2}$, making the redefinition
\be
z_{m2}(t) = \psi(t)e^{ \int^{t}\frac{1}{2}\left(\frac{\dot{\alpha}}{\alpha} + \beta + b\right)dt'}
\ee
we have
\begin{equation}\label{testHillLimit}
\begin{split}
\left(\frac{d^2}{d t^2} + \left( \frac{\dot{\alpha}\beta}{\alpha} -\dot{\beta} + b\beta - a\alpha + \frac{1}{2}  {d \over d t}   \left( \frac{\dot{\alpha}}{\alpha} + \beta + b \right) - \frac{1}{4}\left(\frac{\dot{\alpha}}{\alpha} + \beta + b  \right)^2\right)\right)\psi\\
=  \left(e^{ - \int^{t}\frac{1}{2}\left(\frac{\dot{\alpha}}{\alpha} + \beta + b\right)dt'}\right)\left(c\alpha + \alpha\left(\frac{d}{d t} - b\right)\left(\frac{\Gamma}{\alpha}\right)\right) \equiv s(t)\,.
\end{split}
\end{equation}
The corresponding equation for $r_{m2}$ has the roles of $\{a, b, c\}$ and $\{\alpha, \beta, \Gamma\}$ reversed.  The solution is therefore
\be
\psi(t) = \psi_0 (t) + \int^t dt' s(t') G(t,t')
\ee
where $\psi_0$ is the homogeneous solution to the Hill equation, $G(t,t')$ is the Green's function, and $s(t')$ is the term sourcing the Hill equation.  In practice this will still need to be solved numerically, but it is simpler than solving the full system.

\begin{figure}[t]
  \centerline{
  \includegraphics[width=0.4 \textwidth]{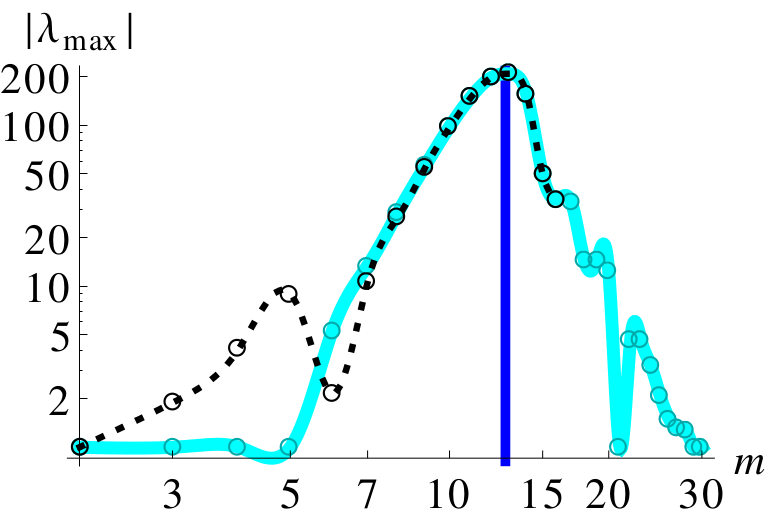}
 \includegraphics[width=0.4 \textwidth]{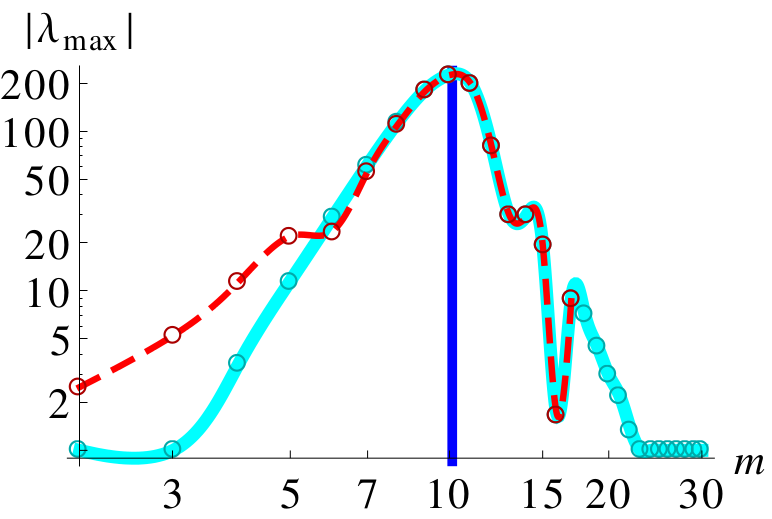}}
 \centerline{
  \includegraphics[width=0.4 \textwidth]{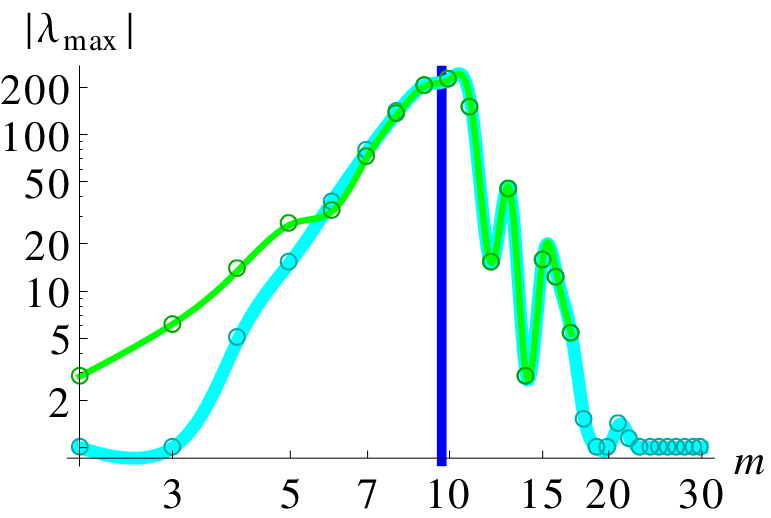} 
  \includegraphics[width=0.4 \textwidth]{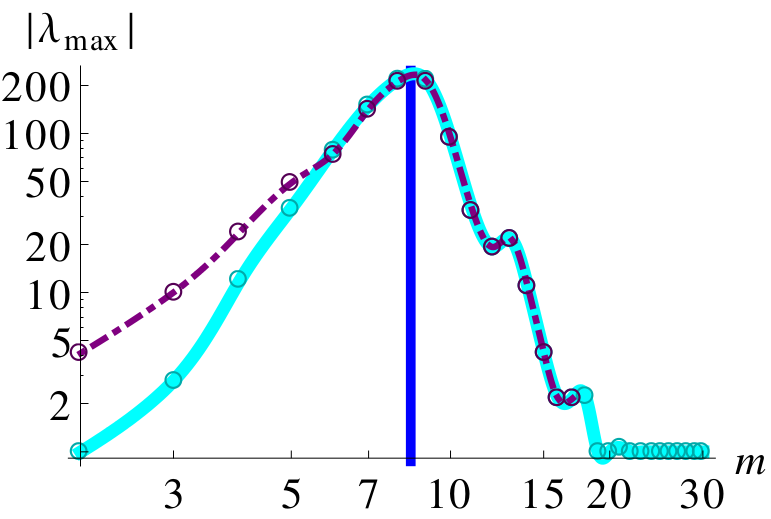}}
  \caption{(Color online.) The maximum eigenvalue and the most unstable mode of the transfer matrix: approximate (cyan), and exact. Note that exact solutions are not shown for the entire range of $m$.  To better guide the eye while comparing, the instabilities have been interpolated to non-integer values of mode $m$ using polynomials of degree $3$.  Initial conditions: $r_1 = 12 \ell_{n_1}\,,\, r_2 = 13 \ell_{n_1}\,,\, \Delta z = 4 \ell_{n_1}\,,\, \gamma = 0.05$. {\bf Top left} (black): $\chi = 1$. {\bf Top right} (red): $\chi = \log 20 \approx 3$. {\bf Bottom left} (green): $\chi = \log 50 \approx 6.21$. {\bf Bottom right} (purple): $\chi = 10$. The blue vertical line marks the estimate for the most unstable mode $m_{\rm Biggest}$ as obtained from \eno{mBiggestEst}.}\label{pertNumerics3}
 \end{figure}

Before going on to estimate the most unstable mode in the next subsection, we note that in most cases when $\gamma$ is small (either positive or negative), 
taking the 2 by 2 matrix $III$ to be vanishingly small at sufficiently large $m$ is a reasonable approximation. This matrix comprises $m$-dependent $S_{ij2}$ functions found by doing the integrals in \eno{Sij2Def}. As noted previously, they fall off exponentially with $m$ at large $m$. As an example, in figure \ref{pertNumerics3} we show a comparison between the approximate values for the highest eigenvalue of the transfer matrix, $|\lambda_{\rm max}|$ as a function of mode $m$ computed by completely neglecting the $S_{ij2}$ functions {\em for all $m$} and thus solving a $2 \times 2$ transfer matrix, and the exact $|\lambda_{\rm max}|$ computed from the full $4 \times 4$ transfer matrix found by solving the exact system. We find that for $\gamma = 0.05$, there is a strong agreement with the exact solution for $m \gtrsim 10$. Numerics show this approximation works at negative $\gamma$ as well, and gets better at even smaller $|\gamma|$. In the Hill equation above, this translates to neglecting the source term for $m \gtrsim 10$.
 
\subsection{The most and first unstable modes}
\label{MOSTUNSTABLE}
In the limit  of small $\gamma$ and large $m$ but $\gamma m^2 \log m \sim 1$, as was discussed in section \ref{MSQRGAMMA1} the transfer matrix reduces to 2 by 2 blocks. For $\gamma > 0$, as we argued, this is the regime in which the instability bulge extends over many values for $m$. This reduction to a two dimensional system allows us to study the instability bulge in more detail.  In this section we will assume $\gamma > 0$ so that the bulge exists, and we will discuss what happens when $\gamma < 0$ in the next subsection.

\begin{figure}
   \centerline{
 \includegraphics[width=0.4\textwidth]{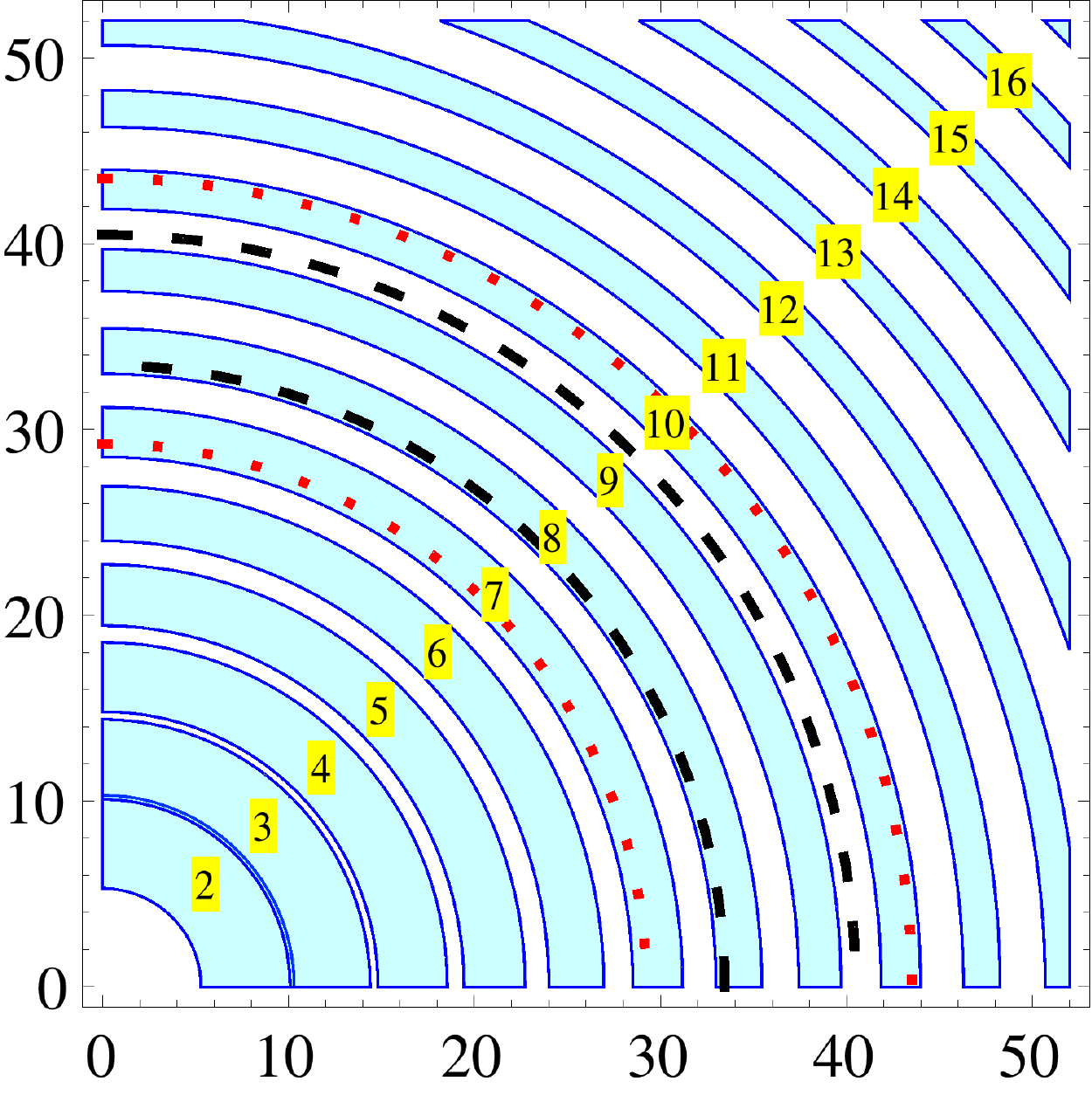}
  \includegraphics[width=0.6\textwidth]{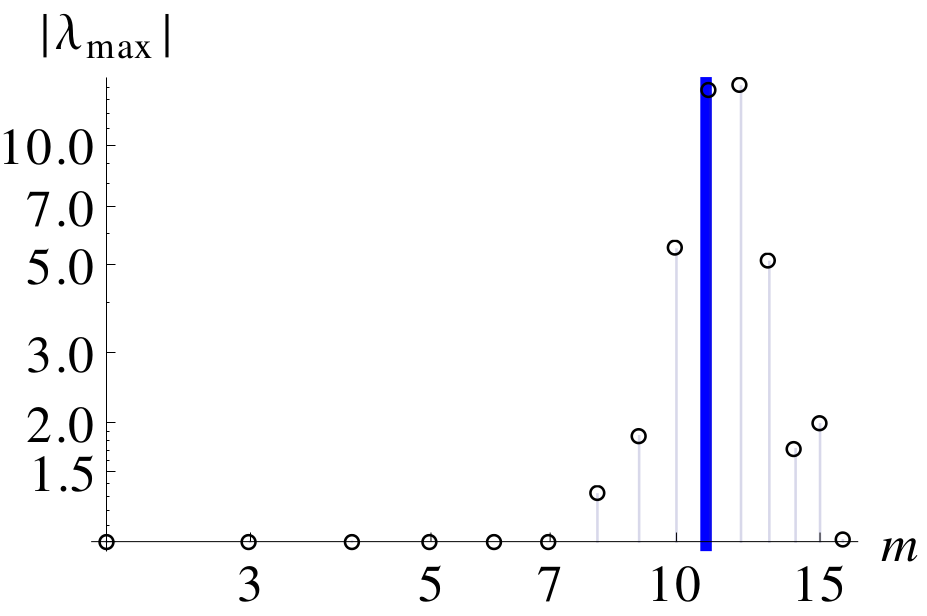}}
   \centerline{
  \includegraphics[width=0.4\textwidth]{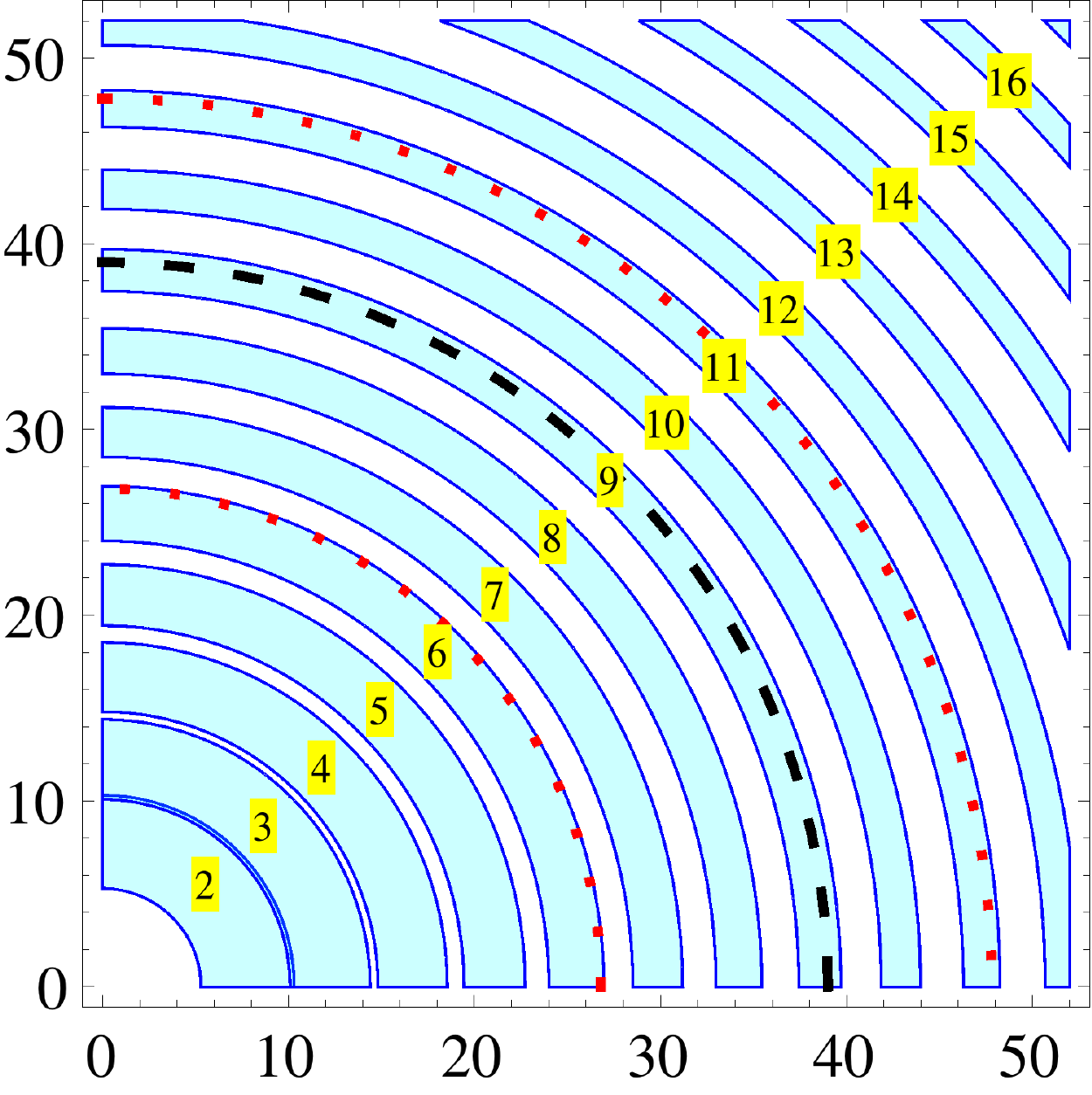}
  \includegraphics[width=0.6\textwidth]{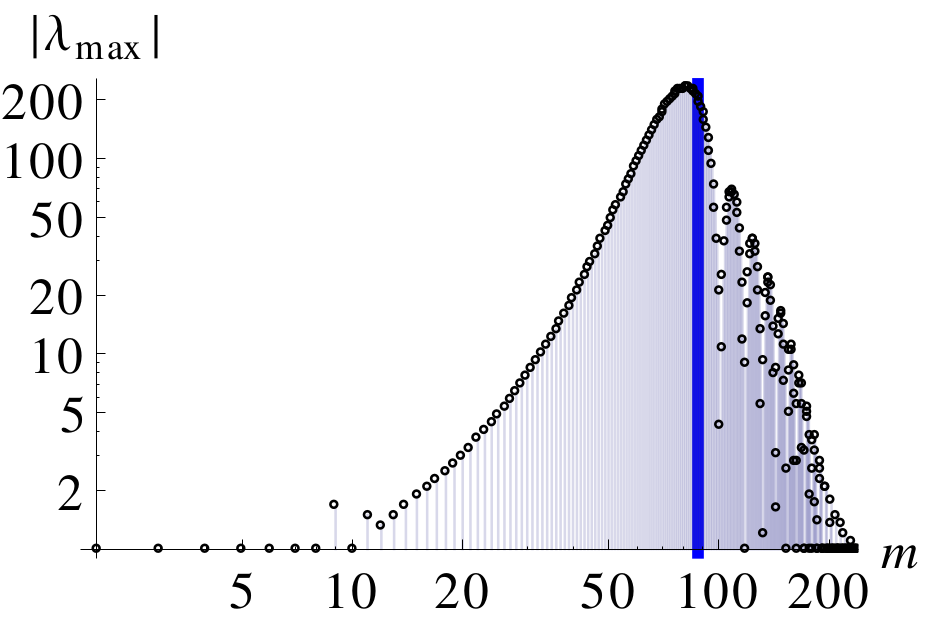}}
  \caption{(Color online.) An example where the rings oscillate across higher Widnall bands. {\bf Left}: the radii corresponding to the Widnall instabilities are in cyan (with the unstable mode $m$ in yellow boxes), the radius of ring 1 varies within the dashed black circles, and the radius of ring 2 varies within the dotted red circles.  {\bf Right}: the maximum eigenvalue of the transfer matrix. Initial conditions are $r_1 = 39 \ell_{n_1}\,,\, r_2 = 33 \ell_{n_1}\,,\, \Delta z = 10 \ell_{n_1}\,,\, \chi=1$. 
  {\bf Top}: $\gamma = 0.5$. {\bf Bottom}: $\gamma = 0.001$. The blue vertical line marks the estimate for the most unstable mode $m_{\rm Biggest}$ as obtained from \eno{mBiggestEst}.
  }\label{pertNumericsLateM}
 \end{figure}

Numerics show that sweeping across a range of values for $m$, an instability bulge develops, rises, reaches a maximum and then falls (see for example figures \ref{pertNumerics}-\ref{pertNumericsLateM}.)  We have previously argued that the bulge ends when $\gamma m^2 \log m \gg 1$. More precisely, the bulge ends when the $m$-{\em dependent} terms in the equations of motion \eno{ExplicitEOM} dominate the $m$-{\em independent} terms. In much the same way, the growth of the bulge at low $m$ corresponds to the case when the $m$-{\em dependent} terms in the equations of motion begin to compete with the $m$-{\em independent} terms. 

We observe that the onset of the bulge tends to occur when $m$ is on or around the number of the Widnall band corresponding to $r_1$ or $r_2$: for example, in figures \ref{pertNumerics} and \ref{pertNumerics3} the bulge rises almost immediately (at or around $m=2$), however, the onset of the bulge can be pushed to higher values of $m$ if the radii of the vortex rings sweep across Widnall bands of higher modes (see figure \ref{pertNumericsLateM}).  This appears to hold for all the numerical examples we have checked, however, it is easy to understand analytically only in limiting cases.  If the radii are so large that the mode number $m_{\rm Widnall}$ corresponding to the Widnall instability is much greater than $1$ (in the sense of section \ref{LARGEM}) the analysis of that subsection makes it clear why the instability appears only around the Widnall value.   

In the regime of interest considered here where $m$ is large but $\gamma$ is small, the 4 by 4 matrix decomposes into 2 by 2 blocks, as described in section \ref{MSQRGAMMA1}. The equations of motion for the perturbations of ring 2, governed by the 2 by 2 matrix $IV$ given in \eno{MatIV} have the form:
\eqn{TestEOMsBulge}{
{r_2 \over 2} \dot{z}_{m2} &= S_{r2z2} z_{m2} + \left(F_z + G_{z}(m) \right) r_{m2} \cr
{r_2 \over 2} \dot{r}_{m2} &= -\left( S_{r2z2} + {\dot{r}_2 \over 2} \right)r_{m2} + \left(F_r + G_{r}(m) \right) z_{m2}\,,
}
where 
\eqn{FGDef}{
F_z = S_{r2r2}  -\frac{\dot{z_2}}{2} & \qquad \qquad G_z(m) = \frac{\gamma}{r_2} \left(\frac{1}{8}R_{rr} - \frac{m^2}{2}\log\frac{r_2}{\ell_{n_2}}\right) \cr
F_r = -S_{z2z2} & \qquad \qquad G_r(m) = \frac{-\gamma}{r_2} \left(\frac{1}{8}R_{zz} - \frac{m^2}{2}\log\frac{r_2}{\ell_{n_2}}\right)\,.
}
The terms $G_z(m)$, $G_r(m)$ vanish for $m \approx m_{\rm Widnall}$ and also for $m=0$.  But for $m = 0$ we know that the system is on the edge of stability: in this case we are expanding perturbatively around the wrong background solution, with a different period and amplitude, so the perturbations grow linearly and the eigenvalues of the transfer matrix are all identically $1$.  At nonzero $m$, the $G_r(m)$ and $G_z(m)$ terms get turned back on when $m \neq m_{\rm Widnall}$ but have different signs above and below $m_{\rm Widnall}$, and so are expected to push in opposite directions -- namely towards stability for $m$ below this value, and instability above. Working with \eno{TestEOMsBulge} numerically, changing the sign of the $m$-dependent terms by hand does indeed alter the behavior of the system from instability to stability and vice versa. For modes with $m > m_{\rm Widnall}$ the $2 \times 2$ block approximation is sufficient for this argument.  For low modes, however, we need to consider the full $4 \times 4$ system to see this explicitly, and we do. Further above $m_{\rm Widnall}$, the instability bulge continues to rise as the $m$-dependent terms become more important.  

It is natural to expect that the bulge peaks when the contributions from $m$-dependent and $m$-independent terms become (more or less) equal.  This expectation can be made more precise as follows. At the peak of the bulge, we expect
\eqn{mBiggestExpect}{
\sqrt{\left< (G_a(m)^2 \right>} \approx \sqrt{\left< ( F_a)^2  \right>} \qquad a= z,r
}
where $\left< \cdots \right>$ stands for time average over one time period of the background solution. When $m \gg 1$, the $m$-dependent functions in \eno{mBiggestExpect} simplify to
\eqn{GApprox}{
G_z(m) = - G_y(m) = {\gamma  m^2 \over 2 r_2} \left( \log  {4 m \ell_{n_2} \over r_2} + \gamma_E -{ 1 \over 2}\right) + O(\log m)\,.
}
We further have $F_z = F_r$ on the $m$-independent side, in the limit of small $\gamma$ -- making use of the $z_2$ equation of motion \eno{UnpertEOMs2}, and the definitions of $Q_0$ and the $S_{ij}$ functions in equations \eno{Q0Def} and \eno{SijDef} respectively, it is easy to confirm the equality holds\footnote{The $z_i$ equations of motion in \eno{UnpertEOMs2} can be rewritten in terms of the $S_{ij}$ functions as
\[\dot{z}_1 =  {1 \over r_1} \log{ \ell_{n_1} \over r_1}  + 2 \gamma (S_{rr} + S_{zz}) \qquad \dot{z}_2 = {\gamma \over r_2} \log{ \ell_{n_2} \over r_2}  + 2 (S_{r2r2} + S_{z2z2})\,. \]
Thus when $\gamma \ll 1$, $F_z \approx F_r$.
} when $\gamma \ll 1$.

\begin{figure}[t]
  \centerline{ \includegraphics[width=0.5 \textwidth]{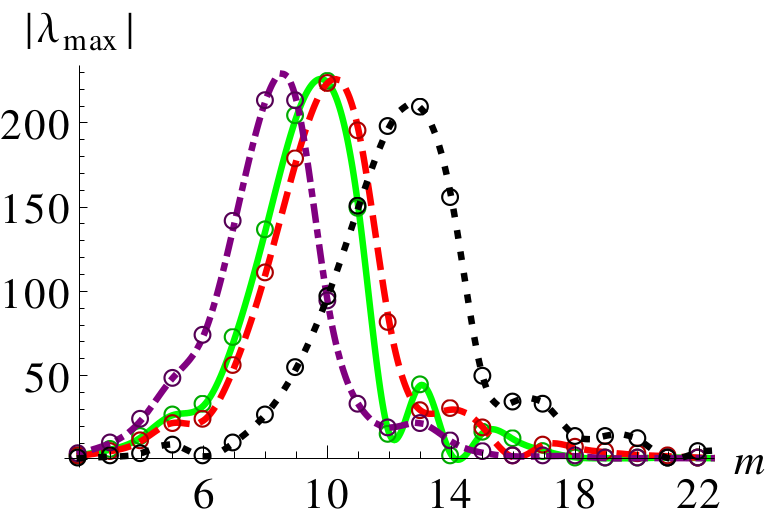} }
  \caption{(Color online.) The variation of the maximum eigenvalue and the most unstable mode of the transfer matrix with $\chi$.  Interpolated to non-integer modes as in figure \ref{pertNumerics3}.  Initial conditions are $r_1 = 12 \ell_{n_1}\,,\, r_2 = 13 \ell_{n_1}\,,\, \Delta z = 4 \ell_{n_1}\,,\, \gamma = 0.05$. {\bf Dotted} (black): $\chi = 1$. {\bf Dashed} (red): $\chi = \log 20 \approx 3$. {\bf Solid} (green): $\chi = \log 50 \approx 6.21$. {\bf Dot-dashed} (purple): $\chi = 10$. 
  }\label{pertNumerics2}
 \end{figure}

To write down an estimate for the most unstable mode $m$, we proceed as follows. First, in the large $m$ limit, approximate the l.h.s. of \eno{mBiggestExpect}  by
\eqn{}{
{\gamma  m^2 \over 2 \left<r_2\right>} \left( \log  {4 m \ell_{n_2} \over \left<r_2\right>} + \gamma_E -{ 1 \over 2}\right) =  F_{\rm rms}\,,
}
where $\left< r_2 \right>$ is the time average of $r_2(t)$ over one time period and $F_{\rm rms} \equiv  \sqrt{\left< ( F_r)^2  \right>} \approx  \sqrt{\left< ( F_z)^2  \right>}$. Then, solving for $m$ we obtain an estimate for the most unstable mode\footnote{When $r_1,\, r_2,\, \Delta z$ are taken to be of the same parametric size $r$, $F_{\rm rms} \sim 1/r$, and \eno{mBiggestEst} reduces to
\[
  m_{\rm Biggest} \sim { 2 \over  \sqrt{ \gamma\, W_0\left( 64  \, e^{2 \gamma_E-1}  \ell_{n_2}^2 / (\gamma\, r^2 ) \right)} } \,.
\]
Then, in the $\gamma \rightarrow 0$ limit, $\gamma m_{\rm Biggest}^2 \log m_{\rm Biggest} \rightarrow 2$, 
which is consistent with the regime  of interest in this subsection.
}
\eqn{mBiggestEst}{
  m_{\rm Biggest} = { \sqrt{ 4\left< r_2 \right> \,   F_{\rm rms}  \over   \gamma\, W_0\left( 64  \, e^{2 \gamma_E-1}  \ell_{n_2}^2   F_{\rm rms} / (\gamma\, \left< r_2 \right> ) \right)} } \,,
}
where $W_0(y)$ is the principal branch of the real valued Lambert-W function. The Lambert-W function $W(y)$ satisfies $y = W(y) e^{W(y)}$, and is defined for $y > -1/e$. The principal branch takes values between $-1$ and $\infty$, and is positive valued for positive $y$. This semi-analytic estimate establishes the functional dependence of the most unstable mode on $\gamma$, in turn explaining why the most unstable modes occur at larger and larger values of $m$ at smaller and smaller $\gamma$. Moreover, the shifting of the most unstable mode to the left as $\chi$ is increased, as shown in figure \ref{pertNumerics2}, can be understood from the dependence of $m_{\rm Biggest}$ on $\ell_{n_2} = \chi \ell_{n_1}$.  

In practice, $m_{\rm Biggest}$ can be computed numerically in a straightforward way, once the time dependences of the background solution are known. These in turn can be computed numerically by solving the integrable system defined by the equations of motion \eno{UnpertEOMs1}-\eno{UnpertEOMs2}. We find close agreement between the numerics and the estimate \eno{mBiggestEst}, to within $10\%$. In fact, quite surprisingly, though there is {\em a priori} no reason for this to be so, the estimate works well at finite $\gamma$ as well, to about $15\%$ accuracy. We show some typical examples in figures \ref{pertNumerics},\ref{pertNumerics3} and \ref{pertNumericsLateM}, where blue vertical lines mark the estimates for $m_{\rm Biggest}$ found using \eno{mBiggestEst} at various values for $\gamma$ and $\chi$.

We find that the reduced wavelength of the most unstable mode is at most an order one multiple of the dynamical length scale. If $\lambda$ is the wavelength of the unstable mode $m_{\rm Biggest}$, then
\eqn{lambdaUnstable1}{
{\lambda \over 2\pi} \approx {\left< r \right> \over m_{\rm Biggest} }\,,
}
where $\left< r \right> \equiv \left< r_2 \right> \approx \left< r_1 \right>$ is the time average of the radii of the rings over one time period. Substituting for $m_{\rm Biggest}$ using \eno{mBiggestEst}, and taking $\left< r \right>  = R \ell_{n_1}$ and $F_{\rm rms} = {\kappa / \left< r\right>}$, we obtain
\eqn{lambdaUnstable1Approx}{
 {\lambda \over 2\pi} \leq 4  e^{\gamma_E -1/2}  \chi \ell_{n_1} \approx 4.321 \ell_{n_2}
 }
where the bound is saturated as $ \gamma R^2 / \kappa \rightarrow \infty$. The bound coincides exactly with the reduced wavelength of unstable modes of a single ring~\cite{Gubser:2014yma} with winding number $n_2 = \gamma n_1$. 
This explains why in the numerics (see for example figure \ref{pertNumericsLateM}) the most unstable modes for a pair of vortex rings with $\gamma > 0$ occur only at values higher than the corresponding Widnall unstable mode for a single ring with the corresponding time averaged radius.

\subsection{\texorpdfstring{$\gamma < 0$}{n2/n1<0}}
\label{NEGATIVEGAMMA}

Even for $\gamma \approx 0$, the case of $\gamma < 0$ is qualitatively very different from $\gamma > 0$.  An example is depicted in figure \ref{negativeGamma}.  The main points to note are:
 \begin{figure}
  \centerline{
  \includegraphics[width=0.3 \textwidth]{case4lite.pdf}
  \qquad \includegraphics[width=0.5 \textwidth]{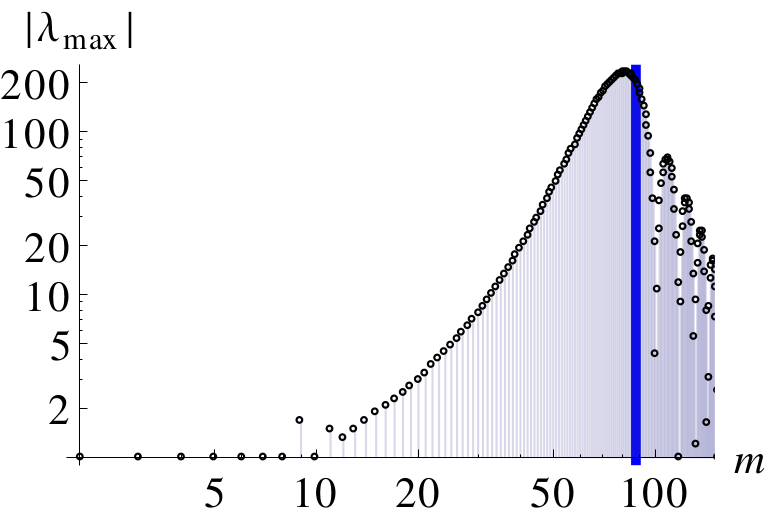}}
    \centerline{
 \includegraphics[width=0.3 \textwidth]{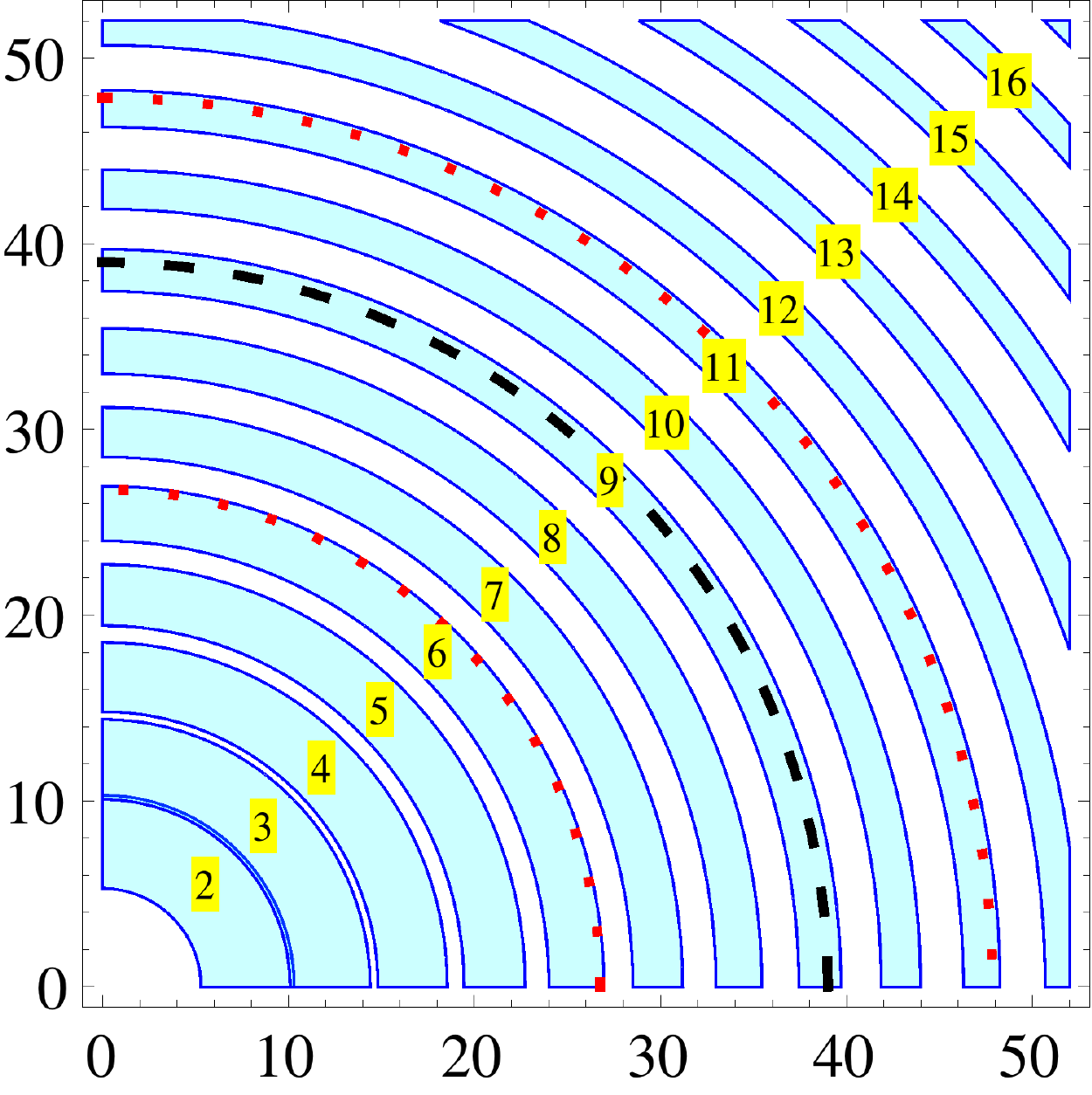}
  \qquad \includegraphics[width=0.5 \textwidth]{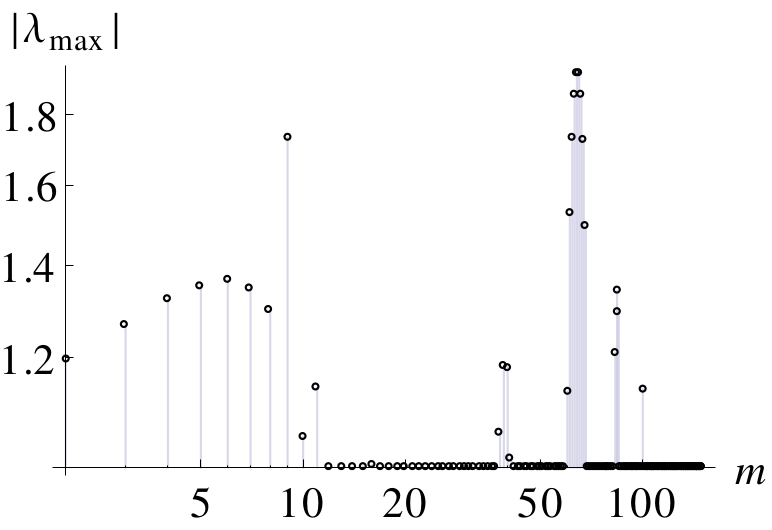}}
         \centerline{
       \includegraphics[width=0.5 \textwidth]{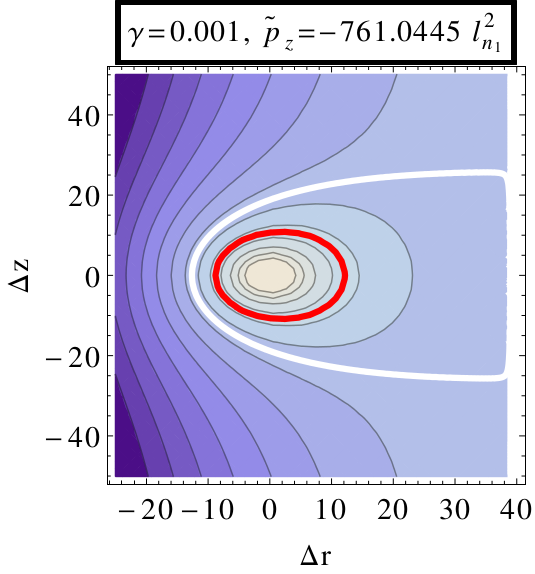}
   \includegraphics[width=0.5 \textwidth]{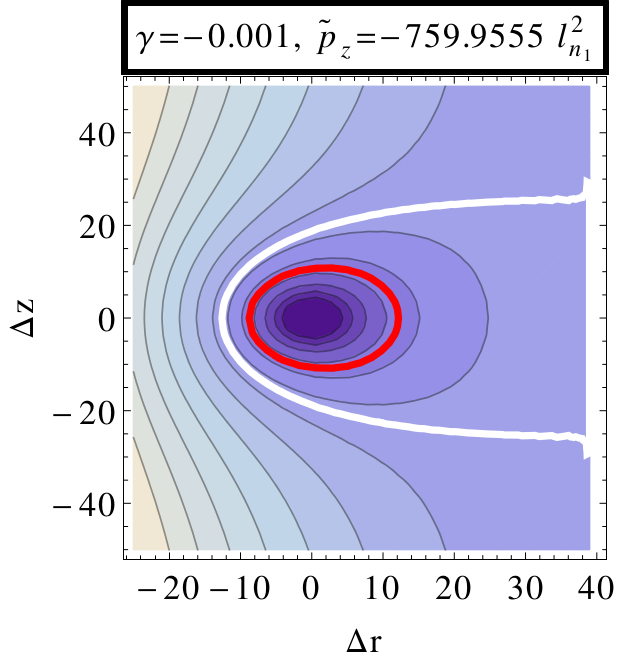}}
  \caption{(Color online.) Initial conditions: $r_1 = 39 \ell_0\,,\, r_2 = 33 \ell_0\,,\, \Delta z = 10 \ell_0$. {\bf Top}: $\gamma = 0.001$. {\bf Middle}: $\gamma= - 0.001$. {\bf Bottom}: Contour plots for top (left panel) and middle (right panel). The red contour marks the trajectory in the background phase space whose instability is being investigated. }\label{negativeGamma}
 \end{figure}
\begin{enumerate}

\item The modes should still be stable for sufficiently large $m$, due to the analysis of section \ref{LARGEM}, which is valid independently of the sign of $\gamma$.  This is borne out by the numerics for very large values of $m$ when $|\gamma| m^2 \log m \gg 1$.

\item The instability bulge analyzed in sections \ref{MSQRGAMMA1} and \ref{MOSTUNSTABLE} is no longer present -- most modes in this range are stable except for a few instability bands, and these have $|\lambda_{\rm max}|$ of order one, rather than in the hundreds.  Although the bounds on potential instabilities derived in section \ref{MSQRGAMMA1} do not a priori depend on the sign of $\gamma$ and are therefore still valid, since they are inequalities, they do not preclude the possibility that the instability does not even occur, in which case the estimate in section \ref{MOSTUNSTABLE} does not apply.  To understand why the qualitative behavior can depend so sensitively on the sign of $\gamma$, notice that the energy landscape for the unperturbed vortices is very different for positive and negative $\gamma$: in the example depicted in figure \ref{negativeGamma}, the periodic trajectory is almost identical regardless of the sign of $\gamma$; however, for positive $\gamma$ it surrounds a global maximum of the energy in phase space, and for $\gamma < 0$ it surrounds a global minimum.  Intuitively, this appears consistent with the observation that the perturbations tend to remain stable for many more values of $m$ in the latter case, and the instabilities that do exist are weaker -- we emphasize, however, that since this argument is based only on the energetics of the background, it is not a proof.

\item For small negative values of $\gamma$, at small length scales the instabilities tend to assemble into  disconnected bands, and the small magnitudes of the eigenvalues suggest that they are on the edge of instability.  Both these points indicate that these instabilities are caused by resonant effects with the time-dependent background.  See figure \ref{negativeGamma} for an example where for $\gamma < 0$ there can be instability bands at intermediate $m$, whereas for $\gamma > 0$ parametric resonances can lead to stability for certain values of $m$ within the instability bulge.

\item Finally, we note that unlike what happened for positive $\gamma$, for negative $\gamma$ we see that for small $m$ and length scales larger than $\ell_{n_1}, \ell_{n_2}$, the perturbations tend to be unstable.  The discussion in section \ref{MOSTUNSTABLE} can be applied here in mirror image: in the limit where $m_{\rm Widnall}$ is large, the system decomposes into 2 by 2 blocks, and since the signs of the $m$-dependent terms are reversed (since they are proportional to $\gamma$), the system will be unstable at small $m$ and mostly stable at large $m$.  The same caveat mentioned earlier applies here as well, though, and to really see that the $m$-dependent terms push the system to instability at small $m$ one needs to keep the off-diagonal interactions,  {\it i.e.}~consider the full $4 \times 4$ system.  Furthermore, as we discussed above, time-dependent effects can alter the fate of some of the modes at short scales, leading to narrow instability or stability bands in spite of the overall pattern. In principle, this may occur at large scales as well, though we have not observed this directly in any of our numerical examples -- there are fewer modes on this side of $m_{\rm Widnall}$, and any resonances that exist will be very narrow.

\end{enumerate}

\subsection{Elliptic fixed point}
\label{EFPPERTURB}

When $\gamma < 0$, there exists an elliptic fixed point (EFP) for values of the rescaled momentum \eno{MomentumTwoV} satisfying an appropriate bound (refer to appendix \ref{FIXEDPOINTS} for a detailed analysis in the special case of $\gamma = -1$).  The EFP occurs at $\Delta z = 0\,,\, \Delta r = (1-\delta_{\rm EFP}) r_1$ where $0 < \delta_{\rm EFP} < 1 $. The background dynamics are very simple, with  $\Delta \dot{z} = \Delta \dot{r} = \dot{r_1} = 0$, and $\dot{z}_1 = {\rm constant}$.  At the level of the background solution, this represents a pair of nested vortices moving together with a constant axial velocity.

The calculation presented in appendix \ref{ELLIPTIC} for computing $\delta_{\rm EFP}$ when $\gamma = -1$ can easily be generalized to any {\em negative} $\gamma$. The perturbative expression for $\delta_{\rm EFP}$ in \eno{DeltaEEfp} generalizes to
\eqn{DeltaEfpGamma}{
\delta_{\rm EFP} \approx \gamma { W_0 \left( { 2\pi - \log(r_1/\ell_{n_1}) \over \gamma  r_1/(\ell_{n_1}  \chi)} \right)  \over 2\pi - \log (r_1/\ell_{n_1})}\,,
}
where $\chi = \ell_{n_2} / \ell_{n_1}$ and $W_0(y)$ is the principal branch of the real valued Lambert-W function. As was the case when $\gamma = -1$, the existence of a real solution for $\delta_{\rm EFP}$ restricts the radius $r_1$ to\footnote{This in turn translates to a bound on the rescaled momentum $\tilde p_z$, but we do not write it down explicitly here.}
\eqn{r1GammaBoundEfp}{
r_1 \geq -{\chi \over \gamma } e\: W_0 \left(-{\gamma \over \chi} e^{2\pi -1} \right)  \ell_{n_1}\,.
}
Additionally, to ensure $\delta_{\rm EFP} < 1$, equation \eno{DeltaEfpGamma} implies we must have
\be
r_1 > e^{\frac{2\pi}{1-\gamma}}\chi^{\frac{\gamma}{1-\gamma}}\ell_{n_1}\,.
\ee
The second condition is weaker than the first if the dynamical length scale $\chi$ satisfies
\eqn{chiBoundEfp}{
\chi < e^{2\pi - 1 + \gamma}\,,
}
and stronger when the sign of the inequality is reversed.

The $m$-independent $S_{ij}$ functions given in \eno{SijDef} simplify to\footnote{From now on, to avoid clutter we shall call $\delta_{\rm EFP}$ simply $\delta$.}
\eqn{SijEfp}{
S_{rr} &= \delta^2 S_{r2r2} = { 2 \delta^2 E\left(-{4 \delta \over (-1 + \delta)^2 } \right) \over (-1+ \delta )(1+\delta)^2}{1 \over r_1}\,, \cr
S_{rz} &= S_{r2z2} = 0\,, \cr
S_{zz} &= S_{z2z2} = { -(1+\delta^2) E\left(-{4 \delta \over (-1 + \delta)^2 } \right) + (1+\delta)^2 K\left(-{4 \delta \over (-1 + \delta)^2 } \right) \over (-1+\delta)(1+\delta)^2}{ 1 \over r_1} \,.
}
The $m$-dependent $S_{ij2}$ functions given by the integrals \eno{Sij2Def} can be computed for general $m$. They are given by
\eqn{Sij2Efp}{
S_{rr2}  &= \frac{2 \sqrt{\pi }\, \Gamma \left(m+\frac{3}{2}\right) \, _2F_1\left(\frac{3}{2},m+\frac{3}{2};m+1;\delta ^2\right)}{\Gamma (m+1)} {\delta ^{m+1}  \over r_1}\,, \cr
S_{rz2} &= S_{zr2} = 0\,, \cr
S_{zz2} &= \frac{\sqrt{\pi } \left(\delta ^2+1\right)  \Gamma \left(m+\frac{1}{2}\right) \, _2F_1\left(\frac{3}{2},m+\frac{1}{2};m+1;\delta ^2\right)}{\left(\delta ^2-1\right) \Gamma (m+1)}{ \delta ^m \over r_1} \cr
  & -\frac{\sqrt{\pi }  \left(\delta ^2 (m-1) m-m (m+1)+1\right) \Gamma \left(m+\frac{1}{2}\right) \, _2F_1\left(\frac{1}{2},m+\frac{1}{2};m+1;\delta ^2\right)}{\left(\delta ^2-1\right) \Gamma (m+1)} {\delta ^m \over r_1}\,,
}
where $_2F_1(a,b;c;z)$ is the hypergeometric function and $0 < \delta < 1$.

Note that \eno{SijEfp}-\eno{Sij2Efp} and $\dot{r}_1=\dot{r}_2 = 0$ imply that all the diagonal entries of the $2 \times 2$ matrix blocks $I, II, III$ and $IV$ written in \eno{MBlocksExplicit} vanish, and the non-vanishing entries are time-independent. The perturbations form an integrable system, governed by the following evolution equation (written in a different basis than in \eno{EvolutionEqn})
\eqn{zmrmEoms}{
{d \over dt}  \begin{pmatrix} {z}_{m1} \\ {z}_{m2} \\ r_{m1} \\ r_{m2} \end{pmatrix}  = \begin{pmatrix} 0 & A \\ B & 0 \end{pmatrix}  \begin{pmatrix} {z}_{m1} \\ {z}_{m2} \\ r_{m1} \\ r_{m2} \end{pmatrix} \,.
}
Here the $2\times 2$  time-independent matrices $A$ and $B$ can be read off of \eno{ExplicitEOM} and \eno{SijEfp}-\eno{Sij2Efp}, but their explicit form is not very illuminating to write down. 
The eigenvalues of the evolution matrix in \eno{zmrmEoms} are given by
\eqn{EigenEfpLambda}{
\pm \sqrt{\frac{1}{2} \left( \tr A B \pm \sqrt{(\tr A B)^2-4 \det A \det B}\right)}\,.
}
The system is stable and undergoes harmonic oscillations if the eigenvalues are purely imaginary, which restricts the matrices $A$ and $B$ to
\eqn{ABstable}{
\tr A B < 0 \qquad {\rm and } \qquad 0 < \det A \det B \leq {1 \over 4} (\tr A B)^2\,.
}
When $\delta \ll 1$, the $S_{ij}$ and $S_{ij2}$ functions in \eno{SijEfp}-\eno{Sij2Efp} provide only subleading corrections to the eigenvalues \eno{EigenEfpLambda}. The leading contribution to the eigenvalues just gives the Widnall frequencies for isolated rings with radii $r_1$ and $r_2$,
\eqn{WidnallFreq}{
\pm \frac{i \gamma_i}{4 r_i^2} \sqrt{\left(4 \left(m^2-1\right) \log \frac{r_i}{\ell_{n_i}} - R_{rr} \right) \left(4 m^2 \log \frac{r_i}{\ell_{n_i}} - R_{zz}\right)}\,, \qquad \gamma_i = \begin{cases} 1\,, & i=1 \\ \gamma & i=2 \end{cases}\,,
}
with no sum over the index $i$. Instabilities arise when the radii fall in a window which makes the eigenvalues above real, corresponding to Widnall instability bands~\cite{Gubser:2014yma}. Turning on $\delta$  modifies the boundaries of the Widnall bands, but these effects are small at small $\delta$.  Numerical exploration reveals the (modified) Widnall instabilities, which occur when $\det A \det B < 0$, are the only instabilities which arise at the EFP.

\section{Pair of vortex lines}
\label{LINES}

In the limit $r_1, r_2 \to \infty$, the system becomes effectively a pair of straight lines circling one another, and the distance between the vortex lines,
 \eqn{DeltaDef}{
 \Delta^2 \equiv  (\Delta r)^2 + (\Delta z)^2
 }
is constant by symmetry.  It is convenient to use $x, y, z$ coordinates instead, where $z$ is  now treated as a parameter running along the length of a vortex line:
\eqn{xCoordLine0}
{
\vec{X}(t,z) = \left(
\begin{array}{c}
x(t)\\
y(t)\\
z\end{array}
\right)\, ,
}
and the rescaled Lagrangian $S =  \rho_0 \mu_1 \int dt L_{\textrm{two lines}}$ 
is given by
\eqn{LDefTwoLines0}
{
L_{\textrm{two lines}} = \left( L_{\textrm{one line}} +  {\tilde{\lambda} \over 2}  n_1 n_2 \hat{L}  \hat{S}_0  \right) + (1 \leftrightarrow 2)\,.
}
Here $\hat{L}$ is the total length of the string.  The terms in the Lagrangian can be evaluated either by inserting \eno{xCoordLine0} directly into the Lagrangian \eno{Action}, or by taking the appropriate limit of the expressions in \eno{LDefTwoV0}.  The terms  on the r.h.s.~in \eno{LDefTwoLines0} are
\eqn{L1S0Def}
{
L_{\rm one\ line} = {\hat{L} n_1 \over 2} (x_1 \dot{y_1} - y_1 \dot{x_1}) -   n_1^2 \tilde\lambda \hat{L} \log {e \hat{L} \over 8{\ell}_{n_1}} \,, \qquad 
\hat{S}_0 = 2  \log\left({\Delta \over \hat{L}}\right)\,,
}
where the dynamical length scales are given by
 \eqn{ellHatDef}{
 {\ell}_{n_i} = {a\over 8} e^{1-\eta_{n_i,\rm bare}/(n_i^2\tilde\lambda)} \qquad i=1,2\,.
}
The vortex line limit of a circular ring corresponds to $\hat{L} \gg {\ell}_{n_i}, \Delta$. Deriving the equations of motion for the background, a single line will not move, but a pair of lines revolves according to 
\eqn{TwoLinesEom}
{
\dot{y}_1 + n_2\frac{2\tilde{\lambda}(x_1 - x_2)}{\Delta^2} = 0 \, , \quad  -\dot{x}_1 + n_2 {2\tilde{\lambda}(y_1 - y_2) \over \Delta^2} = 0\,, \quad (1 \leftrightarrow 2)
}
where $\Delta^2 = (x_1 - x_2)^2 + (y_1 - y_2)^2$ is constant by conservation of energy.  The quantities $(n_1 x_1 + n_2 x_2)$, $(n_1 y_1 + n_2 y_2)$ are constant in time, while for $n_1 + n_2 \neq 0$, $(x_1 - x_2)$ and $(y_1 - y_2)$ undergo simple harmonic motion with frequency
\eqn{OmegaTwoLines}
{
\Omega = \frac{2\tilde{\lambda}(n_1 + n_2)}{\Delta^2}\,.
}
The case where $n_1 + n_2 = 0$ does not correspond to periodic motion and will be treated separately.

We can study perturbations about the background solution just discussed by parametrizing them as follows:
\be\label{xCoordLine}
\vec{X}_\alpha(t,z) = \left(
\begin{array}{c}
x_\alpha(t) + \epsilon \delta x_\alpha(t) \cos(k z)\\
y_\alpha(t) + \epsilon \delta y_\alpha(t) \cos(k z)\\
z\end{array}
\right)\qquad \alpha = 1,2\,.
\ee
Here $k$ is the wavenumber of the perturbation, taken to be $\geq 0$ without loss of generality, and cosine and sine perturbations decouple as for the ring.  Then to $O(\epsilon^2)$ the rescaled Lagrangian $S =  \rho_0 \mu_1 \int dt L_{\textrm{two lines}}$ is given by
 \eqn{LDefTwoLines}{
 L_{\textrm{two lines}} &= \Big[ L_{\rm two\ lines} \Big]_{O(\epsilon^0)}  + \epsilon^2 \Big[ L_{\rm two\ lines} \Big]_{O(\epsilon^2)} \,,
}
where $\Big[ L_{\rm two\ lines} \Big]_{O(\epsilon^0)}$ was written down earlier in \eno{LDefTwoLines0}, and
 \eqn{LDefTwoLines2}{
\hskip-0.05in \Big[ L_{\rm two\ lines} \Big]_{O(\epsilon^2)} \!\!\! =& \! \left( \!\Big[ L_{\rm one\ line} \Big]_{O(\epsilon^2)} \!\!\! +  \frac{\tilde{\lambda}}{2}n_1 n_2 \hat{L} (S_{xx}\delta x^2_{1} + 2S_{xy}\delta x_{1}\delta y_{1}+S_{yy}\delta y^2_{1}) + (1 \leftrightarrow 2)\!\!  \right) \cr
 & + \tilde{\lambda} n_1 n_2 \hat{L} \Bigg(S_{xx2}\delta x_{1}\delta x_{2} + S_{xy2}\delta x_{1}\delta y_{2} + S_{yx2}\delta x_{2}\delta y_{1} + S_{yy2}\delta y_{1}\delta y_{2}\Bigg).
}
 Here $\hat{L}$ is the total length of the string, and $k$ is the wavenumber of the perturbation.  The terms in the Lagrangian can be evaluated either by inserting \eno{xCoordLine} directly into the Lagrangian, or by taking the appropriate limit of the expressions in \eno{LDefTwoV}.  The perturbative terms at $O(\epsilon^2)$ are given by 
\eqn{LOL2Def}{
\Big[L_{\rm one\ line} \Big]_{O(\epsilon^2)} = {\hat{L} n_1 \over 4}\! \left( \delta x_1 \delta \dot{y}_1 - \delta \dot{x}_1\delta y_1 +  n_1\tilde{\lambda}k^2 \! \left(\log {4 k \ell_{n_1}}  + \gamma_{E} + \frac{1}{2}\right)\!(\delta x_1^2 + \delta y_1^2)\right)
} 
and
\begin{equation}
\begin{split}
S_{xx}&= \frac{2}{\Delta^2}\left(\frac{1}{2} - \frac{(x_1 - x_2)^2}{\Delta^2}\right)\\
S_{xy}&= -\frac{2}{\Delta^4}(x_1 - x_2)(y_1 - y_2)\\
S_{yy}&= \frac{2}{\Delta^2}\left(\frac{1}{2} - \frac{(y_1 - y_2)^2}{\Delta^2}\right)\\
S_{xx2}&= \frac{1}{\Delta^2}(-k^2 \Delta^2 K_0 (k\Delta) - k\Delta K_1 (k\Delta) + k^2 (x_1 - x_2)^2 K_2(k\Delta))\\
		&= -{k^2 \over \Delta^2} {(y_1-y_2)^2 } K_0(k\Delta) + {k \over \Delta}  \left(1-{2(y_1-y_2)^2 \over \Delta^2}\right) K_1(k\Delta) \\
S_{xy2} &= S_{yx2} = \frac{ k^2}{\Delta^2}(x_1 - x_2)(y_1 - y_2)K_2(k\Delta)\\
S_{yy2} &= \frac{1}{\Delta^2}(-k^2 \Delta^2 K_0 (k\Delta) - k\Delta K_1 (k\Delta) + k^2 (y_1 - y_2)^2 K_2(k\Delta)) \\
		&= -{k^2 \over \Delta^2} {(x_1-x_2)^2  }  K_0(k\Delta) + {k \over \Delta}  \left(1-{2(x_1-x_2)^2 \over \Delta^2}\right) K_1(k\Delta)\,,
\end{split}
\end{equation}
where $K_n(z)$ are the modified Bessel functions of the second kind.
Note that the final result is symmetric under the exchange of the two lines, as it should be.

It is straightforward to find the equations of motion.  In the rest of the subsection we will take $n_1 = n_2$ for simplicity, since for this case the system can be decomposed into 2 by 2 blocks.  Taking the sums and differences of the equations of motion for each line, we have
\begin{equation}\label{eomTwoLinesPert}
\begin{split}
\left(\frac{\partial}{\partial \tau} - A \sin 2\tau \right)\delta x_{12} = (A \cos 2\tau - B)\delta y_{12}\\
\left(\frac{\partial}{\partial \tau} + A \sin 2\tau \right)\delta y_{12} = (A \cos 2\tau + B)\delta x_{12}
\end{split}
\end{equation}
where $\tau = \Omega t =  4\tilde\lambda n_1  t/\Delta^2$, and 
\eqn{ABDef}{
A &= \frac{1}{2}  \pm  {k^2 \Delta^2 \over 4} K_{2}(k\Delta)\, , \cr
B &= - {k^2 \Delta^2 \over 4} \left( \log {4 k \ell_{n_1}} +\gamma_E + \frac{1}{2} \pm K_0(k\Delta)\right)
 }
for the differences (sums) $\delta x_{12} = \delta x_1 \mp \delta x_2$, $\delta y_{12} = \delta y_1 \mp \delta y_2$. Here the background solution which solves \eno{TwoLinesEom} has been taken to be $(x_1-x_2) = \Delta \cos \tau$ and $(y_1 - y_2) = -\Delta \sin \tau$.

In the limit $k\Delta \ll 1$, $\Omega =   4\tilde\lambda n_1  /\Delta^2$ is the only scale in the problem, and we have:
\begin{equation}
\begin{split}
{\partial \over \partial \tau}(\delta x_1 + \delta x_2 ) = {\partial \over \partial \tau}(\delta y_1 + \delta y_2) = \mathcal{O}(k\Delta)^2 \\
\left(\frac{\partial}{\partial \tau} - \sin 2\tau \right)(\delta x_1 - \delta x_2) = \cos 2\tau(\delta y_1 - \delta y_2) \\ 
\left(\frac{\partial}{\partial \tau} + \sin 2\tau\right)(\delta y_1 - \delta y_2) = \cos 2\tau(\delta x_1 - \delta x_2)\,.
\end{split}
\end{equation}
The solution can be found analytically, and it is
\begin{equation}\label{twiningLimit}
\begin{split}
(\delta x_1 - \delta x_2) = \alpha \sin \tau + \beta(2\tau \sin \tau + \cos \tau) \\
(\delta y_1 - \delta y_2) = \alpha\cos \tau + \beta(2\tau \cos \tau - \sin \tau) \,,
\end{split}
\end{equation}
which is marginally unstable and grows linearly in time under generic initial conditions.  This can also be seen from the the transfer matrix directly in the basis ${\delta y_1 - \delta y_2, \delta x_1 - \delta x_2}$:
\be
\left( \begin{array}{c} \delta x_{12}(2\pi) \\ \delta y_{12}(2\pi) \\ \end{array} \right) = \left( \begin{array}{cc} 1 & 0 \\ 4\pi & 1 \end{array} \right)\left( \begin{array}{c} \delta x_{12}(0) \\ \delta y_{12}(0)\\ \end{array} \right)\, ,
\ee
which has a double eigenvalue at 1.  This limit is known as the twining instability:\ regions of varying separation will wind around one another at different rates, and eventually the phase difference may be of order one even though the gradient terms in the equations of motion are still small.  To confirm this quantitatively, consider sending $\Delta \to \Delta + \delta \Delta$ while keeping the rings straight.  In this case the orbital angular frequency becomes $\Omega + \delta \Omega = \Omega\left(1 - \frac{2\delta \Delta}{\Delta}\right)$, and so then measuring the deviation from the background solution,
\be
\delta(y_1 - y_2) = \delta \Delta (2\tau \cos \tau - \sin \tau)\,, \delta(x_1 - x_2) = \delta \Delta (2\tau \sin \tau + \cos \tau)\,
\ee 
consistent with \eqref{twiningLimit}.

When $k \Delta$ is large, the leading order behavior is the same for $\delta x_1 \pm \delta x_2$ and for $\delta y_1 \pm \delta y_2$, and obeys
\be
\delta(x,y)'' + \left(\frac{k^4 \Delta^4}{16}\left(\log 4 k \ell_{n_1}  + \gamma_E + \frac{1}{2}\right)^2 \right)\delta(x,y) = 0
\ee
up to $O((k\Delta)^0)$ corrections.  These are the familiar Kelvin waves~\cite{Endlich:2013dma}, and the equations of motion for each string decouple in this limit.  They are in general stable, however, this is not the whole story, since the system may develop a narrow parametric resonance between the free oscillations and the forcing term.  Let us see how this works in some more detail.

\subsection{Hill's equation and parametric resonances}

The equations of motion \eqref{eomTwoLinesPert} can be recast in the form of a second-order Hill equation, with a forcing term of period $\tau = 2\pi$:
\eqn{HillEqn}{
 \psi_{\delta x}'' + \left(\frac{3(B^2 - A^2)}{(A\cos 2\tau - B)^2} + \frac{2(B(B+2)-A^2)}{(A\cos 2\tau - B)} + (B+1)^2 - A^2\right)\psi_{\delta x} &= 0\,,\cr
\psi_{\delta y}'' + \left(\frac{3(B^2 - A^2)}{(A\cos 2\tau + B)^2} - \frac{2(B(B+2)-A^2)}{(A\cos 2\tau + B)} + (B+1)^2 - A^2\right)\psi_{\delta y} &= 0\,.
}
Here, 
 \eqn{psiDef}{
 \psi_{\delta x} = { \delta x_1 - \delta x_2 \over \sqrt{ A \cos 2\tau - B}} \qquad \psi_{\delta y} = { \delta y_1 - \delta y_2 \over \sqrt{ A \cos 2\tau + B}}\,.
}
The stability of this class of equations has been extensively studied in the literature (see e.g. \cite{Mamode:2004, Brown:2012}).  When $k \Delta \gg 1$ Hill's equation reduces to the Mathieu equations:
\be\label{MathieuLimit}
\psi''_{\delta x,\delta y} + \left(\frac{k^4 \Delta^4}{16}\left(\log {4 k \ell_{n_1}}  + \gamma_E + \frac{1}{2}\right)^2 - \frac{1}{4} \mp \cos 2\tau\right)\psi_{\delta x,\delta y} = 0
\ee
up to $\mathcal{O}(1/(k\Delta)^2)$ corrections.  The time-dependent terms can lead to parametric resonances when the free oscillation (which is due to the constant part of the forcing term, and which is a function of $k$) is an integer multiple of the background frequency $\Omega$ (which has been rescaled to 1).  Considering the behavior of the system as a function of $k$ in the regime where $k \Delta \gg 1$ and the angular frequency of the free oscillation is large, the resonances will be exponentially narrow.  Subleading corrections in higher powers of $1/k\Delta$ will modify the location and width of the resonances, but only perturbatively.

Can we find values of $\ell_0$, $\Delta$ such that the system is stable for all values of $k$?  We have just shown that the answer is no in the limit where the radius $r$ of the vortex ring goes to infinity -- in this case a generic perturbation will include contributions from the entire continuum of values of $k$, some of which will lie within the narrow resonances.  At finite radius, however, but still preserving $\Delta \ll r < \infty$, the spectrum is discrete, 
\be\label{discreteSpectrum}
k \Delta  = \frac{m \Delta}{r}\,\qquad   m = 2,3,\ldots\,,
\ee
and we have only countably many points to worry about.  In fact, we have only finitely many ($m \lesssim r/a$) points to worry about before the effective field theory we have been using breaks down.  Two regions deserve special attention.  For $k\Delta \lesssim 1$ we are close to the twining instability, and we need to check the stability of the first finitely many points (say $m \lesssim 10r/\Delta$ or so) numerically.  For $\ell_{n_1} \ll \Delta$, numerical studies indicate that the first instability bands are already exponentially narrow.  The region where $k\Delta \gg 1$ and $\log(4k\ell_{n_1}/2) + \gamma_E + 1/2 \approx 0$ (corresponding to $k\ell_{n_1} \approx 0.08$)\footnote{Note that we must assume that $\ell_{n_1} \gg a$ in order for the effective field theory to be valid.  See Ref.~\cite{Gubser:2014yma} for a longer discussion of this point.} is more troubling, since here the Mathieu equation \eqref{MathieuLimit} is unstable.  However, if $r$ is not too large, the spacing of the spectrum \eqref{discreteSpectrum} is so large that there is no value of $m$ that gets close enough to the value where the free oscillation should vanish.  This occurs for
\be
\label{StableMathieu}
\frac{d}{dk} \left(\frac{k^2 \Delta^2}{4}\left(\log (4k \ell_{n_1})  + \gamma_E + \frac{1}{2}\right)\right)\Bigg|_{k\sim 0.08/\ell_{n_1}} \times \frac{1}{r}\gg 1 \quad \longrightarrow \quad \frac{\Delta^2}{r \ell_{n_1}} \gg 1\,.
\ee
If this parameter is much less than one, on the other hand, the spacing of the discrete spectrum is too small, and so there are values of $m$ such that the free oscillation vanishes and the driving term makes the system unstable.  This particular instability occurs at a length scale which is the same parametric size ($k \ell_{n_1} \sim 1$) as the Widnall instability.  Here, however, we emphasize that it is the time-dependent driving terms, and not the circular shape of the ring, that is responsible for the instability.

\subsection{Crow's instability}

We return to the case of vortex lines with $n_1 + n_2 = 0$, corresponding to a pair of counter-rotating vortices with equal and opposite circulation.  We take the initial positions to be $(0,0)$, $(\Delta,0)$ in the $xy$-plane, respectively.  This configuration and its perturbations were first studied in Ref.~\cite{C70}, and they are easy to address in our formalism as well.  The background equations of motion are given by
\be
\dot{y}_1 - \frac{2\tilde{\lambda}n_1(x_1 - x_2)}{\Delta^2} = 0\, , \qquad -\dot{x}_1 - \frac{2\tilde{\lambda}n_1(y_1 - y_2)}{\Delta^2} = 0
\ee
and similarly for $(1 \leftrightarrow 2)$.  The combinations $x_1 - x_2$, $y_1 - y_2$ are constant under time evolution, while for this particular choice of initial conditions we have
\be
(\dot{x}_1 + \dot{x}_2) = 0 \, ,  \qquad (\dot{y}_1 + \dot{y}_2) = - \frac{4\tilde{\lambda}n_1}{\Delta}\,.
\ee
Note that we can think of this solution as a limiting case of the elliptic fixed point discussed in section \ref{EFPPERTURB}, with $\gamma = -1$, $\delta_{EFP} \approx 1$.\footnote{The analysis of perturbations around the elliptic fixed point for $\gamma < 0$ therefore interpolates between two classic problems in the stability of vortices -- Widnall's instability for $\delta_{EFP} \approx 0$, and Crow's instability for $\gamma = -1$ and $r_1 \to \infty$ with $(1-\delta_{EFP})r_1$ held fixed.}  The perturbations obey
\eqn{CrowMatrix}{
\frac{1}{\tilde{\lambda}n_1}\frac{d}{dt}\begin{pmatrix} \delta x_1 \\ \delta x_2 \\ \delta y_1 \\ \delta y_2 \end{pmatrix} &= \left(\begin{pmatrix} 0 & 0 & - 2S_{yy} & -2S_{yy2} \\ 0 & 0 & 2S_{yy2} & 2S_{yy} \\  2S_{xx} & 2S_{xx2} & 0 & 0 \\ -2S_{xx2} & -2S_{xx} & 0 & 0 \end{pmatrix} \right. \cr 
 & \qquad \qquad \left. + \; k^2 \left(\log 4k \ell_{n_1} + \gamma_{E} + \frac{1}{2}\right)\begin{pmatrix} 0 & 0 & 1 & 0 \\ 0 & 0 & 0 & -1 \\ -1 & 0 & 0 & 0 \\ 0 & 1 & 0 & 0 \end{pmatrix}\right) \begin{pmatrix} \delta x_1 \\ \delta x_2 \\ \delta y_1 \\ \delta y_2 \end{pmatrix}\,,
}
where the interaction terms are given by
\begin{equation}
\begin{split}
S_{xx} &= -S_{yy} = -\frac{1}{\Delta^2}\, ,\\
S_{xx2} &= \frac{1}{\Delta^2}\left(k\Delta\, K_{1}(k\Delta)\right)\,,\\
S_{yy2} &= \frac{1}{\Delta^2}(-k^2 \Delta^2 K_{0}(k\Delta) - k\Delta K_{1}(k\Delta))\,.
\end{split}
\end{equation}
Eq.\eno{CrowMatrix} can in principle be derived as a limiting case of \eno{zmrmEoms}, and will therefore have the same 2 by 2 block structure.  Writing the matrix in  \eno{CrowMatrix} in block form,
\be
M = \begin{pmatrix} 0 & A \\ B & 0\end{pmatrix}\,,
\ee
where now
\begin{equation}
\begin{split}
A &= \begin{pmatrix} -2S_{yy} & -2S_{yy2} \\ 2S_{yy2} & 2S_{yy}\end{pmatrix} + k^2 \left(\log 4 k \ell_{n_1} + \gamma_E + \frac{1}{2}\right)\begin{pmatrix} 1 & 0 \\ 0 & -1 \end{pmatrix}\,, \\
B &=\begin{pmatrix} 2S_{xx} & 2S_{xx2} \\ -2S_{xx2} & -2S_{xx}\end{pmatrix} + k^2 \left(\log 4 k \ell_{n_1} + \gamma_E + \frac{1}{2}\right)\begin{pmatrix} -1 & 0 \\ 0 & 1 \end{pmatrix}\,,
\end{split}
\end{equation}
and the eigenvalues of $M$ are given by
\be
\pm \sqrt{\frac{1}{2} \left(\tr A B \pm \sqrt{(\tr AB)^2 - 4 \det A \det B}\right)}\,.
\ee
\begin{figure}[t]
  \centerline{\includegraphics[width=0.5\textwidth]{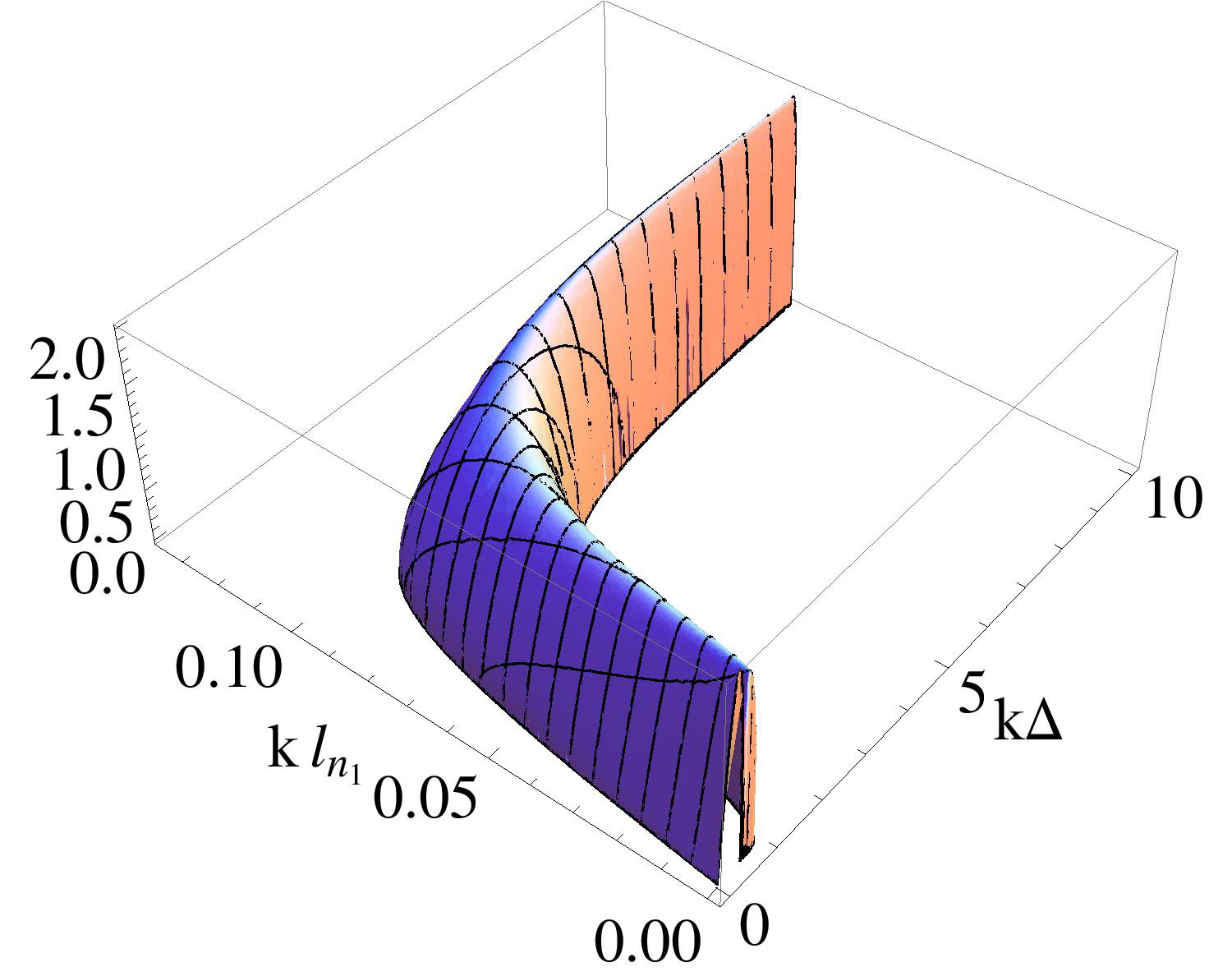}
  \includegraphics[width=0.5\textwidth]{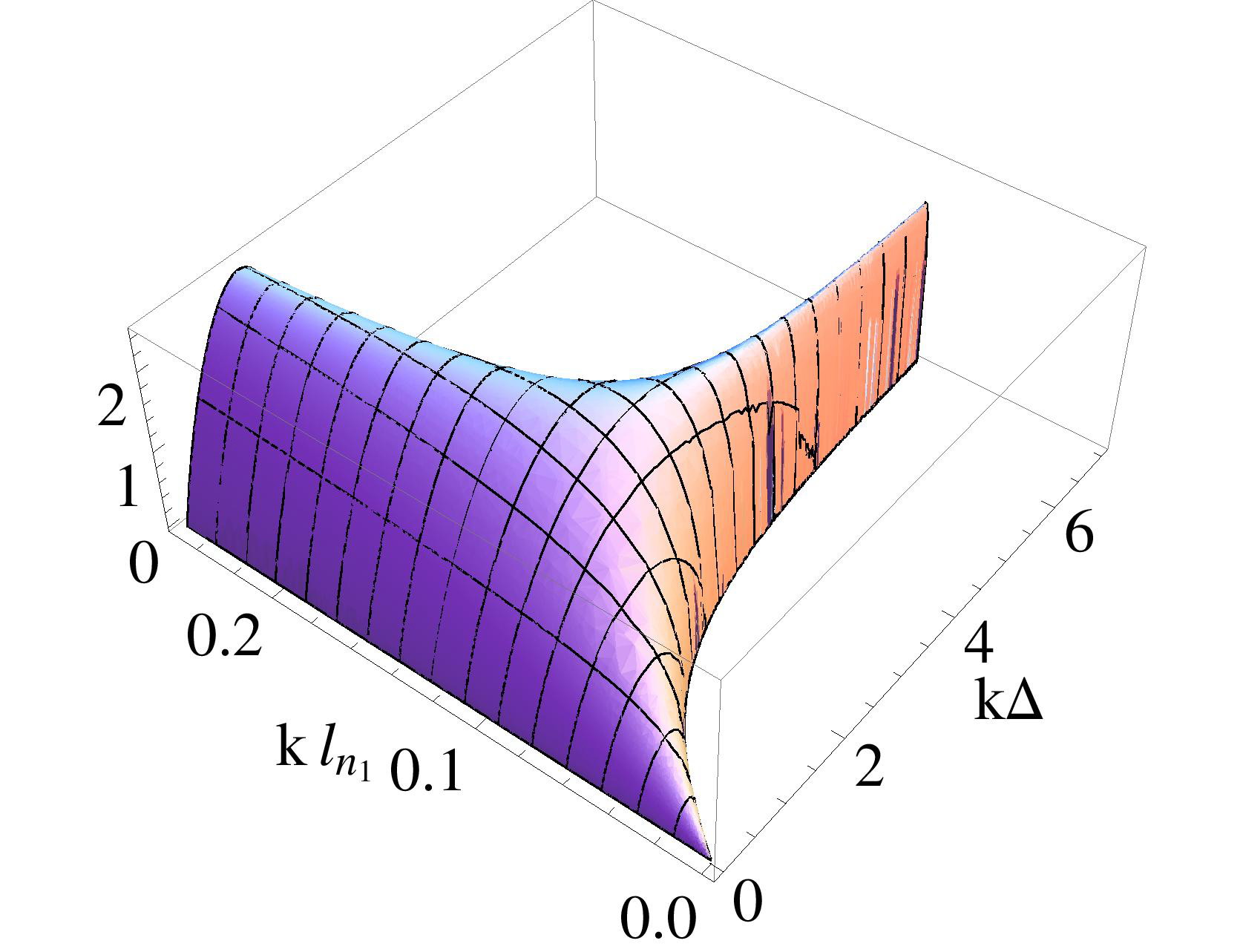}}
  \caption{(Color online.) The positive real eigenvalues of the matrix in \eno{CrowMatrix}, plotted as a function of $k \Delta$ and $k\ell_{n_1}$.}\label{CrowPlots}
 \end{figure}
As in the case of the general elliptic fixed point in section \ref{EFPPERTURB}, writing the explicit expressions for the eigenvalues is straightforward but not particularly illuminating.  Note that the eigenvalues come in pairs whose members differ by a relative minus sign, so there are only two quantities to calculate.  It is also straightforward to show, using the properties of $A$ and $B$, that the argument of the inner radical is equal to $(A_{11}B_{12} - A_{12}B_{11})^2$, and since this is always positive, each pair of eigenvalues is either purely real (corresponding to an instability) or purely imaginary (corresponding to stability).  The location of the unstable region corresponding to each pair, and the magnitude of the (positive) real eigenvalue is shown in figure \ref{CrowPlots}.
To compare this to the original analysis \cite{C70}, we need to start with a particular choice of value for $\ell_{n_1}/\Delta$, (either calculated or observed experimentally in a wind tunnel) and it is simple to use the general expression for the eigenvalue to find the maximally unstable value of $k$. 

\section{Discussion}
\label{DISCUSS}

In this paper we have studied the behavior of a pair of coaxial vortex rings.  At the level of the background evolution we have generalized the phase diagram to include the possibility of vortices having different core sizes when the ratio of their winding numbers is not $\pm 1$, by introducing a parameter $\chi = \ell_{n_2} / \ell_{n_1} \neq 1$. We have also filled in a gap in the phase diagram analysis of Ref.~\cite{Borisov2013}, by including the case of vortices with opposite circulation, {\em i.e.} $\gamma = -1$.

At the level of perturbations, we have analyzed the stability of linearized perturbations around periodic, axially symmetric background solutions.  The system simplifies dramatically in certain limits where the coupled 4 dimensional evolution reduces to two independent 2 dimensional ones, and the $4 \times 4$ transfer matrix decomposes into $2 \times 2$  blocks.  Similar to what was found for the stability of a single ring~\cite{Gubser:2014yma}, for a ratio of circulations $\gamma > 0$ the pair of vortex rings are in general found to be stable when all length scales are larger than the dynamical scales $\ell_{n_1}, \ell_{n_2}$; below this scale, the Widnall instability of a single ring is widened into an instability band for a vortex pair, which may include many modes for the test ring in the limit of small $\gamma$.  More precisely, when $\gamma > 0$ the most unstable mode is estimated by \eno{mBiggestEst}. The reduced wavelength of this mode was found to satisfy
\eqn{}{
 {\lambda \over 2\pi} \leq 4  e^{\gamma_E -1/2}  \chi \ell_{n_1} \approx 4.321 \ell_{n_2}\,.
 }
The upper limit corresponds to the reduced wavelength of unstable modes of a single ring~\cite{Gubser:2014yma} with winding number $n_2 = \gamma n_1$. Thus the largest instabilities for a pair of vortex rings with $\gamma > 0$ arise only at wavelengths comparable to or shorter than the dynamical length scale, and in particular at wavelengths shorter than the corresponding unstable wavelengths for isolated rings.

In addition, unlike the single ring, the paired vortex rings exhibit a novel class of instability sourced by the time dependence of the background.  We have analyzed this issue in detail for narrow parametric resonances when the rings are close together.

We have presented a qualitative picture of the instabilities when $\gamma < 0$. The cases of positive and negative $\gamma$ are very different, primarily because of the different phase space structure. In particular we found that whereas for $\gamma > 0$ modes with $m$ below the corresponding Widnall unstable mode for a single ring are prohibited from becoming unstable, when $\gamma < 0$ such modes do become unstable, signalling instabilities at wavelengths larger than the dynamical length scale. 

It seems worth investigating the effects of the background phase space on the stability of perturbations in more detail, especially near the separatrices in the phase space such as near trajectories corresponding to chasing or nesting vortices.  Although we have not analyzed these in detail, preliminary studies of nesting vortices indicate that the behavior is moderated at low $m$ in the nesting limit, becoming less unstable for positive $\gamma$ and less stable for negative $\gamma$. Additionally, it would be interesting to investigate in more detail periodic motion near the limiting case of $\gamma = -1$ where it was found~\cite{Gubser:2014yma} that the most unstable mode had a wavelength much larger than the dynamical length scale.

Another exercise which may be of interest is to study the stability of the rings beyond the linearized level: the twining instability is an example of a motion which is perturbatively unstable and yet remains bounded, and it would be interesting to understand whether this may be true for other instabilities as well.  We also cannot discount the possibility that the evolution at linearized level may become very large, so that a small but finite initial perturbation becomes nonlinear during its evolution.  

An effective action starting from the Gross-Pitaevskii action (but without including the Nambu-Goto term) was recently used to study instabilities to vortex-sound interactions of a pair of point vortices in a two dimensional superfluid~\cite{Lucas:2014tka}.  This work also briefly discusses the so-called `dynamical instability' of vortices with large winding number to decay into vortices of unit circulation, investigated in \cite{Pu:1999, Simula:2002, Mottonen:2003, Shin:2004}: this arises due to quantum mechanical effects at core sizes, and it would be interesting to generalize our formalism to include such effects.  At the classical level, it would also be of interest to use effective field theory techniques, such as developed in Ref.~\cite{Horn:2015zna}, to study long-range interactions of bound states of vortex rings, either with one another or with external sound waves.

\section*{Acknowledgements}
We thank Martin Kruczenski, Revant Nayar, Alberto Nicolis, and Riccardo Penco for very helpful discussions, and we are particularly indebted to Alberto Nicolis and Riccardo Penco for collaboration on related issues.  This work was supported in part by the United States Department of Energy under contracts DE-FG02-91-ER40671 and DE-FG02-92-ER40699, and by the National Science Foundation under grant No. PHY-1316033.  BH was also supported in part by National Science Foundation grant No. PHYS-1066293 and the hospitality of the Aspen Center for Physics.  

\clearpage
\appendix

\section{Fixed points in \texorpdfstring{$\gamma=-1$}{n2/n1=-1} phase space}
\label{FIXEDPOINTS}

In order to analyse the phase space when $\gamma=-1$, we focus on finding its fixed points in this section. To do so, first rewrite the equations of motion \eno{UnpertEOMs1}-\eno{UnpertEOMs2} as
\eqn{UnpertEOM}{
{d v_i \over dt}
 = f_i(v_j) \,,
}
where $v_i = \left(\Delta z,\,\Delta r\right)$ and $f$ is a $2 \times 1$ vector which is a non-linear function of $\Delta z$ and $\Delta r$.\footnote{Working with coordinates $\left( r_1,\,\Delta r,\,z_1,\,\Delta z\right) $, the constant of motion \eno{MomentumTwoV} gives an algebraic relation between $r_1$ and $\Delta r$. The constant of motion \eno{EnergyTwoV} in turn yields an (implicit) algebraic relation between $\Delta r$ and $\Delta z$. Were the relation between $\Delta r$ and $\Delta z$ invertible, one would need only solve a single first order non-linear ODE to determine the dynamics of the background solution. Since that is not the case, instead of directly making use of the constant of motion \eno{EnergyTwoV}, we rewrite the problem as two coupled first order ODEs, as written in \eno{UnpertEOM}. } 
We can classify all the fixed points of the system by evaluating eigenvalues of the Jacobian matrix, defined as $J_{ij} \equiv \partial f_i / \partial v_j$, at each fixed point. 

 \begin{figure}
  \centerline{\includegraphics[width=0.55\textwidth]{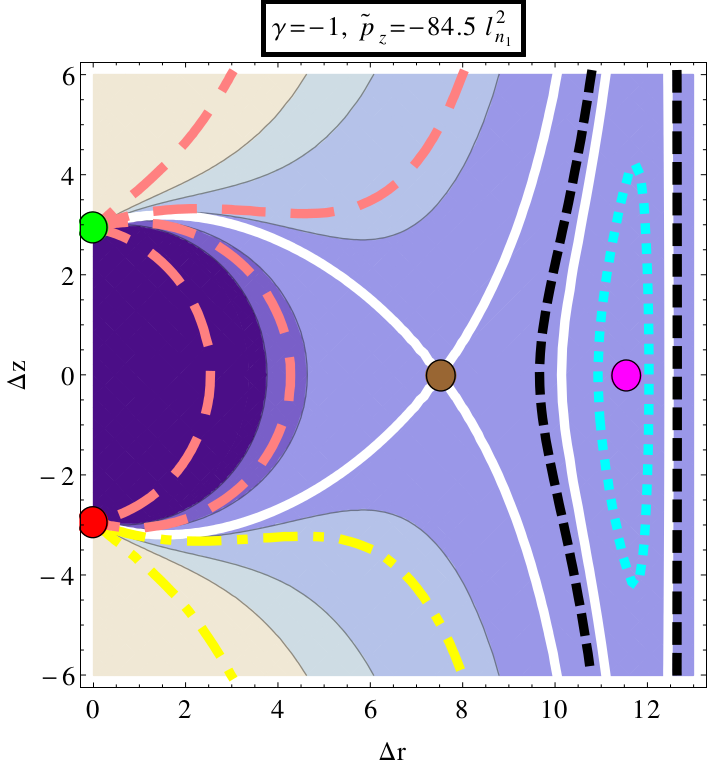}}
  \caption{(Color online.) Fixed points of the phase diagram when $\gamma = -1$. {\bf Green} dot: attractor point (see subsection \ref{ATTRACT}). {\bf Red} dot: repeller point (see subsection \ref{ATTRACT}). {\bf Brown} dot: Saddle point (see subsection \ref{SADDLE}). {\bf Magenta} dot: Elliptic fixed point (see subsection \ref{ELLIPTIC}). Refer to figures \ref{GammaGrtLessZero}-\ref{GammaMinusOne} for the color key for the special contours shown.}\label{GammaMinusOneFixedPoints}
 \end{figure}

Without loss of generality, take $n_2 < 0$. We will focus on the case $r_2(t=0) \le r_1 (t=0)$. Then $\tilde p_z \le 0$. Thus the relation \eno{MomentumTwoV} and the positivity of the radii imply $0 \le \Delta r<r_1$. The case of $r_2(t=0) > r_1(t=0)$ will just be the mirror of the analysis presented below. In particular, the phase space diagram will be the mirror image of figure \ref{GammaMinusOneFixedPoints}, reflected about $\Delta r =0$. Note that at $\gamma = \pm 1$, $\ell_{n_1} = \ell_{n_2} \equiv \ell_0$.

\subsection{Attractor/repeller \texorpdfstring{at  $\Big\{\Delta z = \pm {8\ell_0 / e}\,, \Delta r=0\Big\}$}{}}
\label{ATTRACT}
In this subsection we identify the attractor (repeller) fixed points of the $\gamma = -1$ phase diagram, marked in figure \ref{GammaMinusOneFixedPoints} with a green (red) dot. 

The equations of motion \eno{UnpertEOMs1} imply $\Delta \dot{r} = \Delta r\Delta z\, g(\Delta z,\,\Delta r)$ where $g$ is a non-singular function in the entire domain of $\left(\Delta z,\, \Delta r\right)$ except for the point $(0,0)$. Thus $\Delta \dot{r} = 0$ at $\Delta r=0$ and $\Delta z \neq 0$. Additionally assuming $r_1 \gg \ell_0$ at the fixed point described by $\Delta z \equiv \beta_z \ell_0/e$ and $\Delta r=0$, the equations of motion \eno{UnpertEOMs2} yield
\eqn{zDotAttractor}{
\Delta \dot{z} = {   \left(16 e^2 r_1^2-3 \beta_z^2 \ell_0^{2} \right)  \over 16 e^2 r_1^3 } \log \left({64 \over \beta_z ^2}\right)
+ {1  \over 8 r_1^3 }{\beta_z ^2 \ell_0^2  \over e^2}\left(3 \log \left({\ell_0 \over r_1 }\right) + 1\right) + O\left(\left({\ell_0 \over r_1 }\right)^5\right)\,.
}
Thus the leading order contribution to $\Delta \dot{z}$ vanishes when $\beta_z^2 = 64$, and the sub-leading contribution becomes vanishingly small, provided $r_1/\ell_0 \rightarrow \infty$ as $\Delta z \rightarrow \pm 8 \ell_0/e$ and $\Delta r \rightarrow 0$.
This proviso can indeed be (numerically) verified using \eno{EnergyTwoV}. Thus we conclude $\Big\{\Delta z = \pm 8\ell_0 /e\,, \Delta r=0\Big\}$ are fixed points of the equations of motion.\footnote{This evolution of the background solution for a head-on collision between co-axial vortices of opposite circulations is well known~\cite{Dyson1893,Shariff89}, and the late time stability of linearised perturbations in this case was studied in Ref.~\cite{Gubser:2014yma}, in the special limit when $\tilde p_z = 0$. Starting with initial conditions that lie in the basin of attraction, the radii of the vortices at late times is given by a large multiple of the length scale $\ell_0$, and tends to infinity linearly with time. 
 As the radii grow large, $s_p$ given by \eno{SpQpDef} tends to zero. 
The expression for conserved energy \eno{EnergyTwoV} in this late time limit indeed gives,
\eqn{DeltazLateTimes}{
\Delta z \xrightarrow{\rm late\ times} {8 \ell_0 \over e} \exp \left({\tilde\epsilon \over 2  \tilde\lambda n_1^2 r_1}\right) \approx {8 \ell_0 \over e}  \equiv  \Delta z_{\rm min}\,,
} 
where the sign of the energy fixes whether $\Delta z \rightarrow \Delta z_{\rm min}$ from above or below. }

The eigenvalues of the Jacobian matrix evaluated at the two fixed points 
 are given by 
\eqn{EigenAttRep}{
\lambda_J = \Big\{ \mp {e \over 2 \ell_0 r_1},\, \mp {  e \over 4 \ell_0 r_1} \Big\} \qquad {\rm at} \qquad  \Big\{\Delta z = \pm {8\ell_0 \over e}\,, \Delta r=0\Big\}\,,
}
where $r_1/\ell_0$ (which depends on $\tilde p_z$ and $\Delta r$ through \eno{MomentumTwoV}) tends to infinity, and we have imposed momentum conversation when evaluating the Jacobian matrix. Thus $\Delta z = 8\ell_0/e$ ($\Delta z=-8\ell_0/e$) is an attractor (repeller) fixed point as both the eigenvalues of the Jacobian evaluated at this point are real and negative (positive).

\subsection{Saddle point \texorpdfstring{at $ \Big\{\Delta z = 0 \,, \Delta r = (8-\delta_s) \ell_0 \Big\}$}{}
}
\label{SADDLE}
We now proceed to locate the lone saddle point in the $\gamma = -1$ phase diagram, marked in figure \ref{GammaMinusOneFixedPoints} with a brown dot.

Similar to the analysis in subsection \ref{ATTRACT}, we have $\Delta \dot{r} = 0$ when $\Delta z=0$ and $\Delta r \neq 0$. Differentiating the relation \eno{MomentumTwoV}, we obtain,
\eqn{r1DotSaddle}
{
\dot{r}_1 = \left(1-{r_1 \over \Delta r}\right) \Delta \dot{r}\,.
}
Thus for $\Delta r \neq 0$, $\Delta \dot{r}=0$ implies $\dot{r}_1=0$. 

In the simplifying limit $r_1 \gg \ell_0$, the equations of motion \eno{UnpertEOMs2} at $\Delta z = 0$ and $\Delta r \equiv \beta_s \ell_0$ yield
\eqn{zDotSaddle}{
\Delta \dot{z} =  {1 \over r_1} \log {64 \over \beta_s ^2} 
+{\beta_s   \ell_0  \over 2 r_1^2}   \log {64 \over \beta_s ^2} + \cdots\,,
}
Thus the leading and sub-leading contributions to $\Delta \dot{z}$ vanish when $\beta_s^2 = 64$. 
Since $\Delta r \ge 0$, $\beta_s = 8$ to leading order.
Away from the $r_1 \gg \ell_0$ limit, $\beta_s$ gets corrected to $\beta_s \equiv 8 - \delta_s$, where the first correction to $\delta_s = 0 $ is given by
\eqn{DeltaS}{
\delta_s \approx {32 \left(3 \log \left({r_1 / \ell_0}\right)- 5\right) \over \left({r_1 / \ell_o}\right)^2 + 4 \left({r_1 / \ell_o}\right) + 24 \log \left({r_1 / \ell_0}\right)-20} \,.
}
At even smaller $r_1$, sub-leading contributions to $\delta_s$ become important. An excellent estimate for $\delta_s$ which works for any $\alpha_r \equiv r_1/\ell_0 \geq 8$ is 
 \small
\eqn{DeltaS2}
 {
\delta_s \approx\frac{2 (\alpha_r -8) (\alpha_r -4) \left((\alpha_r -8) \alpha_r  K\left(-\frac{1}{16} (\alpha_r -8) \alpha_r \right)+16 \log (\alpha_r )-2 \alpha_r  \log ((\alpha_r -8) \alpha_r )\right)}{\alpha_r  \left((\alpha_r -4) \left((\alpha_r -8) K\left(-\frac{1}{16} (\alpha_r -8) \alpha_r \right)-4 \log (\alpha_r -8)+4\right)-4 (\alpha_r -8) E\left(-\frac{1}{16} (\alpha_r -8) \alpha_r \right)\right)}\,.
 }
\normalsize
There are no real solutions for $\delta_s$ for $\alpha_r <8$.  In fact, numerics show no fixed point exists for $\alpha_r \lesssim 11$. In terms of the conserved momentum, this corresponds to $\tilde p_z \gtrsim - 55 n_1 \ell_0^2$. 

The eigenvalues of the Jacobian matrix evaluated at the fixed point are of the form $\pm \sqrt{b c}$ where $b$ and $c$ are the off-diagonal elements of the matrix, since the diagonal elements are zero. Thus as long as the product of the off-diagonal elements is positive, the traceless matrix has a positive and a negative eigenvalue, signalling an unstable saddle point. 

In the limit $r_1 \gg \ell_0$, the eigenvalues are 
\eqn{EigenSaddle}{
\lambda_J = \pm {1 \over 4 \ell_0 r_1}\,.
}
Away from the $r_1 \gg \ell_0$ limit, the product of the off-diagonal elements is still (numerically) found to be positive for $r_1 \gtrsim 11 \ell_0$.  Thus for $\tilde p_z \lesssim -55 n_1 \ell_0^2$, there exists a saddle point at $ \Big\{\Delta z = 0 \,, \Delta r = (8-\delta_s) \ell_0 \Big\}$ where $\delta_s$ is given by \eno{DeltaS2}. The brown dot in figure \ref{GammaMinusOneFixedPoints} was plotted on the phase space by employing \eno{DeltaS2}. Vortices at this fixed point are in an  unstable nested  configuration.\footnote{Note that at this saddle point, at large (negative) momentum $\tilde p_z \ll -n_1 \ell_0^2$, the dispersion relation given by
\eqn{DispRelSaddle}
{
\tilde \epsilon = -\frac{\tilde\lambda n_1\tilde p_z}{4 \ell_0} 
-\frac{32 \tilde\lambda \ell_0^3 n_1^3 }{\tilde p_z} \left(2+\log \left({64 \ell_0^4 n_1^2 \over \tilde p_z^2}\right)\right) + \cdots
}
is linear to $O(1/\tilde p_z)$.}

\subsection{EFP  \texorpdfstring{at $ \Big\{\Delta z = 0 \,, \Delta r = (1-\delta_{\rm EFP})r_1 \Big\}$}{}}
\label{ELLIPTIC}
Finally, in this subsection we identify the elliptic fixed point (EFP) of the $\gamma = -1$ phase diagram, marked in figure \ref{GammaMinusOneFixedPoints} as a magenta dot. 

The equations of motion \eno{UnpertEOMs1} imply $\Delta \dot{r}=\dot{r}_1=0$ for $\Delta z=0$ and $\Delta r \neq 0$.
Define $\Delta r \equiv (1-\delta_{\rm EFP}) r_1$ for small $\delta_{\rm EFP}>0$, and then
\eqn{zDotEfp}
{
\Delta \dot{z} = {1 \over r_1} \left( -{1  \over \delta_e}\log \left(\frac{r_1  \delta_{\rm EFP} }{\ell_0}\right) + \left(2 \pi - \log {r_1 \over \ell_0 }\right) + {\pi  \delta_{\rm EFP}^2 \over 2}  \right) + O(\delta_{\rm EFP}^4)\,.
}
Thus for $\delta_{\rm EFP} \ll 1$,  $\Delta \dot{z}$ vanishes when
\eqn{DeltaEEfp}
{
\delta_{\rm EFP} \approx  {W_0\left({ \log (r_1 / \ell_0) - 2 \pi \over   {r_1 / \ell_0} }\right) \over  \log (r_1 / \ell_0) - 2 \pi  }\,,
}
where $W_0(y)$ is the principal branch of the real valued Lambert-W function, which takes values between $-1$ and $\infty$ for $y \geq -1/e$ and is positive valued for positive $y$. This means the fixed point exists only if 
\eqn{r1BoundEfp}
{
r_1 > e W_0\left( e^{2 \pi -1}\right) \ell_0 \approx 11 \ell_0 \,.
}
The estimate \eno{DeltaEEfp} agrees with numerics with great accuracy for radii away from the bound given in \eno{r1BoundEfp}. However, going ahead and making use of \eno{DeltaEEfp} when the bound is saturated, we deduce a bound on the rescaled momentum. We find, 
\eqn{MomBoundEfp}
{ \tilde p_z \lesssim \tilde p_{z, {\rm EFP}} =  -{e^2 n_1 \ell_0^2 \over 2}  \left(-1 + W_0\left(  e^{2\pi-1}\right)^2 \right) \approx -53 n_1 \ell_0^2\,.
}
Numerically the bound was found to be near $-55 n_1 \ell_0^2$, which is not very far from the analytic estimate.

The Jacobian matrix evaluated at the fixed point is traceless. Its eigenvalues are given by
\eqn{EigenEfp}
{
\lambda_J \approx \pm {i \sqrt{3 \pi } \over \sqrt{\delta_{\rm EFP} } r_1^2} \sqrt{  1 - \log {r_1\:\delta_{\rm EFP}  \over \ell_0}  } \,,
}
where $\delta_{\rm EFP}$ is given by \eno{DeltaEEfp}. Note that in the domain \eno{r1BoundEfp} in which $\delta_{\rm EFP}$ is defined, $0<\delta_{\rm EFP} < e \ell_0/r_1$, thus the eigenvalues \eno{EigenEfp} are purely imaginary. We conclude $ \Big\{\Delta z = 0 \,, \Delta r = (1-\delta_{\rm EFP})r_1 \Big\}$ is an EFP. The magenta dot in figure \ref{GammaMinusOneFixedPoints} was plotted by making use of \eno{DeltaEEfp}. Orbits in phase space about the EFP correspond to pseudo-leapfrogging vortices~\cite{Borisov2013}.

\section{Initial conditions and periodic behavior \texorpdfstring{ for $\gamma = 1,0,-1$}{}}
\label{PhaseBoundaries}

We present the set of initial conditions corresponding to periodic behavior, for the special cases $\gamma = 1, 0$ and $-1$.  The boundaries between different phases correspond to one of the limiting behaviors (chasing, nested, crushed), which are deterimined by the equations of motion.  These must in general be solved numerically; for certain limiting values of $\gamma$, however, the system can be solved analytically.  

We will express the initial conditions in terms of the quantities
\be
r = r_1\, , \qquad x = r_2/r_1
\ee
evaluated at the point when $\Delta z = 0$.  While this notation is more intuitive than that used in Ref.~\cite{Borisov2013}, which discusses the region of periodic behavior in terms of the bifurcation complex of the Hamiltonian system, the disadvantage is that more than one set of initial conditions may correspond to the same trajectory. 

\begin{enumerate}

\item For $\gamma = 1$, we may also assume that $\chi = 1$, so that $\ell_{n_1} = \ell_{n_2} = \ell_0$ are equal, and the region corresponding to periodic behavior is given in figure \ref{fig:GammaOne}.  The upper boundary corresponds to the chasing limit, and is determined by
\be
\log \frac{r}{e \ell_0} \approx \frac{\left(x\log(x) - 2\sqrt{x}Q_0\left(\frac{1-x}{\sqrt{x}}\right) - \sqrt{\frac{1+x^2}{2}}\log\left(\frac{1+x^2}{2}\right)\right)}{\sqrt{2 + 2x^2}-x-1}
\ee
The lower boundary is given by the nested limit, and corresponds to 
\eqn{}{
\log \frac{r}{\ell_0} & = \left(\frac{x}{x-1}\right)\Bigg\{\! \left(\sqrt{x}-\frac{1}{x^{3/2}}\right)Q_{0}\left(\frac{1-x}{\sqrt{x}}\right) + \frac{(1+x)(1+x^2)}{x^2}Q'_{0}\left(\frac{1-x}{\sqrt{x}}\right)  \cr 
 & \qquad \qquad \qquad +  \frac{1}{x}\log{x}\Bigg\}
}
\begin{figure}[t]
\centering
\includegraphics[scale = 0.8]{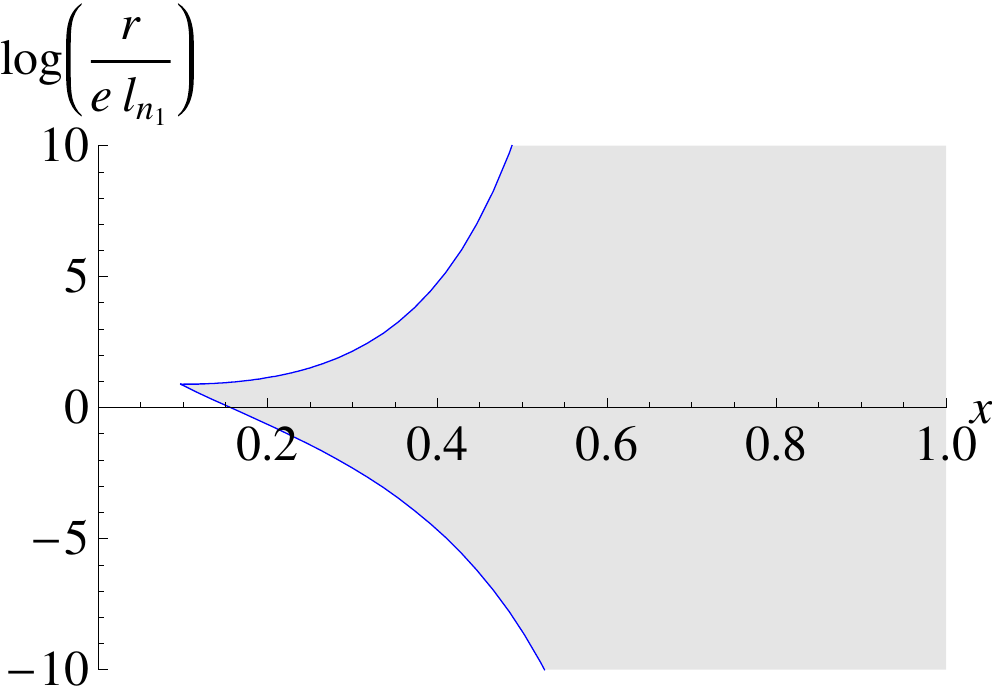}
\caption{The allowed region for the initial conditions $x, \log\left(\frac{r}{e \ell_{n_1}}\right)$.  Periodic solutions exist within the shaded blue region.  The upper boundary represents the chasing limit, and the lower boundary represents the nested limit.}
\label{fig:GammaOne}
\end{figure}
Without loss of generality, for $\gamma = 1$ we need only consider the region $x < 1$.  No periodic solutions exist for values of $x \lesssim 0.097$.  As $x \to 1$, however, all values of $r$ lead to periodic (leapfrogging) behavior.

\item For $\gamma = 0$, we must consider all values $x \neq 1$.  The region corresponding to periodic solutions is shown in figure \ref{fig:GammaZero}.  For $x < 1$, the upper boundary in parameter space is given by the nested limit, which for $\gamma = 0$ becomes
\be
\log \frac{r}{\ell_{n_1}} = - \frac{\sqrt{x}Q_0 \left(\frac{1-x}{\sqrt{x}}\right) + (1 + x)Q'_0 \left(\frac{1-x}{\sqrt{x}}\right)}{x^2}\,.
\ee
When $x$ is small, the maximum value of $\log\left(\frac{r}{\ell_{n_1}}\right)$ is given by $2\pi$, and as $x \to 1^{-}$, $\log\left(\frac{r}{\ell_0}\right)_{max} \to \infty$.
\begin{figure}[t]
\centering
\includegraphics[scale = 0.8]{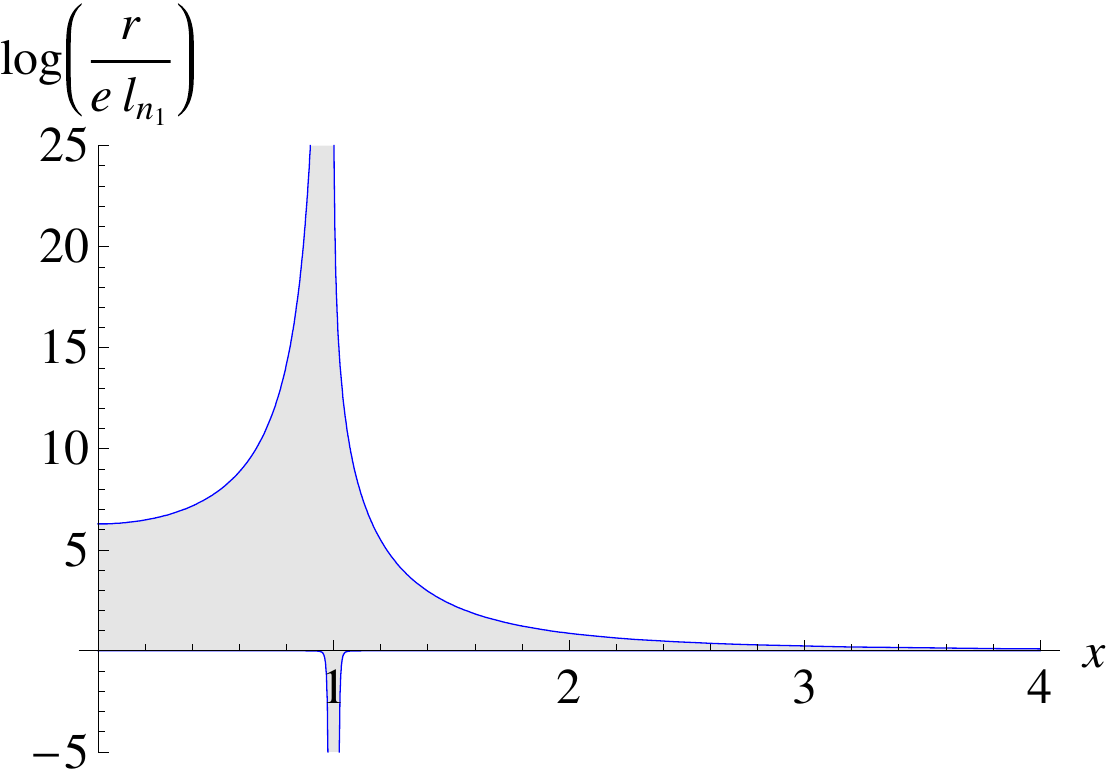}
\caption{The allowed region for the initial conditions $x, \log\left(\frac{r}{\ell_{n_1}}\right)$, for $\gamma = 0$.  The blue region corresponds to periodic solutions, while the upper left boundary and both the lower boundaries are in the nested limit, and the upper right is in the crushed limit.}
\label{fig:GammaZero}
\end{figure}
For $x > 1$, the upper boundary is given by the crushed limit, and obeys
\be
\log \frac{r}{\ell_{n_1}} = -\frac{4}{x^{3/2}}Q_0 \left(\frac{x-1}{\sqrt{x}}\right)\,.
\ee
This approaches zero as $x \to \infty$.  For all values of $x$, the lower boundary of the allowed parameter space corresponds to the nesting limit, though not at the initial conditions, and so this boundary must be found numerically.

\item For $\gamma = -1$, once again we may set $\chi = 1$, and once again it suffices to consider only the region where $x \leq 1$.  The limits must be found numerically and the phase diagram is depicted in figure \ref{fig:GammaMinusOne}.
\begin{figure}[t]
\centering
\includegraphics[scale = 0.8]{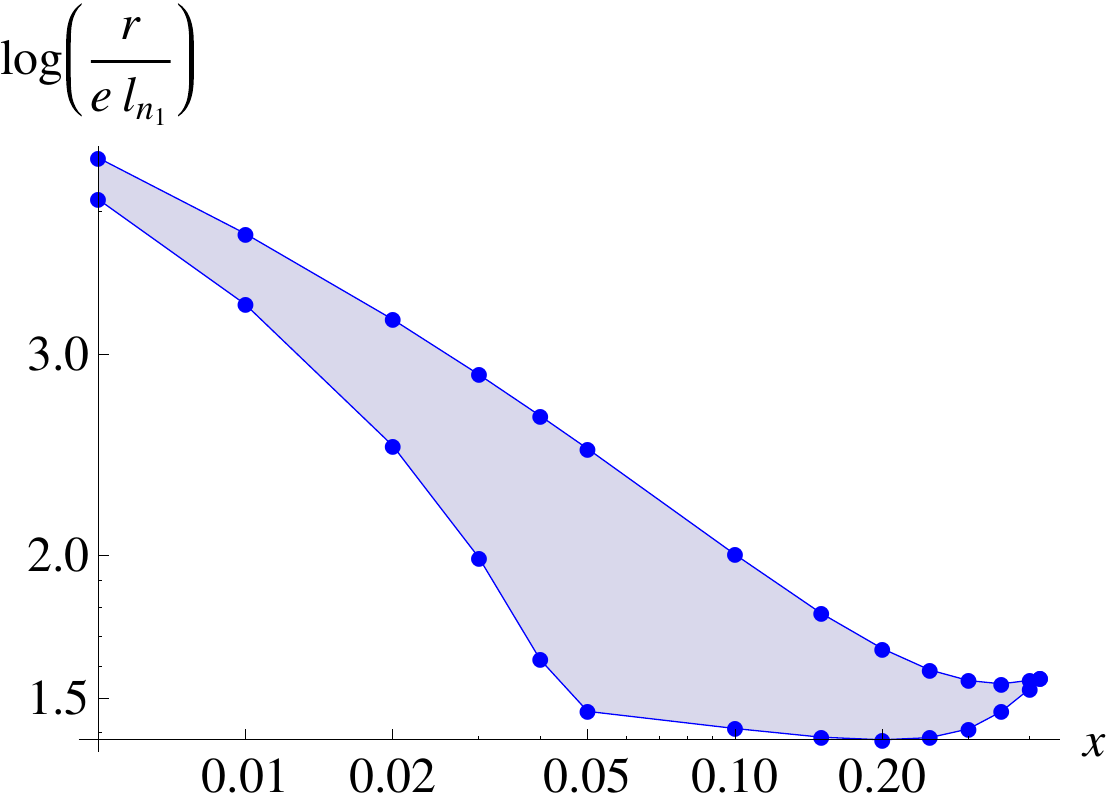}
\caption{The approximate allowed region for $\gamma = -1$, with lines interpolating between known data points.  Periodic pseudo-leapfrogging solutions exist within the shaded blue region.  The upper boundary represents the chasing limit, and the lower boundary represents the nested limit above $x \sim 0.05$ and the chasing limit below.}
\label{fig:GammaMinusOne}
\end{figure}
Note that for $x \gtrsim 0.42$, no periodic solutions are possible.  The region where $x \to 0$ corresponds to the elliptic fixed point discussed in section \ref{ELLIPTIC}.

\end{enumerate}

\section{\texorpdfstring{$S_{ij2}$}{Sij2} functions}
\label{SIJ2FNS}

The $m$-dependent $S_{ij2}$ functions defined by the integrals in \eno{Sij2Def} can be written down explicitly as series expansions in hypergeometric functions, as follows (for $r_2 \leq r_1$)
 \eqn{Sij2n}{
\hskip-0.1in S_{zz2} &= - \sum_{n=0}^\infty \Bigg[ \frac{(m+2 n) \Gamma \left(n+\frac{1}{2}\right) \Gamma \left(m+n+\frac{1}{2}\right)}{ \Gamma (n+1) \Gamma (m+n+1)} {1 \over r_1}\left(r_2 \over r_1 \right) ^{m+2 n} \times  \cr 
\hskip-0.1in & \left(m \; _2F_1\left(n+\frac{1}{2},m+n+\frac{1}{2};\frac{1}{2};-\frac{\Delta z ^2}{r_1^2}\right) + (2 n+1) \; _2F_1\left(n+\frac{3}{2},m+n+\frac{1}{2};\frac{1}{2};-\frac{\Delta z ^2}{r_1^2}\right)\right) \Bigg] \cr
\hskip-0.1in S_{rr2} & = \sum_{n=0}^\infty \frac{4 \Gamma \left(n+\frac{3}{2}\right) \Gamma \left(m+n+\frac{3}{2}\right) \, }{\Gamma (n+1) \Gamma (m+n+1)}\;  {}_2F_1\left(n+\frac{3}{2},m+n+\frac{3}{2};\frac{1}{2};-\frac{\Delta z ^2}{r_1^2}\right){1 \over r_1 } {\left({r_2 \over r_1}\right) ^{m+2 n+1} },
 }
where $\Gamma (m)$ is the Gamma function. For $r_2 > r_1$, switch $r_1 \leftrightarrow r_2$ in the expressions above. 
The $S_{zr2}$ function is given by  (for $r_2 \leq r_1$) 
\eqn{Szr2n1}{
\hskip-0.1in S_{zr2} & = \sum_{n=0}^\infty  \Bigg[ \frac{ 4^{1-n} (2 n+1)  \Gamma (2 n)  \Gamma \left(m+n+\frac{3}{2}\right) }{\pi^{-1/2} \Gamma (n) \Gamma (n+1) \Gamma (m+n+1)} {\Delta z \over r_1^2} \left({r_2 \over r_1} \right)^{m+2 n+1} \times \cr
 & \left(m \; _2F_1\left(n+\frac{3}{2},m+n+\frac{3}{2};\frac{3}{2};-\frac{\Delta z ^2}{r_1^2}\right)+(2 n+3) \; _2F_1\left(n+\frac{5}{2},m+n+\frac{3}{2};\frac{3}{2};-\frac{\Delta z^2}{r_1^2}\right)\right) \Bigg],
 }
while for $r_2 > r_1$,
\eqn{Szr2n2}{
\hskip-0.1in S_{zr2} = -\sum_{n=0}^\infty \frac{4  (m+2 n) \Gamma \left(n+\frac{3}{2}\right)  \Gamma \left(m+n+\frac{3}{2}\right)}{ \Gamma (n+1) \Gamma (m+n+1)}\; _2F_1\left(n+\frac{3}{2},m+n+\frac{3}{2};\frac{3}{2};-\frac{\Delta z ^2}{r_2^2}\right){  \Delta z \over r_2^2}  \left({r_1 \over r_2}\right)^{m+2 n}
}
and the $S_{rz2}$ function is given by $S_{zr2} (r_1 \leftrightarrow r_2, z_1 \leftrightarrow z_2)$.

For most values of $r_1$, $r_2$ and $\Delta z$, summing over a small range of $n$ in the expressions above already yields excellent estimates, {\it i.e.}~the series converge fairly quickly.

\bibliographystyle{ssg}
\bibliography{vortices}
\end{document}